\documentclass{cernyrep}
\usepackage{graphicx,amsmath,amsfonts,amscd,amsthm,amssymb,eucal,psfrag,color}
\usepackage{texnames}
\usepackage[T1]{fontenc}
\usepackage{varwidth}

\usepackage{subfig}

\usepackage{xcolor}
\usepackage[bookmarks, colorlinks=true, linktoc=page, pdftex, linkcolor=black, citecolor=black, urlcolor=blue]{hyperref}
\usepackage[ocgcolorlinks]{ocgx2}

\usepackage{listings}
\definecolor{codegreen}{rgb}{0,0.6,0}
\definecolor{codegray}{rgb}{0.5,0.5,0.5}
\definecolor{codepurple}{rgb}{0.58,0,0.82}
\definecolor{backcolour}{rgb}{0.95,0.95,0.92}
 
\lstdefinestyle{mystyle}{
   frame=tb,
    commentstyle=\color{codegreen},
    keywordstyle=\color{red},
    numberstyle=\tiny\color{codegray},
    stringstyle=\color{codepurple},
    basicstyle=\footnotesize,
    basicstyle=\linespread{0.8}\footnotesize,
    breakatwhitespace=false,         
    breaklines=true,
    postbreak=\mbox{\textcolor{red}{$\hookrightarrow$}\space},
    captionpos=b,                    
    keepspaces=true,                 
    numbers=left,                    
    numbersep=5pt,                  
    showspaces=false,                
    showstringspaces=false,
    showtabs=false,                  
    tabsize=1
}
 
\usepackage{fancyhdr}
\fancyhfoffset{4 mm}
\fancypagestyle{ARTTITLE}{%
\fancyhf{} 
\lhead{\small{Proceedings of the 2018 CERN--Accelerator--School course on\\
\it{Numerical Methods for Analysis, Design and Modelling of Particle Accelerators}, Thessaloniki, (Greece)}} 
\lfoot{Available online at \url{https://cas.web.cern.ch/previous-schools}}
\rfoot{\thepage\hspace*{3mm}}
 
}
\renewcommand{\headheight}{22.38pt}

\begin{document}
\title{Imperfections and corrections}

\author{R.~Tom\'as, 
  X. Buffat, J. Coello, E. Fol and L. Malina} 
\institute{CERN, CH 1211 Geneva 23, Switzerland}

\maketitle
\thispagestyle{ARTTITLE}

\begin{abstract}
  The measurement and correction of optics parameters has been a major concern since the advent of strong focusing synchrotron accelerators. A review of typical imperfections in accelerator optics together with measurement
  and correction algorithms is given with emphasis on numerical implementations. Python examples are shown
  using existing libraries when possible.
  
\end{abstract}


\section{Introduction}\label{intro}

\begin{figure}
\includegraphics[height=7.93cm, angle=-90]{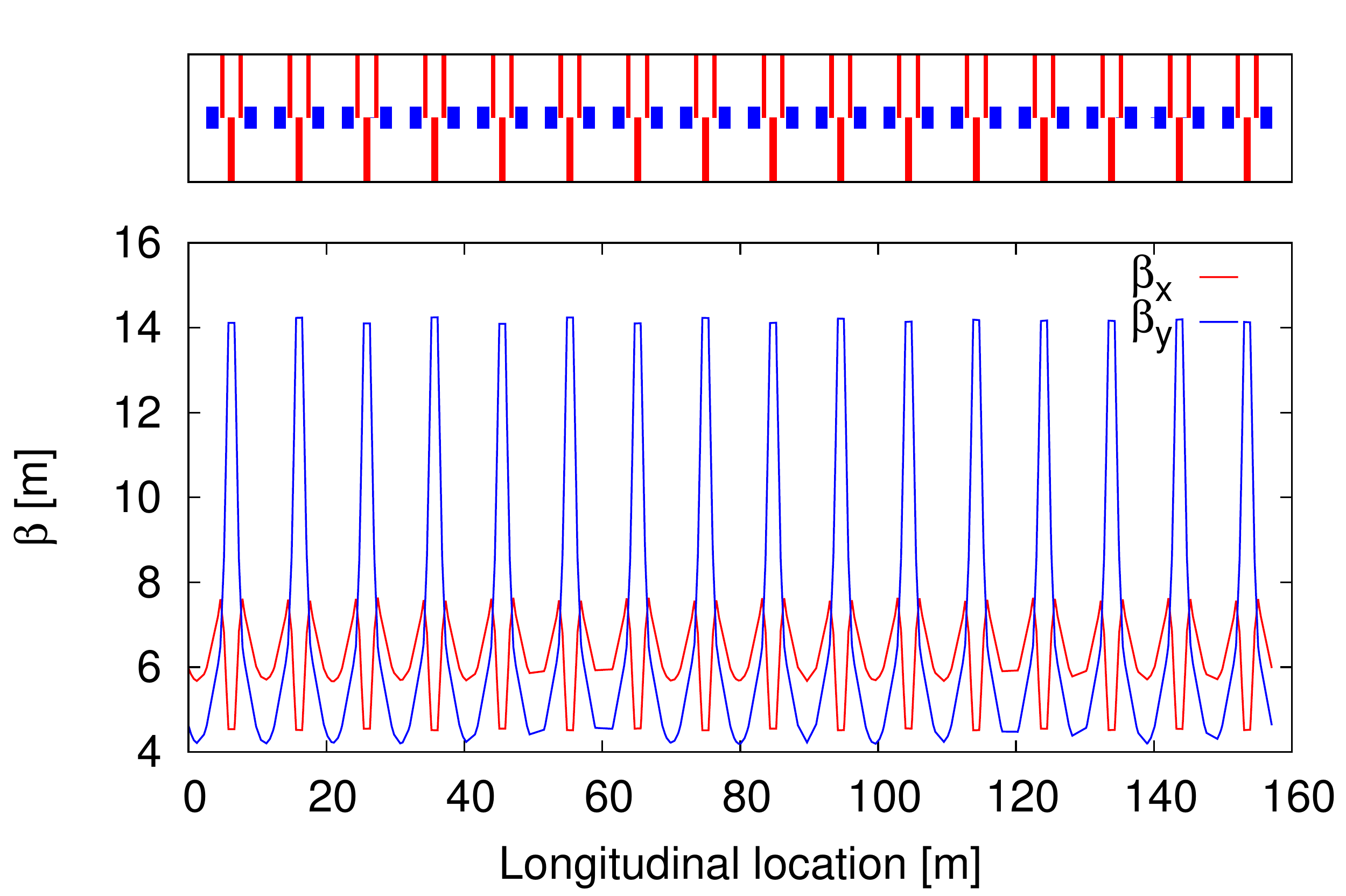}
\includegraphics[height=8cm, angle=-90]{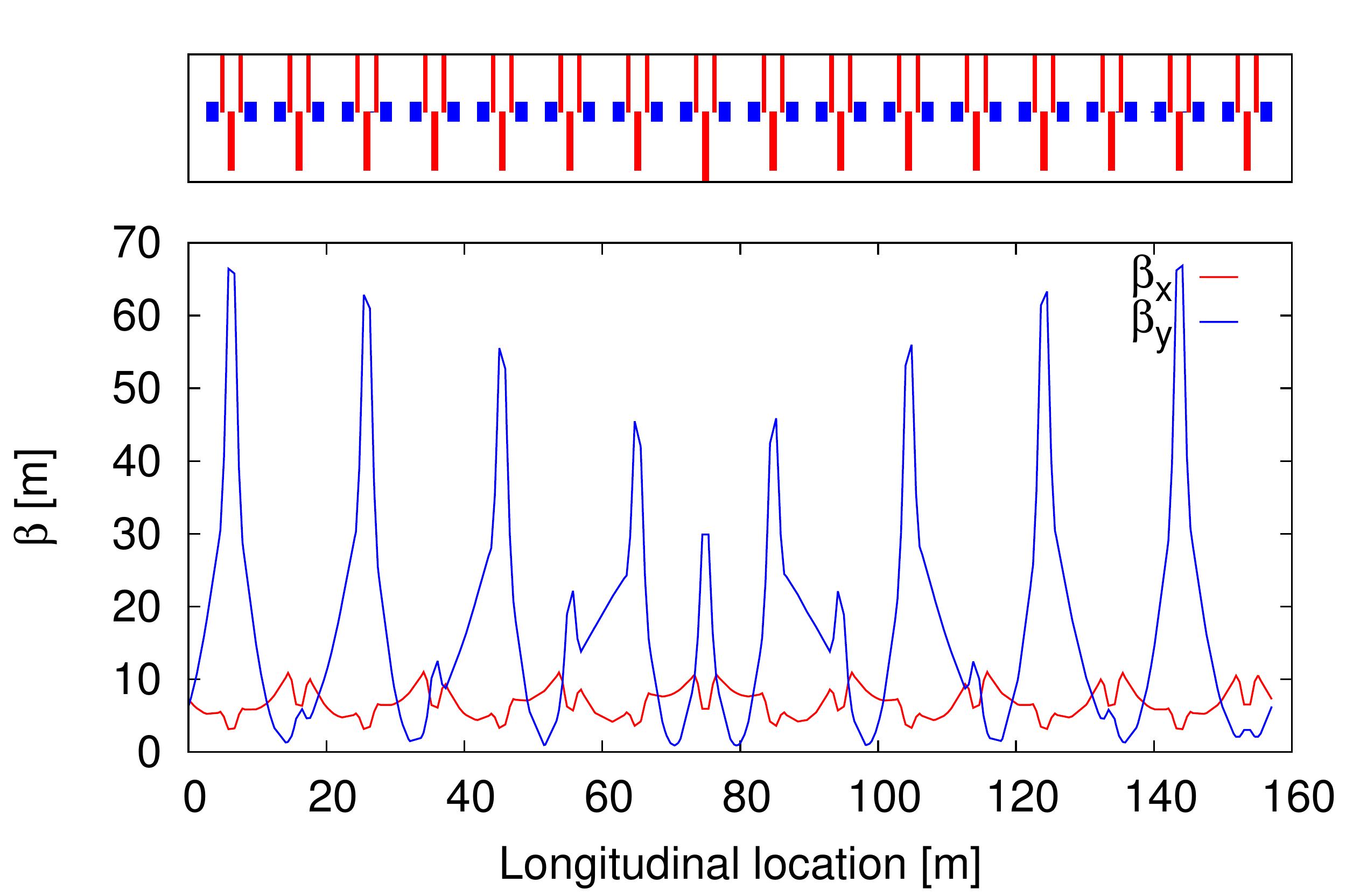}
\caption{Design $\beta$ functions of the CERN Proton Synchrotron Booster
  featuring a triplet lattice (left) and the same lattice
  with a 10\% gradient perturbation in the 8$^{\rm th}$ defocusing quadrupole (right).\label{psb}}
\end{figure}

Imperfections in accelerator lattices cause beam parameters
to deviate from design. An illustration is shown in Fig.~\ref{psb},
where ideal and perturbed $\beta$ functions are shown. The perturbation
assumed is simply a 10\% gradient error in the 8$^{\rm th}$ defocusing quadrupole. This causes large relative deviations in $\beta$ functions of up to 500\%
with respect to the design value. This is usually called $\beta$-beating and represented by $\Delta\beta/\beta$.

\begin{figure}
\centering
\includegraphics[height=10cm, angle=-90]{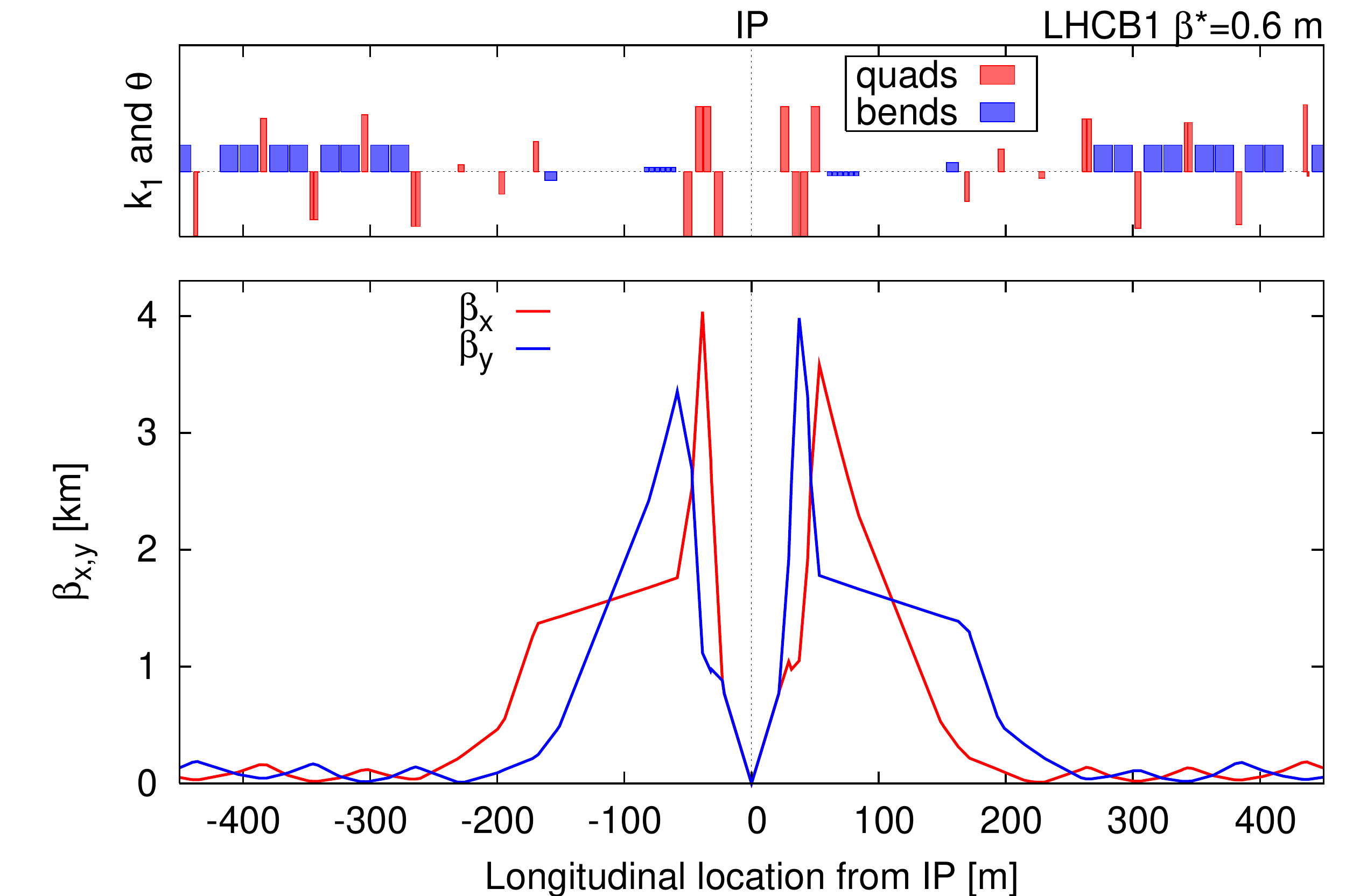}
\caption{Optics functions in the LHC IR for a $\beta$ function
  at the interaction point of 60~cm.\label{LHCIR}}
\end{figure}

Perturbations from field imperfections and misalignments became a concern along 
with the conception of the strong focusing theory in 1957~\cite{courant}.
However, the assumed approach was to specify design tolerances that would not impact
machine performance. For example in~\cite{courant} it is envisaged that with 1\%~rms gradient errors
{\it any particular machine would be unlikely} to have more than 8\% peak $\beta$-beating. 
At that time they did not foresee the great developments in optics
that would push $\beta$ functions to very large values, e.g., in the vicinity
of collision points of collider accelerators. The LHC Interaction Region (IR) optics is shown in Fig.~\ref{LHCIR} as an illustration of optics designs
reaching $\beta$ functions of several km.

\begin{figure}\centering
\includegraphics[height=9.5cm, angle=-0]{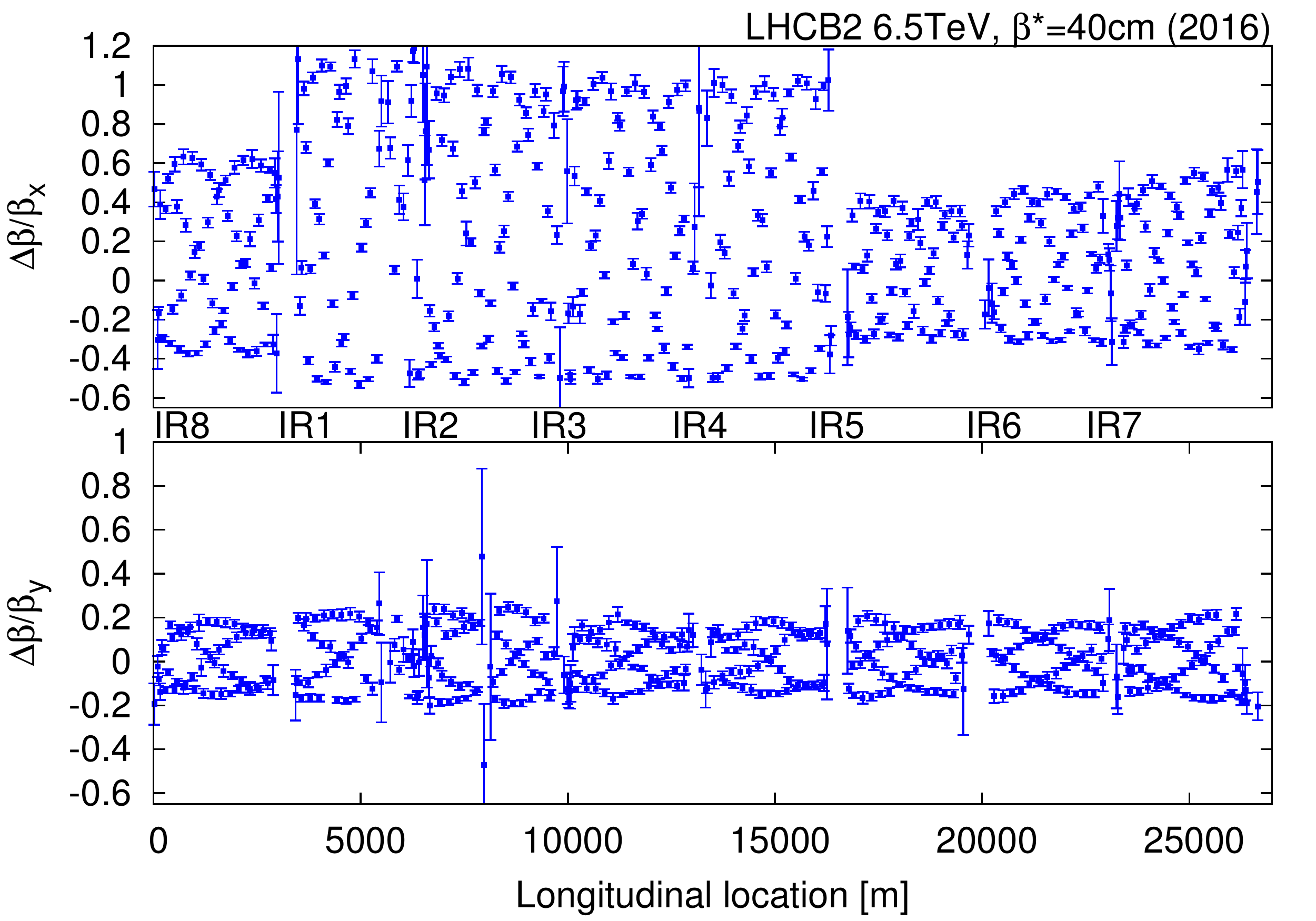}
\caption{$\beta$-beating measured in the LHC commissioning in 2016 with a $\beta$ function at the interaction point of 40~cm.\label{LHCbeat}}
\end{figure}

\begin{figure}\centering
\includegraphics[trim = 0mm 0mm 0mm 0mm, clip,height=7.1cm, angle=-0]{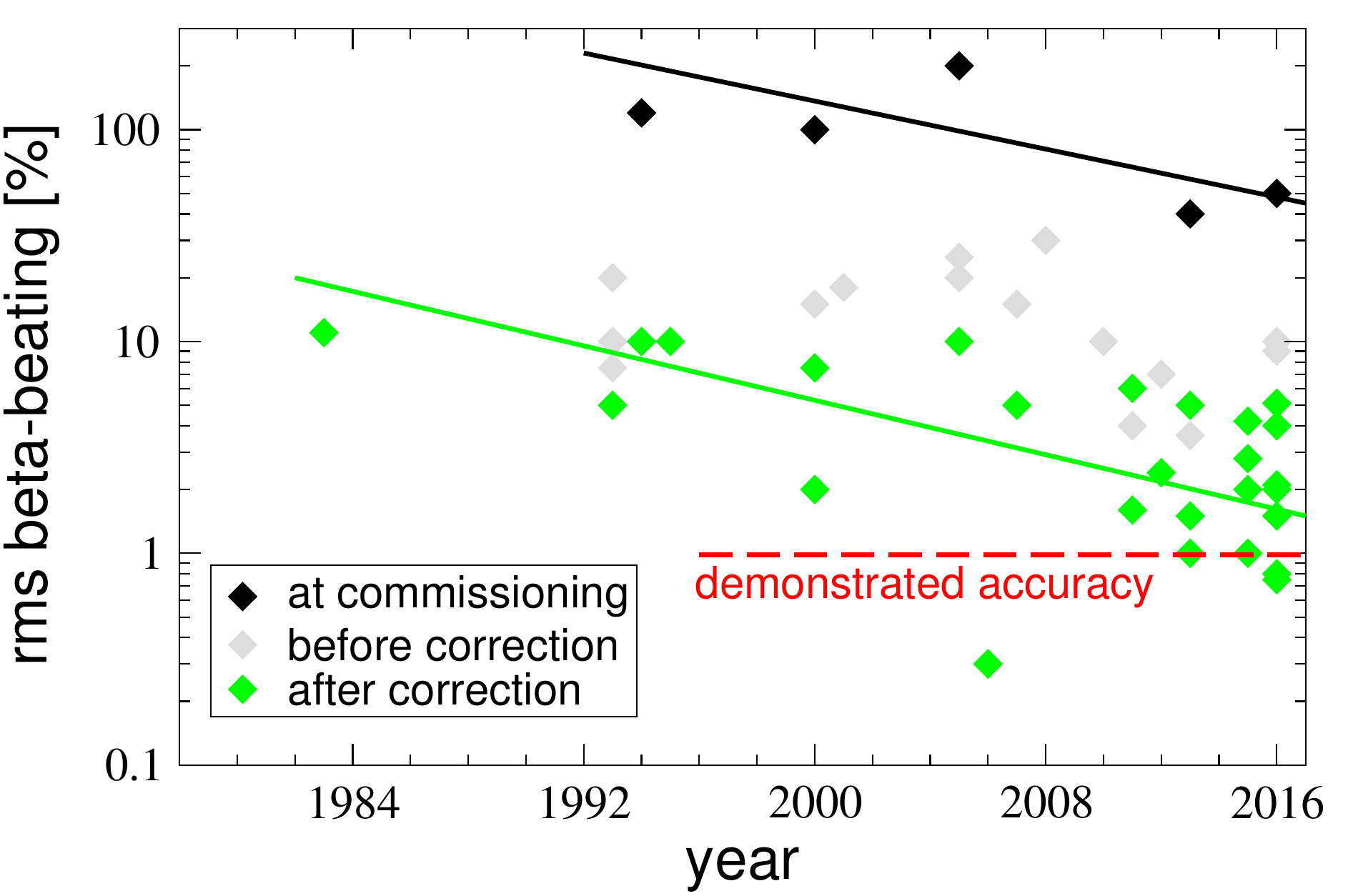}
\caption{Measured or inferred $\beta$-beating versus time for many circular accelerators as found in the bibliography of this paper. Three stages are differentiated: (i) during commissioning when magnet powering mistakes are expected, (ii) after fixing these mistakes but before careful optics corrections and (iii) after optics corrections. Taken from~\cite{review}.\label{fig:beating}}
\end{figure}

Modern accelerators have experienced $\beta$-beating values above
100\%~\cite{Yocky,1st,record}
in the initial commissioning phases.
Figure~\ref{LHCbeat} shows the initial $\beta$-beating measured in the LHC commissioning in 2016 with a peak value of 120\%.
The optics errors need to be corrected
below specified tolerances for safe and efficient operation.
The development of the optics measurement and correction techniques is
illustrated by the evolution of the $\beta$-beating over time for many circular accelerators, see Fig.~\ref{fig:beating}.

The techniques to measure and correct optics are
described in the following with special emphasis on
analysis algorithms and computing aspects.
Section~\ref{sec2} gives the requirements to run the
code examples below. Section~\ref{sec3} describes
the most important imperfections present in accelerator lattices.
Section~\ref{sec4}~describes the key particle dynamics
used in optics measurements.
Section~\ref{sec5} reports on the most used measurement techniques
and data analysis techniques.
Section~\ref{sec6} is an interlude devoted to the Farey sequences
and how they can be used to describe the resonance diagram.
Section~\ref{sec7} reports on optics correction techniques.

\section{Requirements for code examples}\label{sec2}
Code examples below require \href{https://www.python.org/}{Python}.
  The freely available and open source operative system \href{https://www.ubuntu.com/}{Ubuntu} has Python by default. The required plotting and numerical libraries can be installed with, e.g., the following shell command: 
  \begin{lstlisting}[language=Python]
   python -m pip install --user numpy scipy matplotlib ipython  jupyter pandas sympy nose scikit-learn 
  \end{lstlisting}
  Alternatively it is also possible to install  \href{https://www.anaconda.com/download/#linux}{Anaconda} which is a very complete Python free distribution with the required scientific packages.

\section{Accelerator elements and their imperfections}\label{sec3}

\subsection{Dipole}
\begin{figure}\centering
\includegraphics[height=10cm, angle=-90]{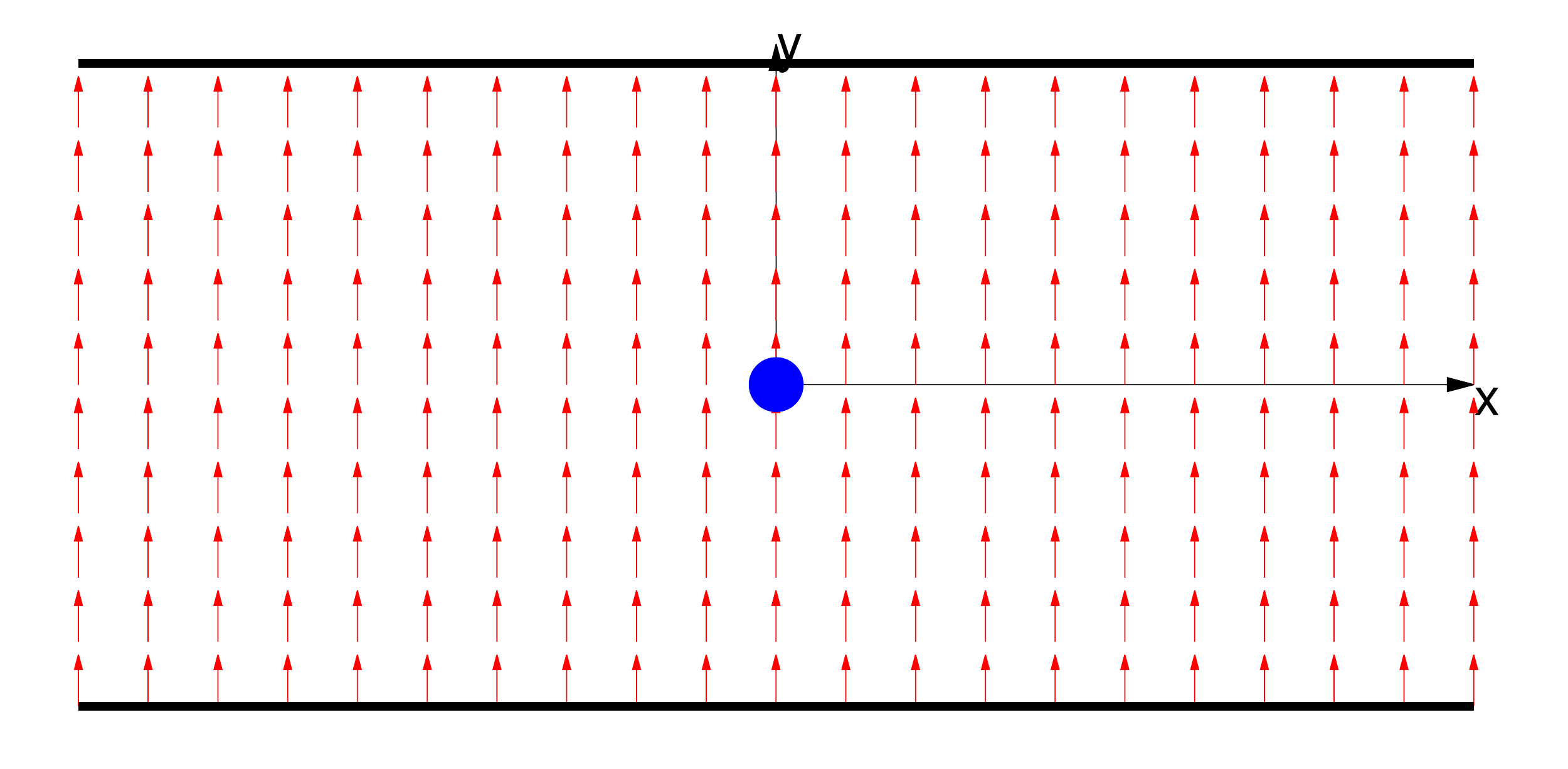}
\caption{The dipole magnetic field. The beam is represented in
  the center with a blue dot traveling perpendicular to the field. \label{dipole}}
\end{figure}

The simplest magnetic element in an accelerator
is the dipole, which provides an homogeneous field as shown
in Fig.~\ref{dipole}.

\begin{figure}\centering
\includegraphics[height=10cm, angle=-90]{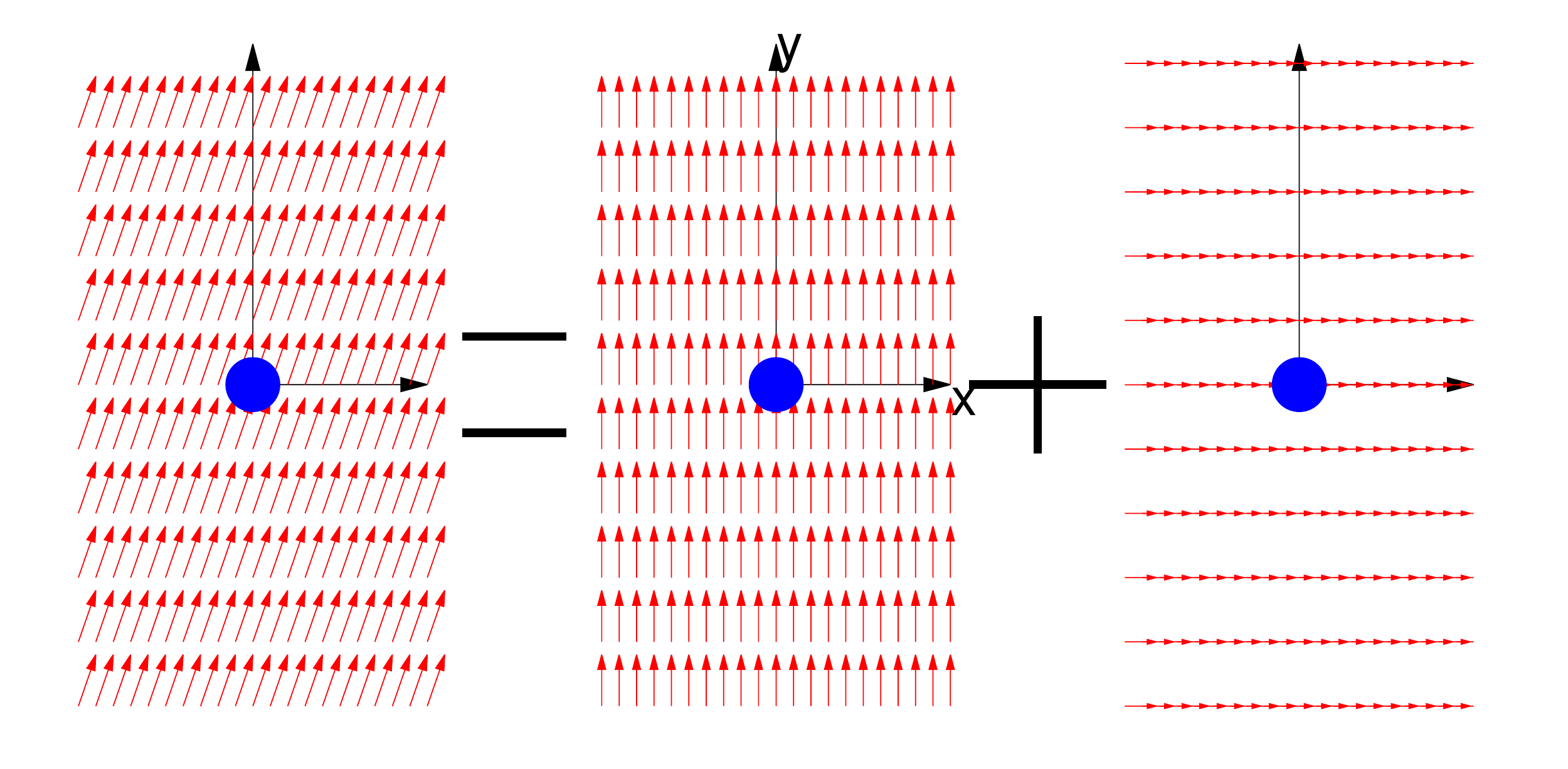}
\caption{A tilted dipolar magnetic field is seen as the
  sum of two orthogonal magnetic fields.\label{dipoletilt}}
\end{figure}

The dipole features two main imperfections: a strength error
and a tilt of the field around the beam axis.
The tilt error is illustrated in Fig.~\ref{dipoletilt}
and can be interpreted as another dipole
with a field orthogonal to the ideal dipole.

Therefore dipole errors are seen as unwanted angular deflections
in both transverse planes that distort the reference trajectories
into  new closed orbits.
Assuming $\theta_i$ to be  unwanted angular deflections
the closed orbit is given by
\begin{equation}
CO(s)=\frac{\sqrt{\beta(s)}}{2\sin\pi Q}\sum_i \sqrt{\beta_i}\theta_i\cos(\pi Q-|\phi(s)-\phi_i|)\ ,
\end{equation}
where $s$ denotes the longitudinal location around the ring, $Q$ is the tune and $\phi$ is the betatron phase advance.
The denominator $\sin(\pi Q)$  makes closed orbit to diverge at the integer resonance $Q\  \in\ \mathbb{N}$.
The effect of longitudinal misalignments is
briefly described in Section~\ref{longmisec}.
Another source of orbit errors is offset quadrupoles which is described
in the following section. 

\subsection{Quadrupole}

\begin{figure}\centering
\includegraphics[height=6cm, angle=-90]{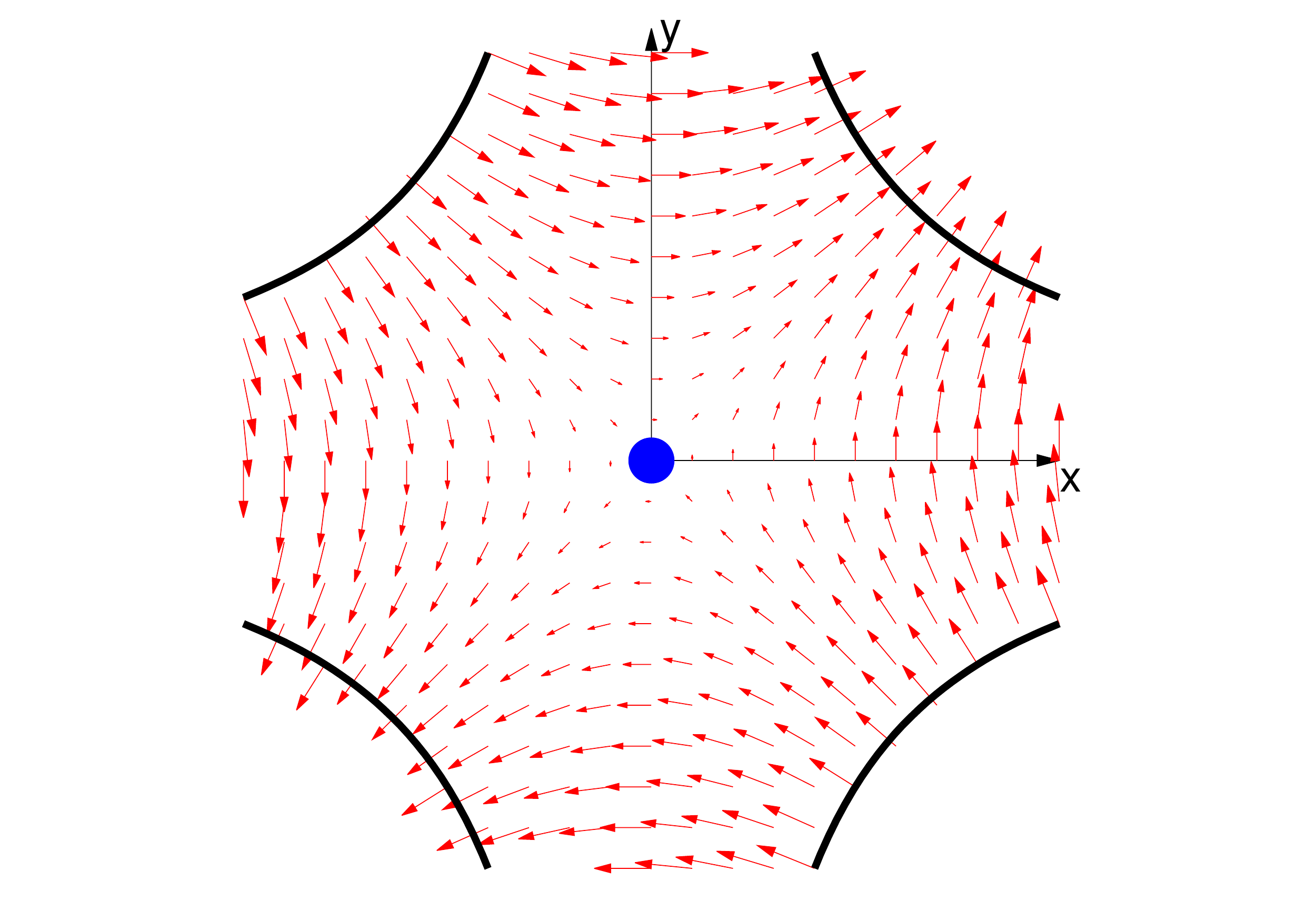}
\put(-45,-20){\color{red} $\vec{B}$}\hspace{-0.5cm}
\includegraphics[height=6cm, angle=-90]{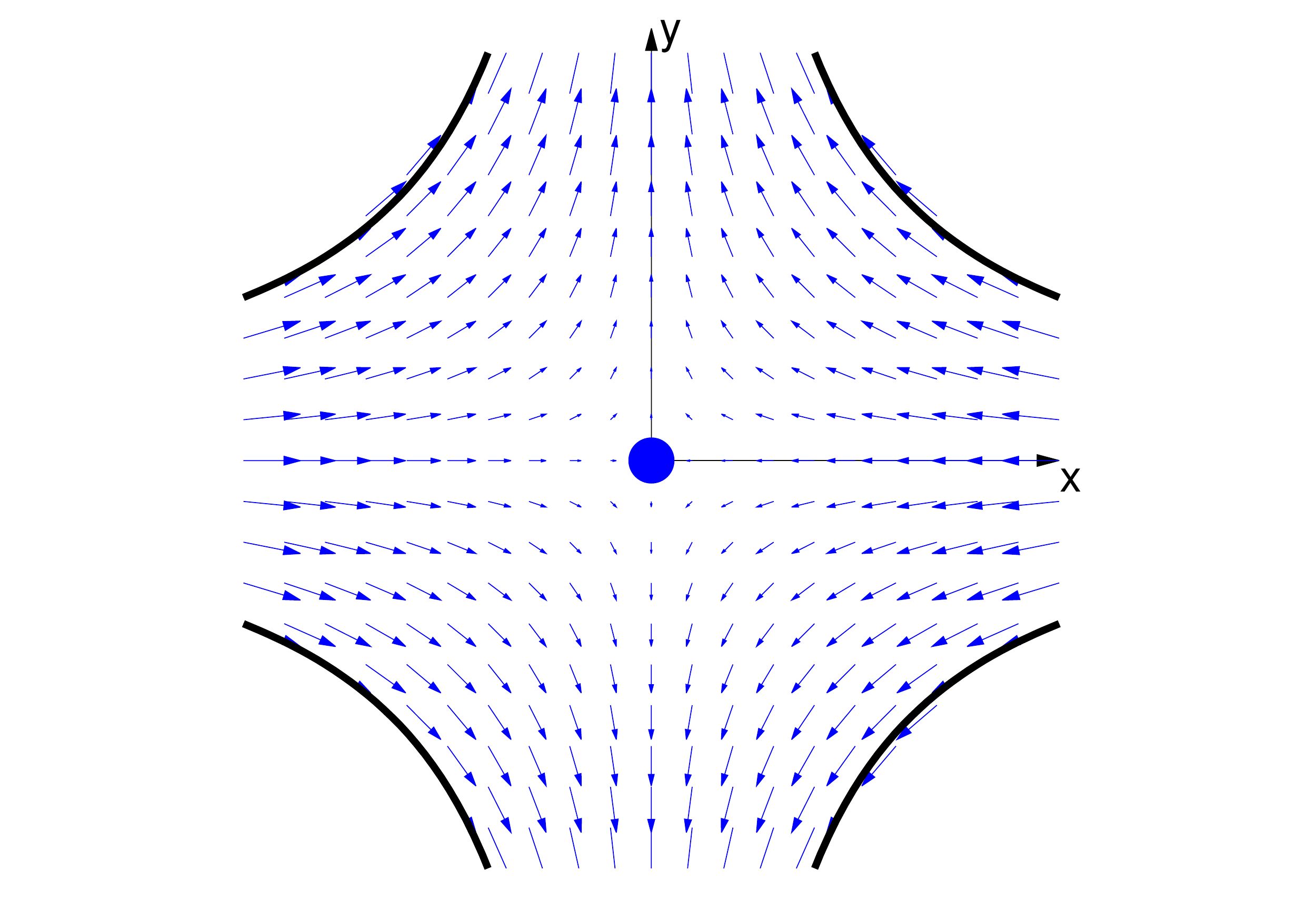}
\put(-45,-20){\color{blue} $\vec{F}$}
\caption{Quadrupolar magnetic field (left) and force imparted
  to a particle traveling perpendicular to the figure (right).\label{quad}}
\end{figure}

\begin{figure}\centering
\includegraphics[height=15cm, angle=-90]{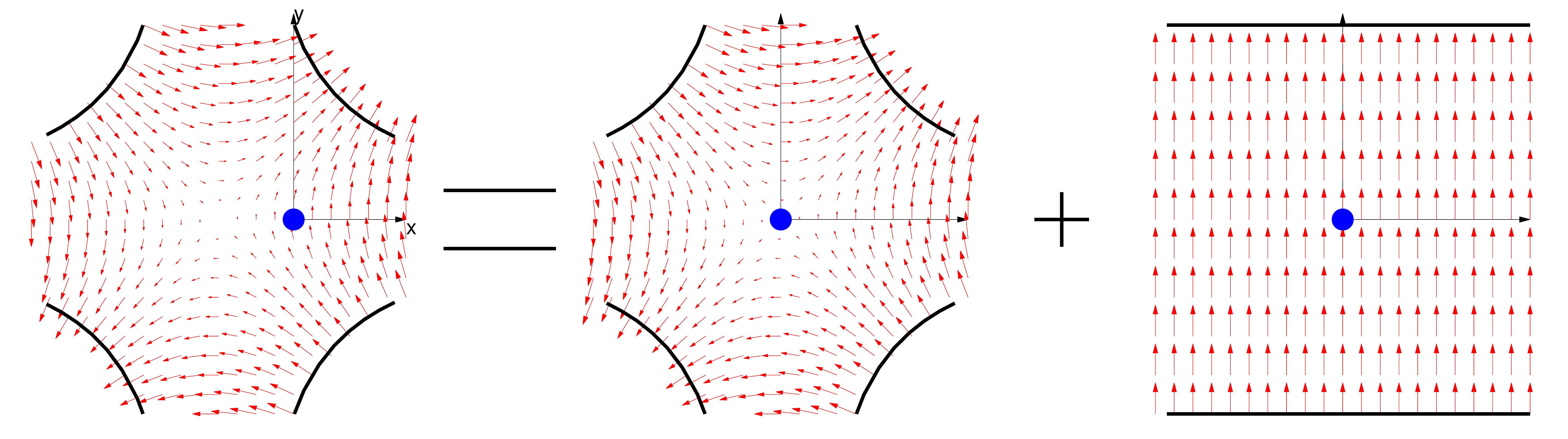}
\caption{An offset quadrupole is seen as a centered quadrupole plus 
a dipole.\label{offsetquad}}
\end{figure}

Figure~\ref{quad} shows the magnetic  and the force fields inside
a quadrupole. An offset quadrupole is seen as the superposition
of a centered quadrupole plus a dipolar field, as shown in Fig.~\ref{offsetquad}, hence introducing orbit deviations.

If the $i^{\rm th}$ quadrupole in a lattice has a gradient error of
$\Delta k_i$ it introduces horizontal and vertical tune deviations given by
\begin{equation}
\Delta Q_x\approx\frac{1}{4\pi}\overline{\beta_x}\Delta k_iL_i, \ \ \ \Delta Q_y\approx-\frac{1}{4\pi}\overline{\beta_y}\Delta k_iL_i \ ,
\end{equation}
where $L_i$ is the length of the quadrupole and $\overline{\beta_{x,y}}$
stands for the average $\beta_{x,y}$ function in the quadrupole.
At the same time the gradient error also introduces $\beta$-beating.
In presence of many quadrupolar errors the $\beta$-beating
can be expressed as
\begin{equation}
\frac{\Delta\beta}{\beta}(s)\approx \pm\sum_i \frac{\Delta k_iL_i\overline{\beta_i}}{2\sin(2\pi Q)}
\cos(2\pi Q - 2|\phi(s)-\phi_i|)\ ,
\end{equation}
where the positive sign stands for the horizontal plane, and negative
for the vertical.
The denominator $\sin(2\pi Q)$ makes the $\beta$-beating diverge at the integer and half integer resonances, $2Q \in \mathbb{N}$.
Betatron phase deviations between two locations
in the accelerator, $s$ and $s_0$, can be computed using the fundamental relation
$1/\beta={\rm d}\phi/{\rm d}s$, yielding
  \begin{eqnarray}
\Delta\phi(s_0,s)
           =  \int_{s_0}^s  \frac{\mathrm{d}s'}{\beta(s')}\left(\frac{1}{1 + \frac{\Delta\beta}{\beta}(s')}-1\right) \ .
\end{eqnarray}
  More explicit first and higher order expansions of the phase beating can be found in~\cite{miyamotothesis,Nicolo,andrea2016}.
  Resonance driving terms, $h_{jklm}$, appear in the expansion of the Hamiltonian and characterize the strength of resonances, $(k-j)Q_x + (m-l)Q_y =  P$, with $P$ any integer. $h_{jklm}$ are connected to the generating function resonance driving terms $f_{jklm}$ via the following relation,
   \begin{eqnarray}
f_{jklm} = \frac{h_{jklm}}{1-{\rm e}^{i2\pi[(k-j)Q_x+(m-l)Q_y]}}\ ,
   \end{eqnarray}
  where the denominator reveals the resonant behaviour.
  One way
  to explore higher order perturbations in the $\beta$-beating is via the
  the generating function resonance driving term $f_{2000}$ (in the horizontal plane), that is defined as 
  \begin{eqnarray}\displaystyle
  f_{2000}(s)=\frac{\sum_j \Delta k_jL_j\overline{\beta_{x,j}}e^{2i\phi_{x,j}}}{1-e^{4i\pi Q_x}} + \mathcal{O}(\Delta k^2)\ ,
  \end{eqnarray}
  where $\phi_{x,j}$ is cycled so to start from 0 at $s$.
  The $\beta$-beating can be expressed as function of $f_{2000}$ via the following equation~\cite{andrea2016},
  \begin{equation}
  \frac{\Delta\beta}{\beta}(s)=2\sinh |f_{2000}|\Big(\sinh |f_{2000}| + \cosh |f_{2000}|\sin\phi_{2000} \Big)\ ,
  \end{equation}
  where $\phi_{2000}$ is the phase of $f_{2000}$.
  Similar equations hold  for the vertical plane with $f_{0020}$.

\begin{figure}\centering
\begin{picture}(240,180)
\put(90,75){\thicklines\line(1,0){120}}
\put(90,150){\thicklines\line(1,0){120}}
\put(90,75){\thicklines\line(0,1){75}}
\put(210,75){\thicklines\line(0,1){75}}
\thicklines\qbezier(30,120)(150,90)(240,165)
\put(90,50){ quadrupole ($k$,$L$)}
\put(64,98){ $\beta_1$}
\put(213,130){ $\beta_2$}
\put(120,120){ $\overline{\beta}$\_\_\_\_\_\_\_\_}
\end{picture}\vspace{-1.2cm}
\caption{Variation of $\beta$ function along a quadrupole, displaying
the the average $\beta$ function as $\overline{\beta}$.\label{fig:avebeta}}
\end{figure}
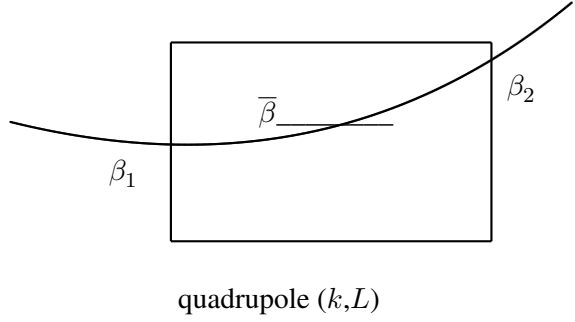

Figure~\ref{fig:avebeta} shows a sketch of the variation of $\beta$ function along a quadrupole, displaying
the the average $\beta$ function as $\overline{\beta}$
and the $\beta$ at the edges as $\beta_{1,2}$.
An approximation of $\overline{\beta}$ is given in~\cite{ISRkMod} as
\begin{equation}
  \overline{\beta}\approx \frac{1}{3}\left(\beta_1+\beta_2+\sqrt{\beta_1\beta_2-L^2}\right)\ .
\end{equation}
Exact equations for $\overline{\beta}$ depend also on the quadrupole
strength $k$ as shown in~\cite{Felix,Paul}.

\begin{figure}\centering
\hspace{-11.8cm}   
\includegraphics[height=6cm, angle=-85]{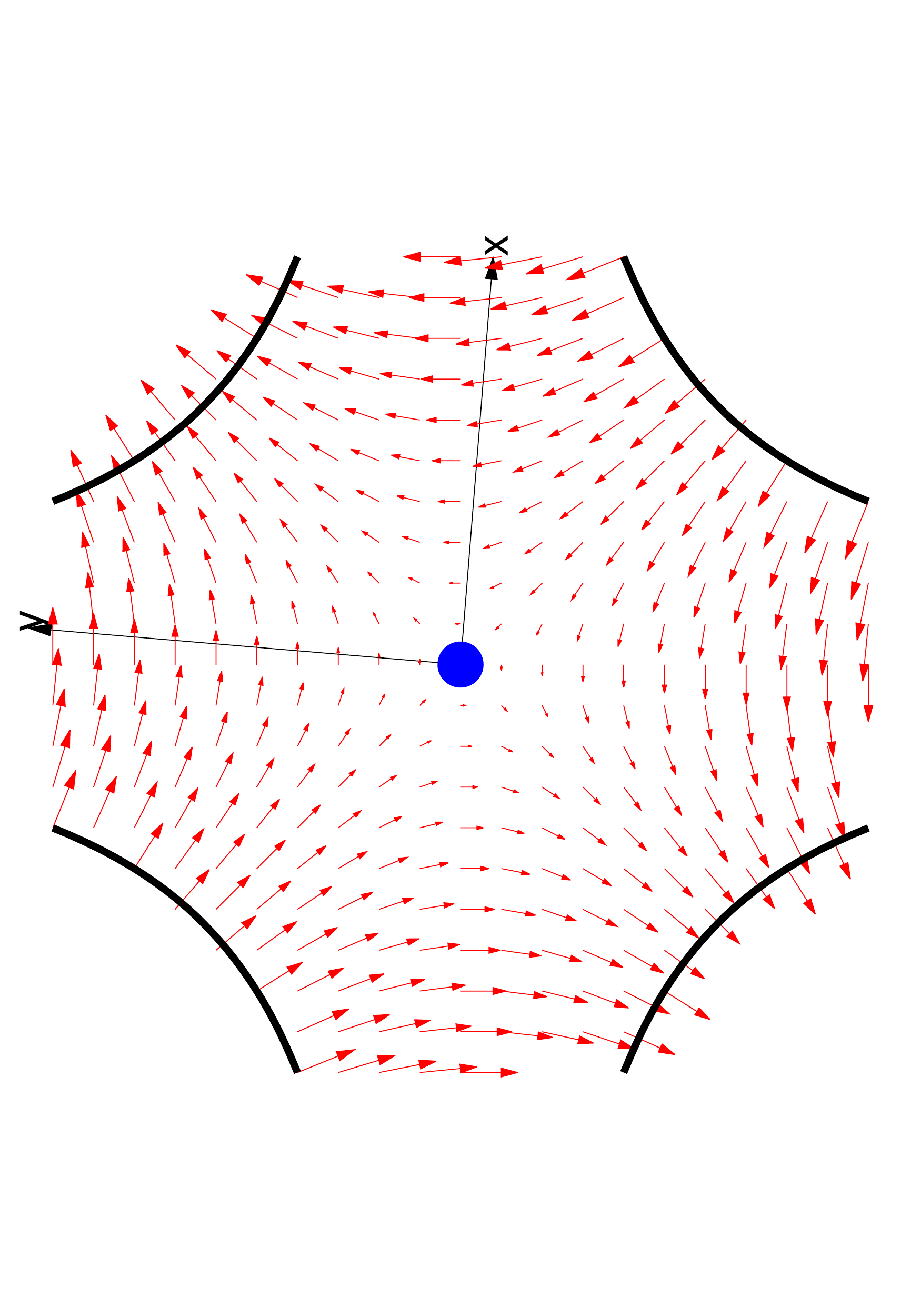}
\put(-37,8){
  \includegraphics[height=12cm, angle=-90]{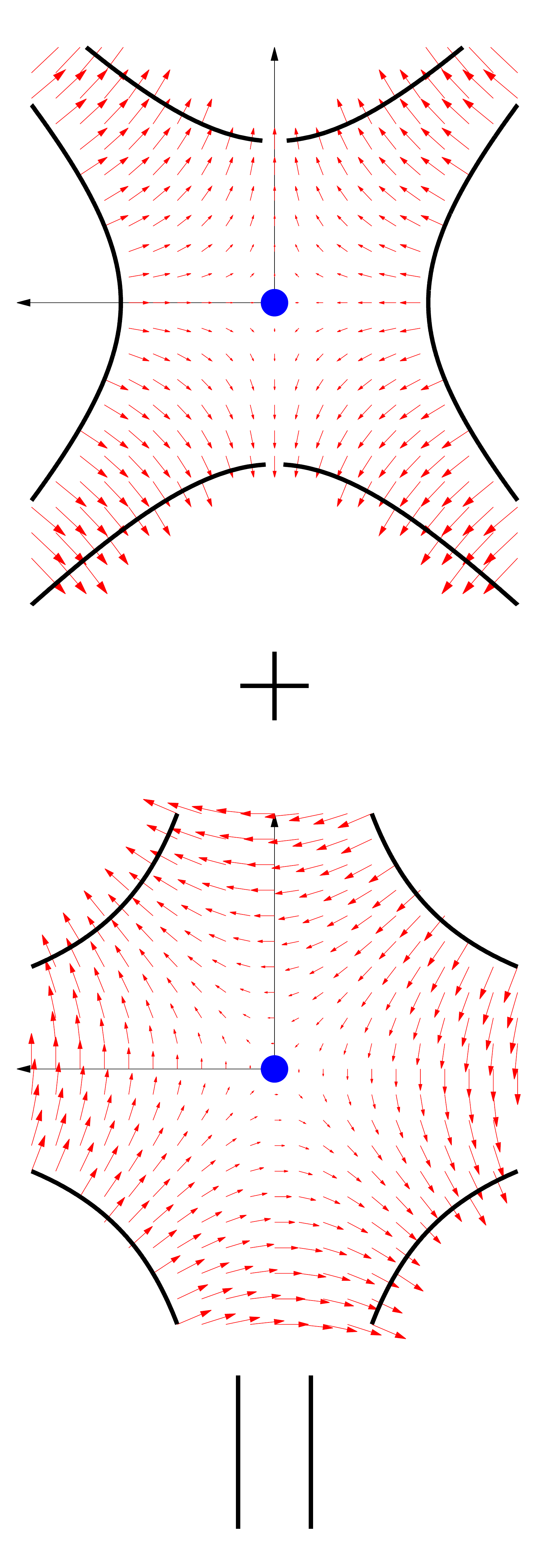}
}
\caption{A tilted quadrupole is seen as a normal quadrupole plus 
another quadrupole tilted by 45$^{\circ}$.\label{tiltquad}}
\end{figure}

\begin{figure}\centering
\includegraphics[height=6.5cm, angle=-90]{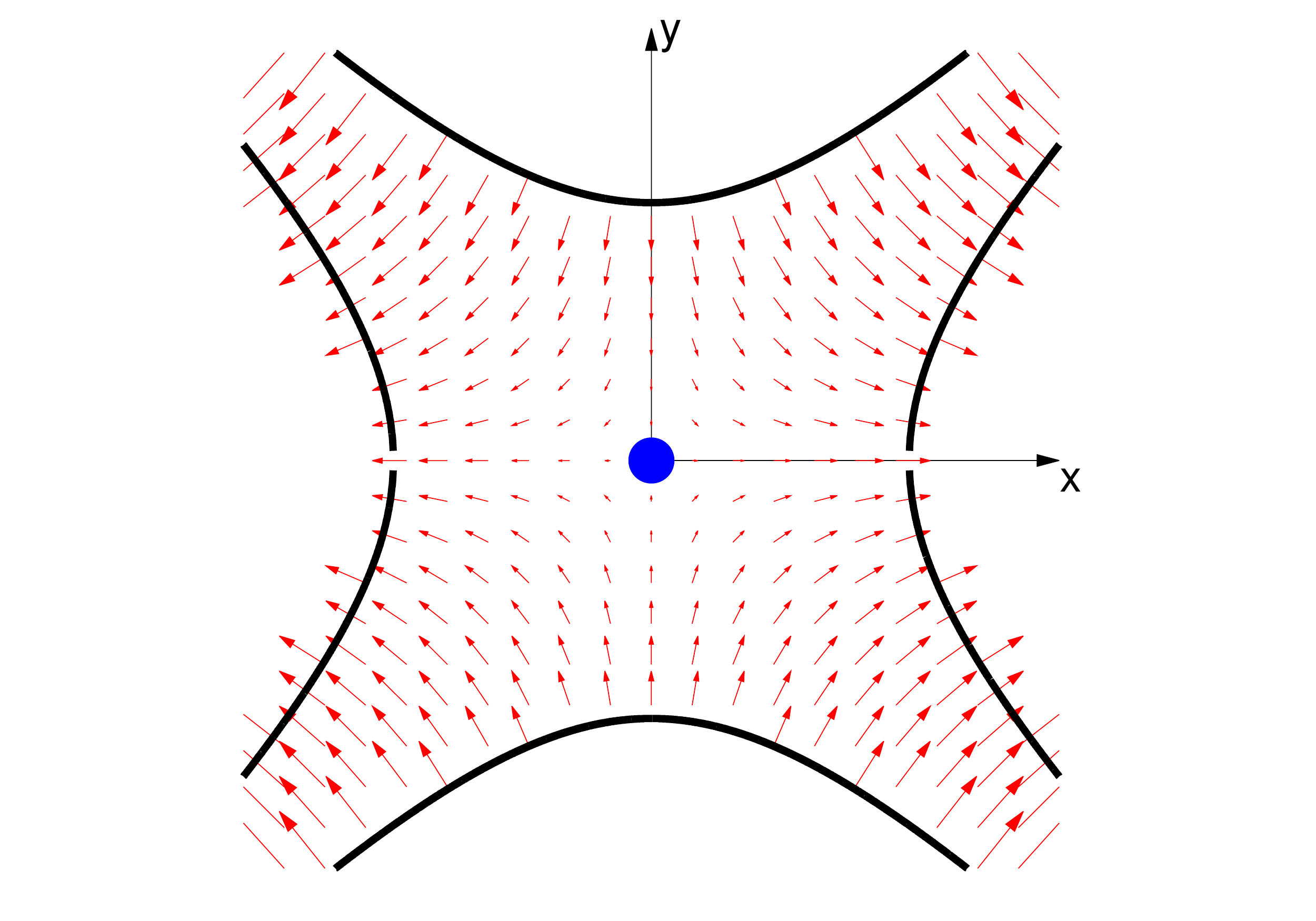}
\put(-80,-20){\color{red} $\vec{B}$}\hspace{-1.5cm}
\includegraphics[height=6.5cm, angle=-90]{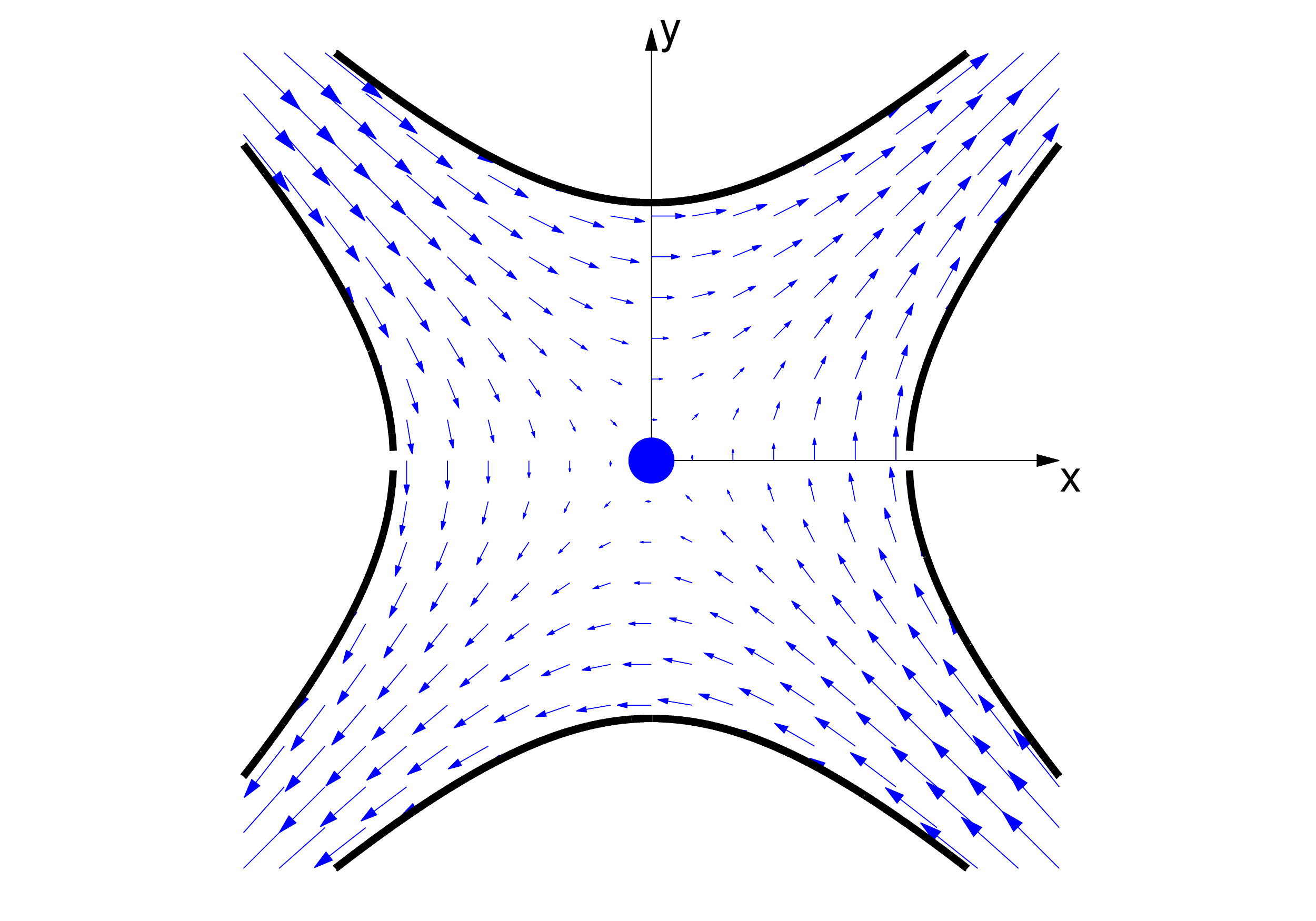}
\put(-80,-20){\color{blue} $\vec{F}$}
\caption{Skew quadrupole magnetic field (left) and force (right). \label{skewquadfield}}
\end{figure}

  A tilted quadrupole is seen as a normal quadrupole plus 
another quadrupole tilted by 45$^{\circ}$, which is called a skew
quadrupole, see Fig.~\ref{tiltquad}.
The magnetic and force fields of a skew quadrupole are shown in
Fig.~\ref{skewquadfield}.

As shown in Fig.~\ref{skewquadfield} particles displaced horizontally in
a skew quadrupole receive a vertical force.
This causes the particle motion to couple between the horizontal and vertical planes.
While the uncoupled betatron motion of the particle
position at turn $N$ and location $s$ is simply expressed as
\begin{eqnarray}
  x(N,s) = \sqrt{\beta_{x}(s)\epsilon_x} \cos(2\pi Q_x N + \phi_{x}(s)+\phi_{x0})\ , \label{xNs}
  \end{eqnarray}
with $\epsilon_x$ being the horizontal single particle emittance and $\phi_0$ the phase
at $N=0$ and $s=0$, the coupled motion in presence of skew quadrupolar fields
can be approximated as~\cite{frankS,bnl}
\begin{eqnarray}
x(N,s) &\approx& \sqrt{\beta_{x}(s)}\Re\Big\{\sqrt{\epsilon_x} e^{i(2\pi Q_x N + \phi_{x}(s)+\phi_{x0})}
 \nonumber\\
 & & \ \   -2if_{1010} \sqrt{\epsilon_y} e^{-i(2\pi Q_y N + \phi_{y}(s)+\phi_{y0})} \nonumber\\
 & & \ \ \nonumber - 2if_{1001} \sqrt{\epsilon_y} e^{i(2\pi Q_y N + \phi_{y}(s)+\phi_{y0})}
 \Big\}\ ,
\end{eqnarray}
where $\Re\{x\}$ stands for the real part of $x$, and $f_{1010}$ and $f_{1001}$ are the sum and difference generating function
resonance driving terms given by
\begin{eqnarray}
 f_{\tiny\begin{array}{c}1010\\1001\end{array}} &=& \frac{\sum_j  k_{s,j}L_j\sqrt{\beta_{x,j}\beta_{y,j}}e^{i(\phi_{x,j} \pm \phi_{y,j})}}{4(1-e^{2\pi i(Q_x\pm Q_y)})} \ ,
\end{eqnarray}
where $k_{s,j}$ represents the $j^{\rm th}$ skew quadrupole gradient in the machine. 
$f_{1001}$ drives the difference resonance $Q_x-Q_y=P$ and $f_{1010}$ drives the sum resonance $Q_x+Q_y=N$, for any $P \in \mathbb{Z}$ and  $N \in \mathbb{N}$.

\begin{figure}\centering
\includegraphics[height=10.2cm, angle=-90]{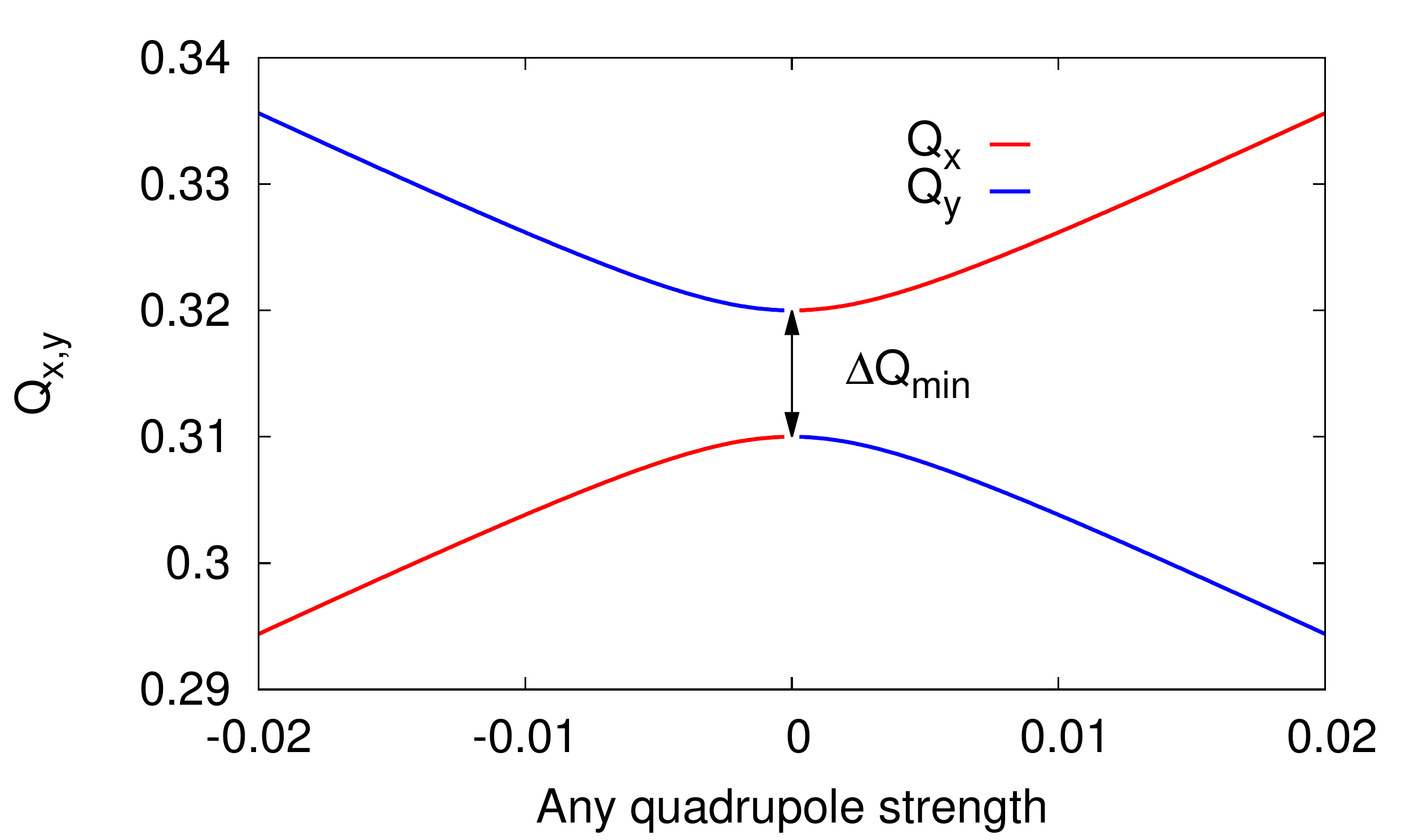}
\caption{Approaching tunes in presence of coupling yielding
  to mode veering. \label{dqmin}}
\end{figure}

\begin{figure}\centering
\hspace{-2cm}
\includegraphics[height=6cm, angle=0]{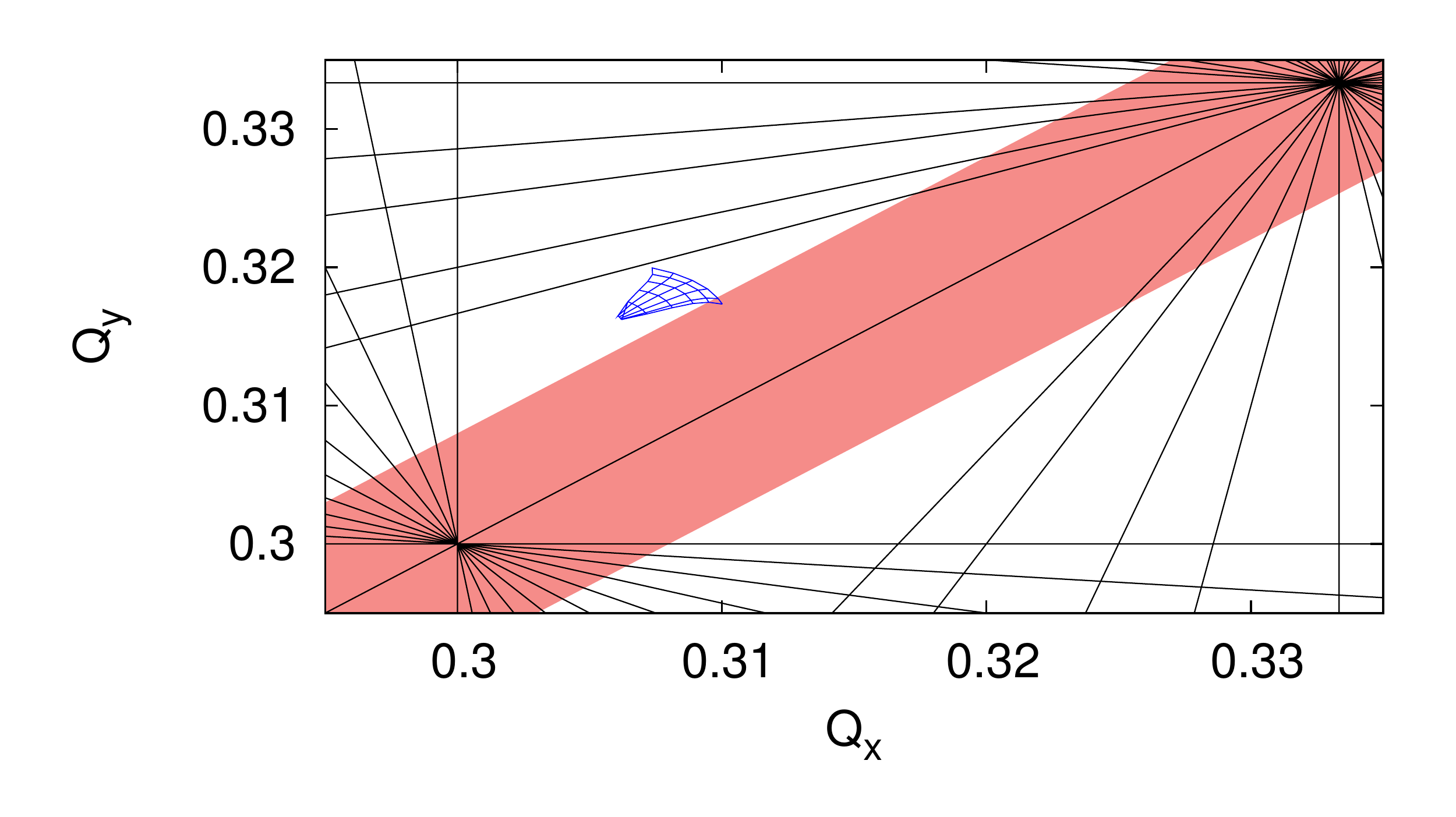}
\caption{A hypothetical large coupling stopband in red limiting the tune space available to the beam in presence of beam-beam tune footprint. Black lines
  represent tune resonances to be avoided.\label{bbcoup}}
\end{figure}

Another important feature of  coupled motion is the appearance of a stopband  
around the difference resonance $Q_x-Q_y=P$,  $P\in\mathbb{Z}$. 
This implies that the fractional tunes cannot get closer than
$\Delta Q_{\rm min}$, the closest tune approach, given by~\cite{Frankbook}
\begin{equation}
\Delta Q_{\rm min}=\left|\frac{1}{2\pi}\sum_{j}{ k_{s,j}L_j\sqrt{\beta_x\beta_y}
   e^{-i(\phi_x-\phi_y)+is(\hat{Q}_x-\hat{Q}_y)/R}}\right| \ ,
\end{equation}
where $k_{s,j}$ represents the skew quadrupolar gradients 
around the ring,  $R$ is the machine radius and $\hat{Q}_{x,y}$ are the fractional tunes. 
 $\Delta Q_{\rm min}$ can also be computed from $f_{1001}$ around the ring by~\cite{alexahin,tobias}
\begin{eqnarray}
\Delta Q_{\rm min} &=& \left| \frac{4(\hat{Q}_x-\hat{Q}_y)}{2\pi R} \oint \mathrm{d}s 
f_{1001} {\rm e}^{-i(\phi_x-\phi_y)+is(\hat{Q}_x-\hat{Q}_y)/R}    \right| 
\lesssim  4|\hat{Q}_x-\hat{Q}_y|\overline{|f_{1001}|}\ ,
\end{eqnarray}
where $\overline{|f_{1001}|}$ represents the ring average of $|f_{1001}|$.
As an illustration,  a hypothetical large coupling stopband (in red) would limit
the tune space available for the LHC beam-beam tune footprint as shown in Fig.~\ref{bbcoup}.

\subsection{Sextupole}
\begin{figure}\centering
\includegraphics[height=6.5cm, angle=-90]{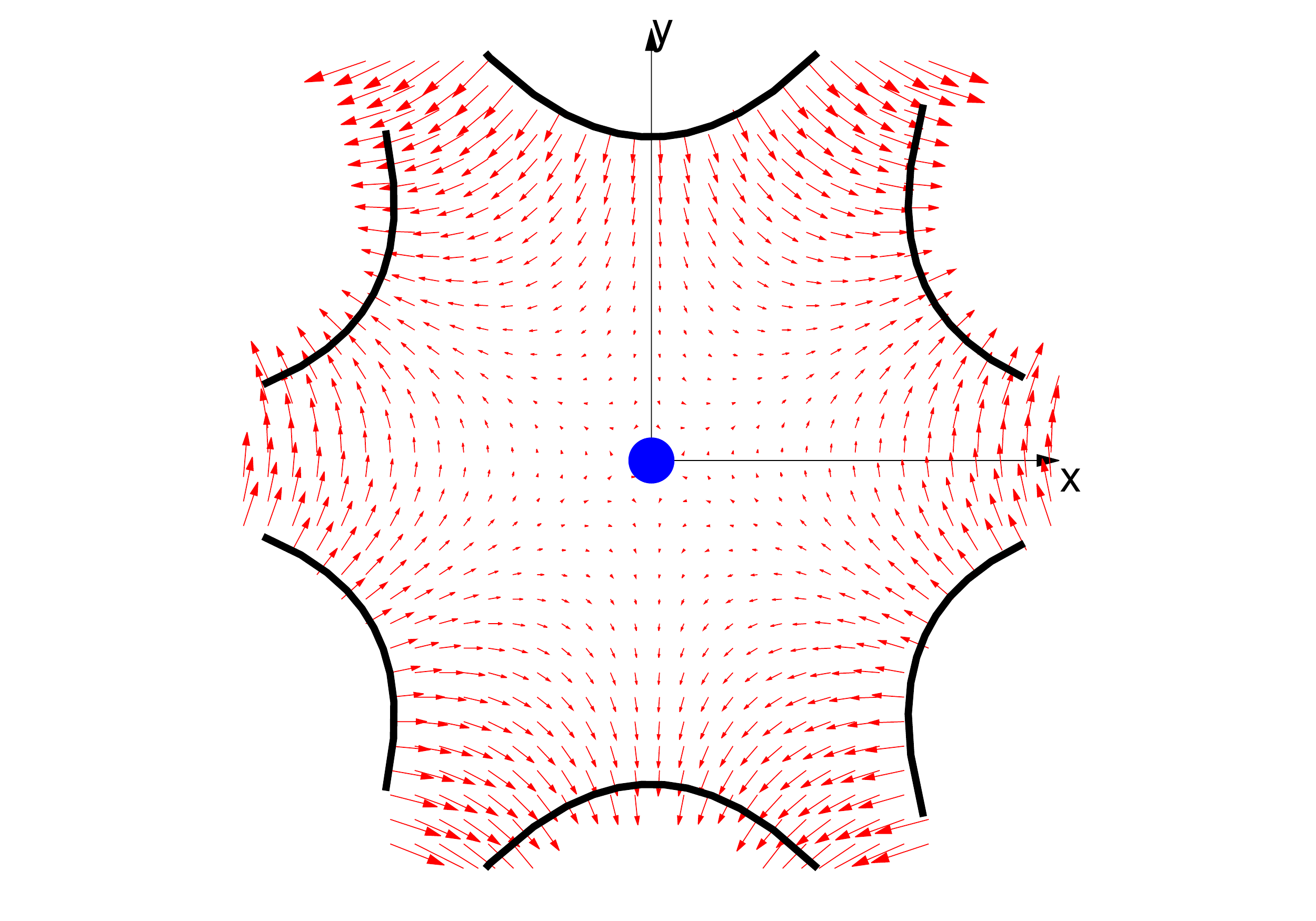}
\put(-90,-5){\color{red} $\vec{B}$}\hspace{-1.5cm}
\includegraphics[height=6.5cm, angle=-90]{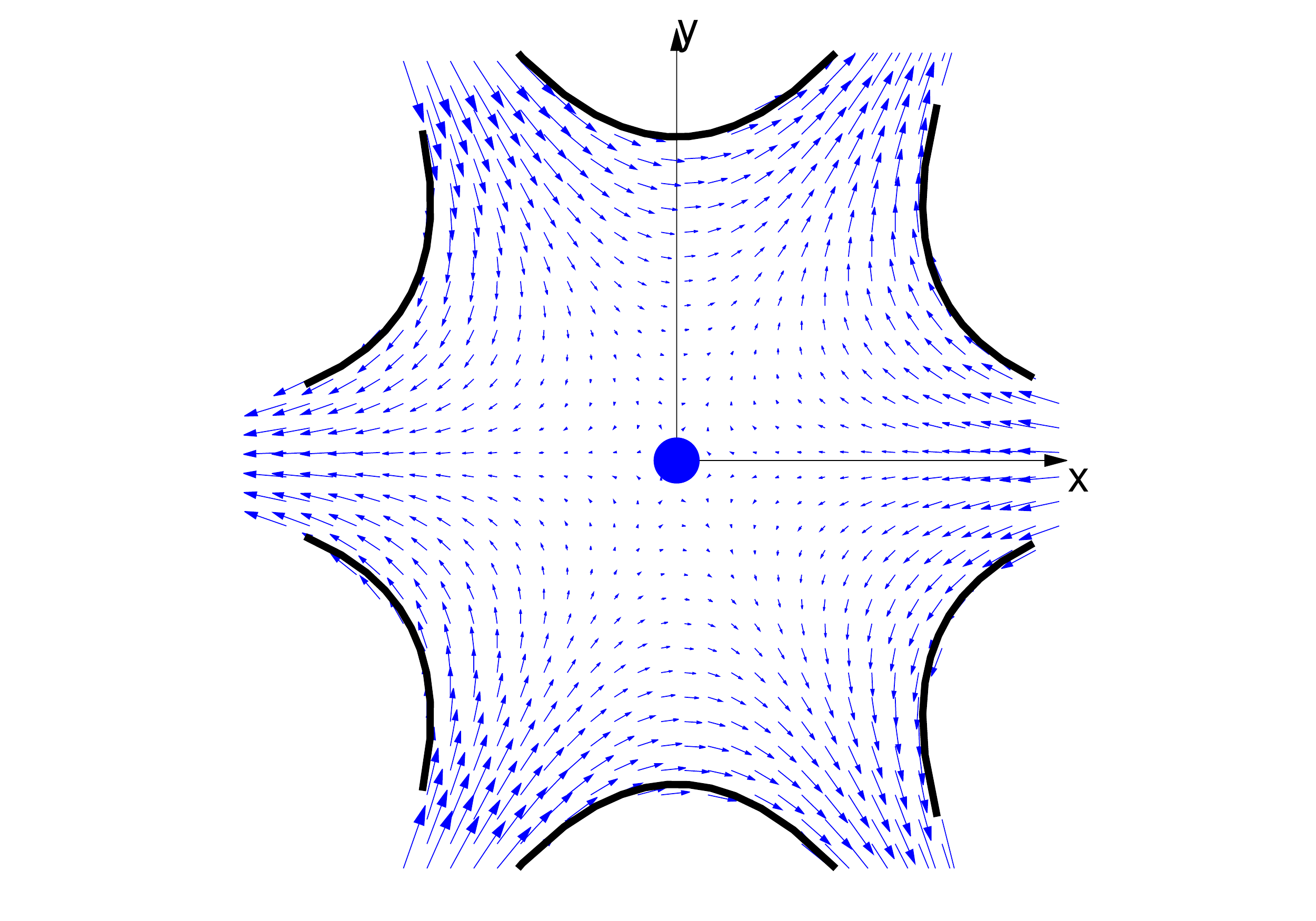}
\put(-90,-5){\color{blue} $\vec{F}$}
\caption{Sextupole magnetic field (left) and force (right). \label{sext}}
\end{figure}

\begin{figure}\centering
\includegraphics[height=15cm,  angle=-90]{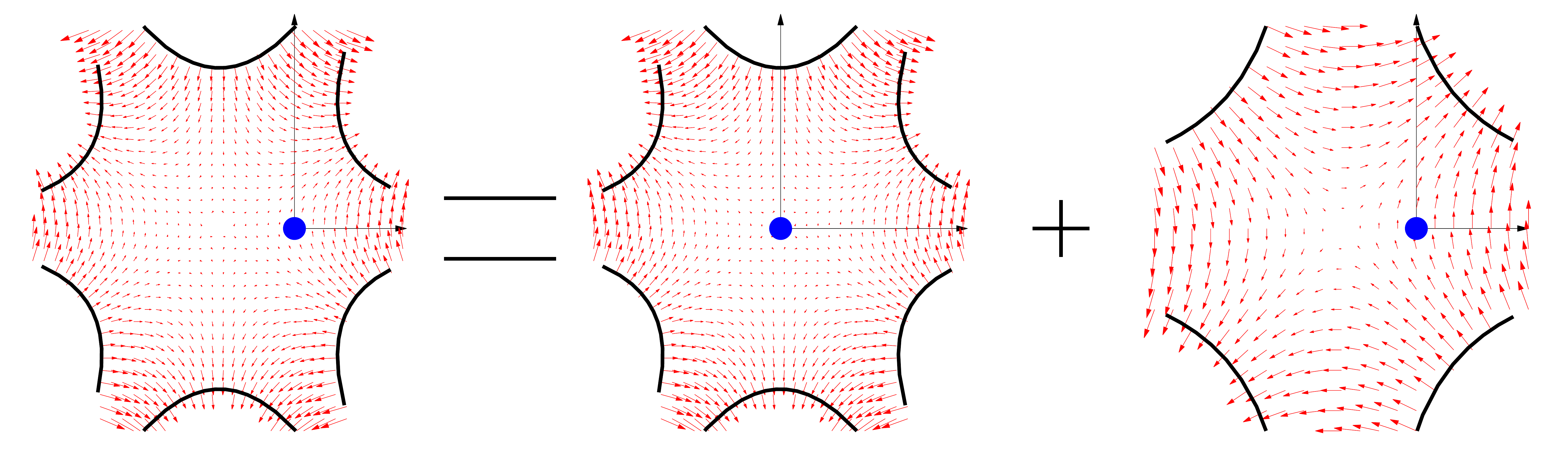}
\caption{A sextupole horizontally (vertically) displaced is seen as a centered sextupole plus
an offset quadrupole (skew quadrupole). \label{offsetsext}}
\end{figure}

Figure~\ref{sext} shows the field and force fields in a sextupole.
The equations describing the horizontal and vertical forces in a
sextupole are given by
\begin{equation}
  F_x=\frac{1}{2}K_2 (x^2-y^2)\ , \ \  F_y=-K_2 xy\ ,
\end{equation}
where $K_2$ is the integrated sextupolar gradient.
Sextupoles are needed in accelerators to compensate
chromaticity ($Q'$), that describes the dependence
of the tune with the relative energy deviation of the particle: $Q'={\rm d}Q/{\rm d}\delta$, with $\delta=(p-p_0)/p_0$.
An offset sextupole is seen as a centered sextupole
together with an offset quadrupole, see Fig.~\ref{offsetsext}.
Horizontal offsets in sextupoles generate normal quadrupole
perturbations while vertical offsets generate skew quadrupolar fields.

\subsection{Longitudinal misalignments}\label{longmisec}
\begin{figure}\centering
\includegraphics[height=12cm,  angle=-90]{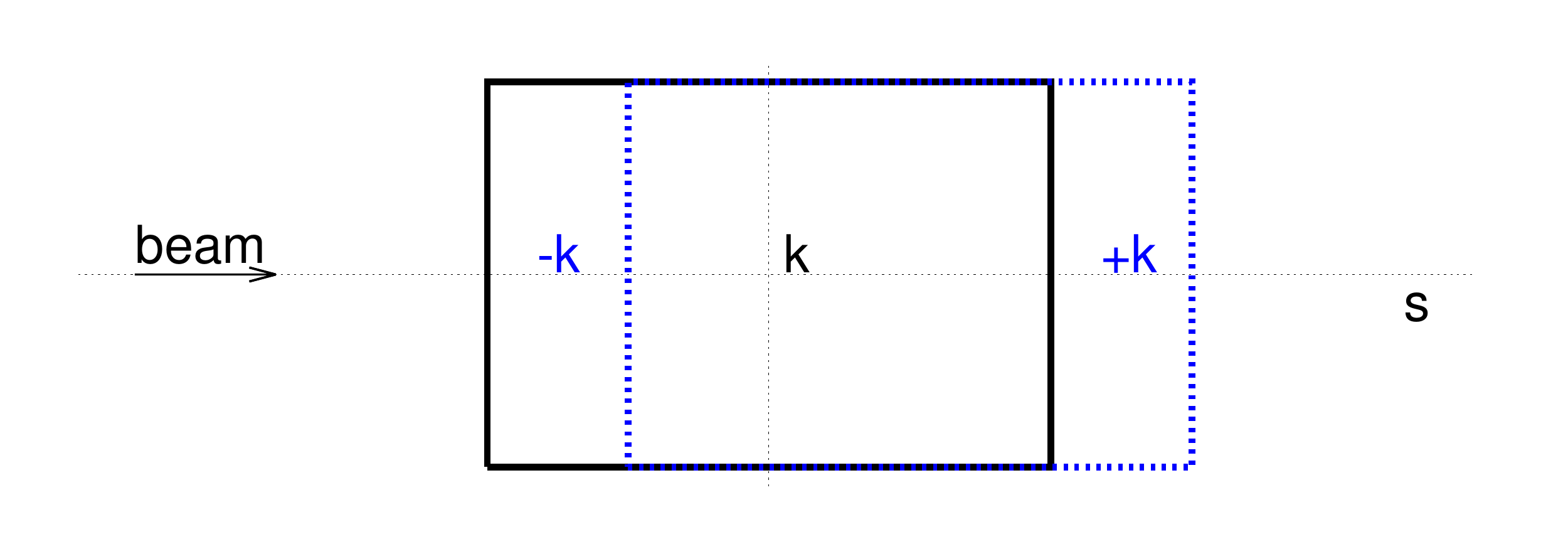}
\caption{Approximating a longitudinal misalignment of an element (blue)
  by kicks at the edges of the unperturbed element (black). \label{longmis}}
\end{figure}

Longitudinal misalignments can be approximated as thin perturbations
at both ends of the ideal magnet with opposite signs as shown in~Fig.~\ref{longmis}. For a longitudinal misalignment by $\delta$ of an element of strength $k$
the thin perturbations have integrated strengths of $\pm \delta  k$.   
Tolerances are generally larger for longitudinal misalignments as
there is usually a partial compensation of the perturbations
generated at both ends thanks to the opposite signs. 

\section{Phase-space and turn-by-turn motion}\label{sec4}
\subsection{The transverse phase-space}
\begin{figure}\centering
\includegraphics[trim = 0mm 0mm 0mm 0mm, clip,height=11.cm, angle=-0]{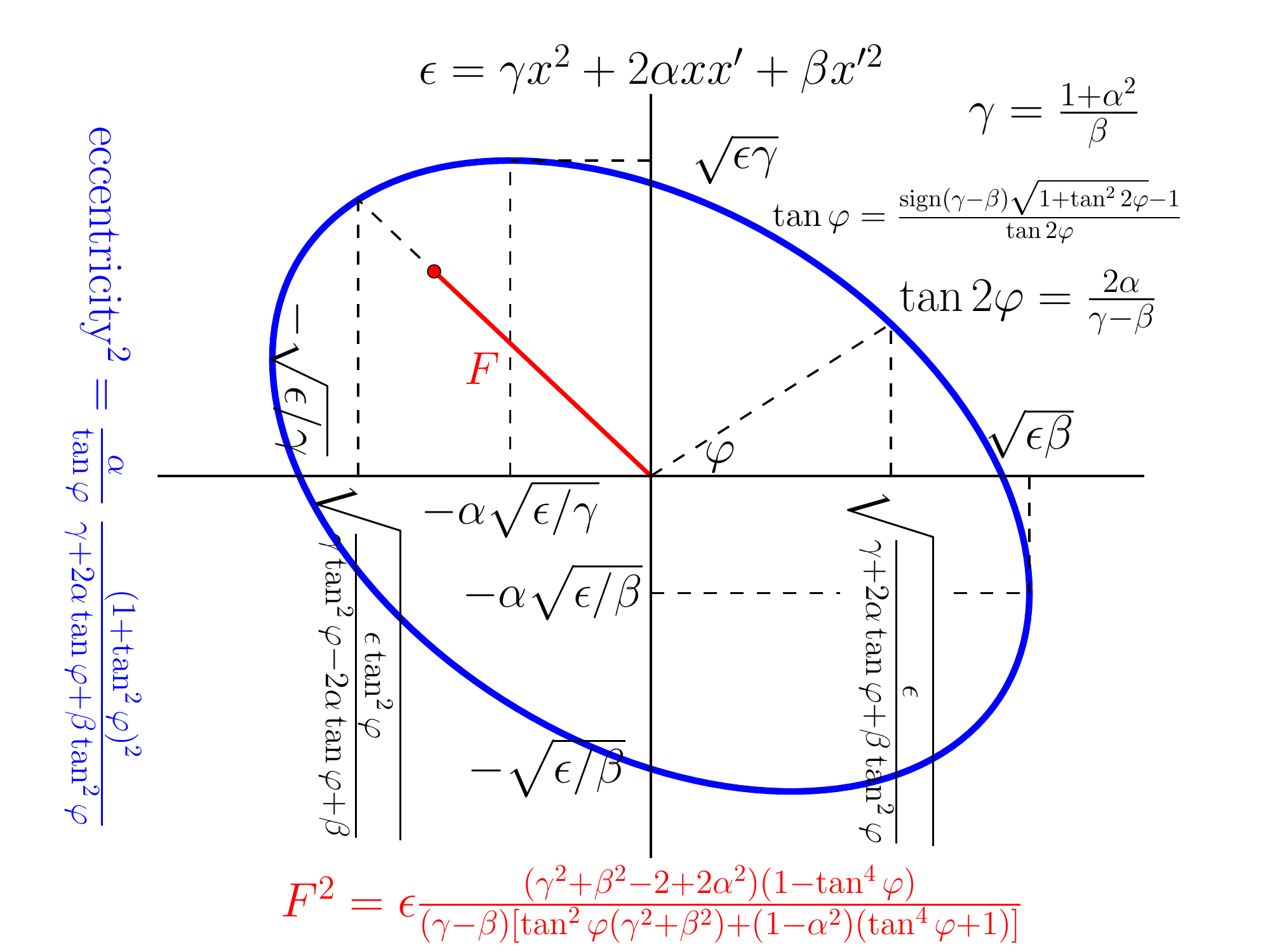}
\caption{Phase-space ellipse $x'$ versus $x$. \label{ellipse}}
\end{figure}

\begin{figure}\centering
\includegraphics[trim = 40mm 10mm 10mm 22mm, clip,height=10.4cm, angle=-0]{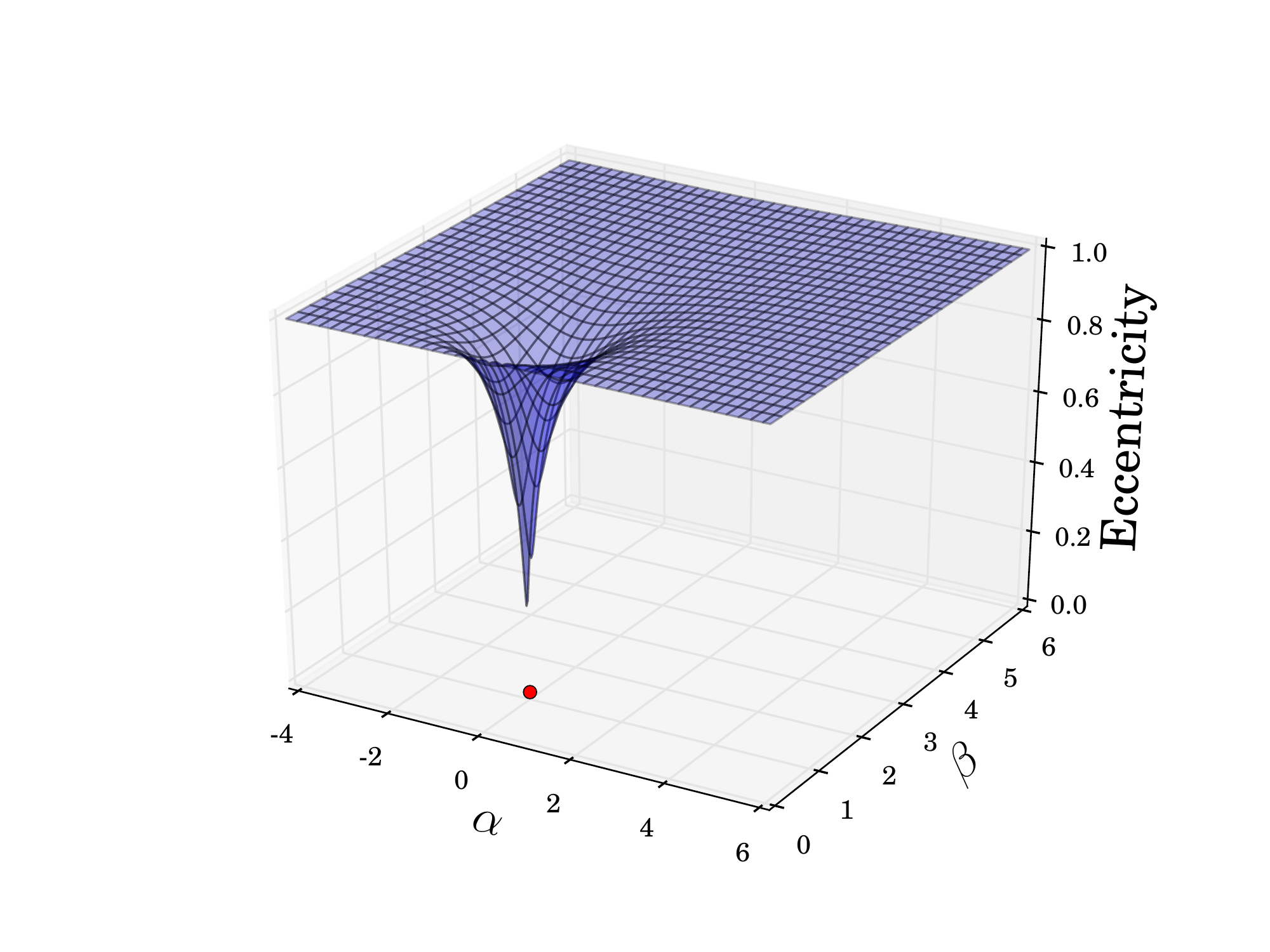}
\caption{Phase-space betatronic ellipse eccentricity versus $\alpha$ and $\beta$. The eccentricity is zero only for $\alpha=0$ and $\beta=1$, marked in red. \label{eccentricity}}
\end{figure}

At any location of the accelerator the turn-by-turn uncoupled motion is represented by
the position of the particle and its angle with respect to the longitudinal
direction, i.e. $x'={\rm d}x/{\rm d}s$, which are parametrized as follows
\begin{eqnarray}
  x(N)&=&\sqrt{\epsilon\beta}\cos(2\pi QN + \phi_0)\ , \nonumber\\
  x'(N)&=&-\alpha\sqrt{\epsilon/\beta}\cos(2\pi QN) + \sqrt{\epsilon/\beta}\sin(2\pi QN + \phi_0)\ , \label{eqsxx'}
\end{eqnarray}
where $\alpha=-\beta'/2$. The particle trajectory stays within an ellipse
in the phase-space $(x,x')$. This trajectory is shown in
Fig.~\ref{ellipse} along with several relevant parameters of the
motion and the ellipse, as the angle of the principal direction
of the ellipse $\varphi$, the coordinates of the intersections
of the ellipse and the axes, the largest excursions in $x$ and $x'$, the eccentricity of the ellipse and its focal length $F$.
The eccentricity of a conic section is a non-negative
real number that uniquely characterizes its shape.
More formally two conic sections are similar if and only if
they have the same eccentricity.
One can think of the eccentricity as a measure of how much a conic section deviates from being circular. In particular
the eccentricity of a circle is zero and
the eccentricity of an ellipse is greater than zero and smaller than 1.
Figure~\ref{eccentricity} shows the eccentricity of the
betatronic ellipse versus $\alpha$ and $\beta$.
It is interesting to note that the motion is a circle only for $\beta=1$
and $\alpha=0$.

\newpage
Expressing Eqs.~(\ref{eqsxx'}) in matrix form
illustrates the transformation that takes a circular motion
into the elliptical one as follows,
\begin{eqnarray}
  \begin{pmatrix} x(N) \\ x'(N) \end{pmatrix} \hspace{0.4cm}=&
  \begin{pmatrix} \sqrt{\beta} & 0  \\ -\alpha/\sqrt{\beta} & 1/\sqrt{\beta} \end{pmatrix}&
  \begin{pmatrix}  \sqrt{\epsilon}\cos(2\pi QN+ \phi_0)\\ \sqrt{\epsilon}\sin(2\pi QN+ \phi_0) \end{pmatrix} \label{Floquet} \\
  \includegraphics[trim = 0mm 0mm 0mm 0mm, clip,height=3cm, angle=-0]{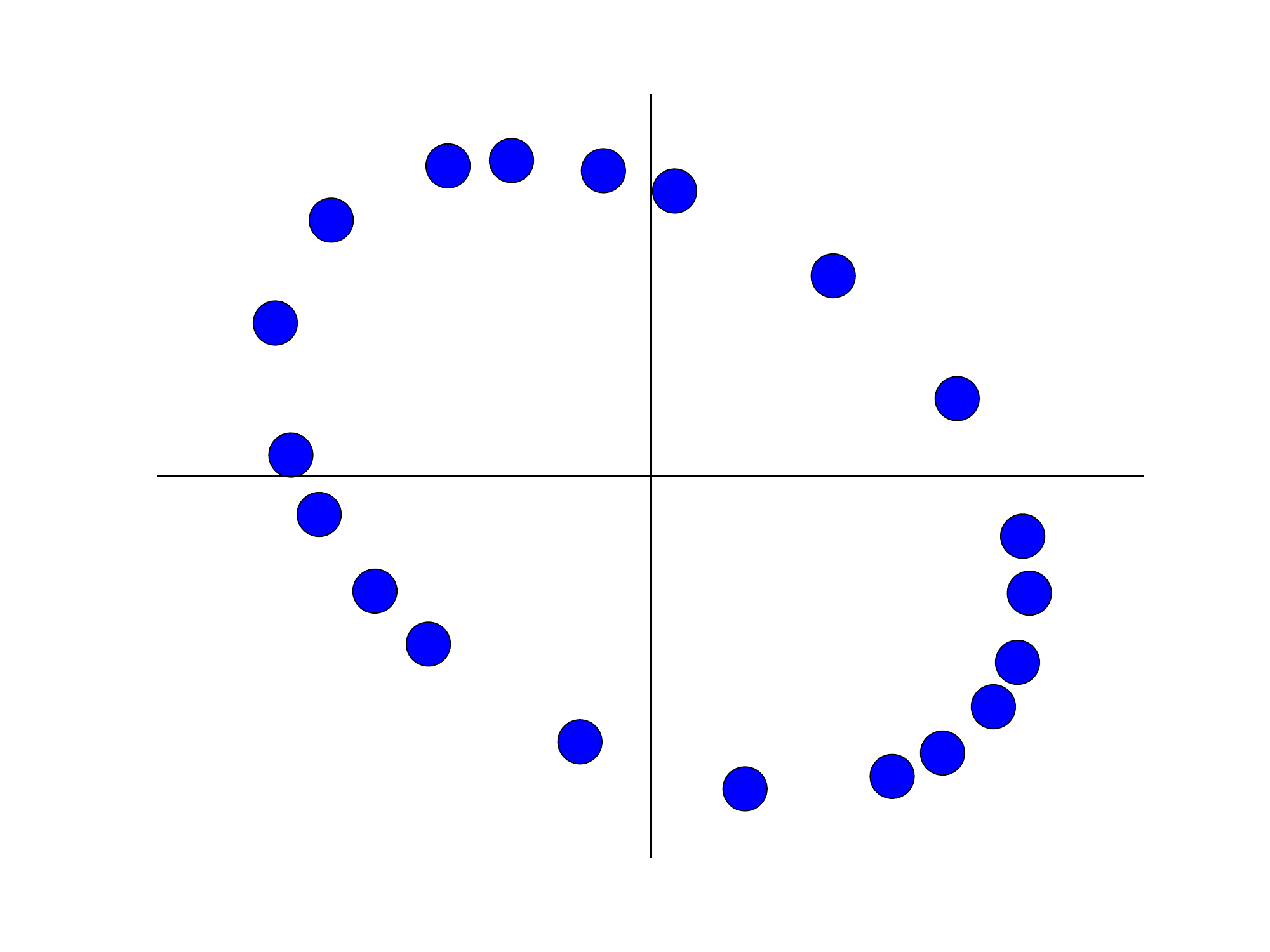}
  &  &
  \includegraphics[trim = 0mm 0mm 0mm 0mm, clip,height=3cm, angle=-0]{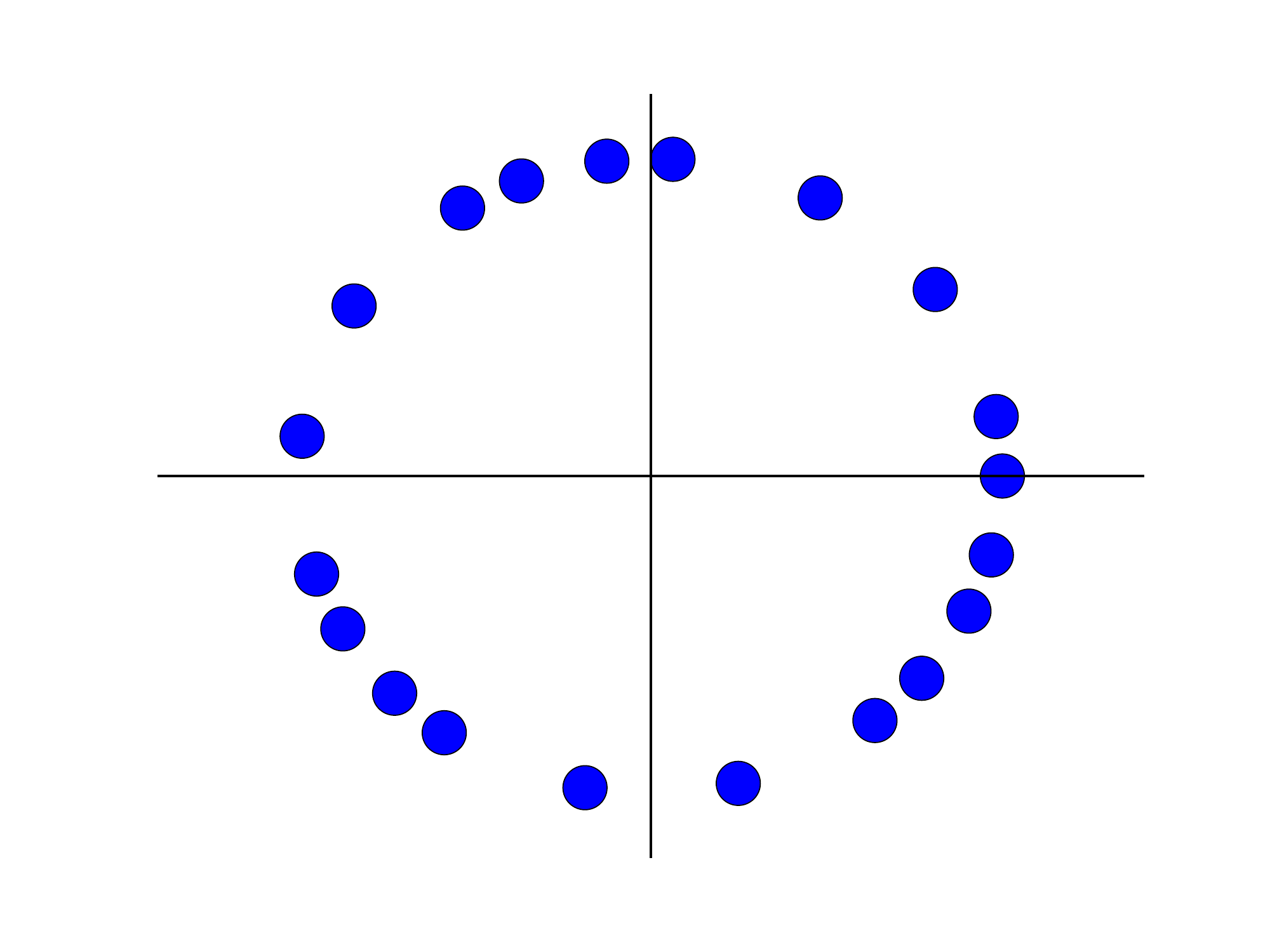}\put(-100,0){\footnotesize Floquet Normal Form}\nonumber
\end{eqnarray}
  The circular motion is preferred in many studies for its simplicity
  and it is referred to as Normal form or Floquet Normal form.

\subsection{Computing \texorpdfstring{$\alpha$, $\beta$ and $\epsilon$}{alpha, beta and epsilon from turn-by-turn data}}
When performing computer simulations of particles traveling
in an accelerator subject to lattice imperfections
or other electrodynamics interactions we have access to the
turn-by-turn coordinates $(x,x')$
and we want to study how the different phenomena perturb
the phase-space ellipse.
Evaluating $\alpha$, $\beta$ and $\epsilon$ is possible
by computing the singular value decomposition of the $2\times n$
matrix composed of the $x$ and $x'$ coordinates for $n$ turns,
  \begin{equation}
    \begin{pmatrix}
      x(1) & x(2) & x(3) & \dots & x(n) \\
      x'(1) & x'(2) & x'(3)  & \dots & x'(n)
    \end{pmatrix}_{2\times n} = U_{2\times2}S_{2\times 2}V^{T}_{2\times n}\ , \label{xx'SVD}
  \end{equation}
  where  $U$ and $V$ are unitary matrices and $S$ is diagonal with non-negative real numbers on the diagonal.
  The diagonal entries of $S$ are known as singular values. The columns of $U$ and the columns of $V$ are called the left-singular vectors and right-singular vectors respectively.  The left-singular vectors are a set
  of orthonormal vectors and similarly for the right-singular vectors.
  
\begin{figure}\centering
  \includegraphics[trim = 0mm 0mm 0mm 10mm, clip,height=7.1cm, angle=-0]{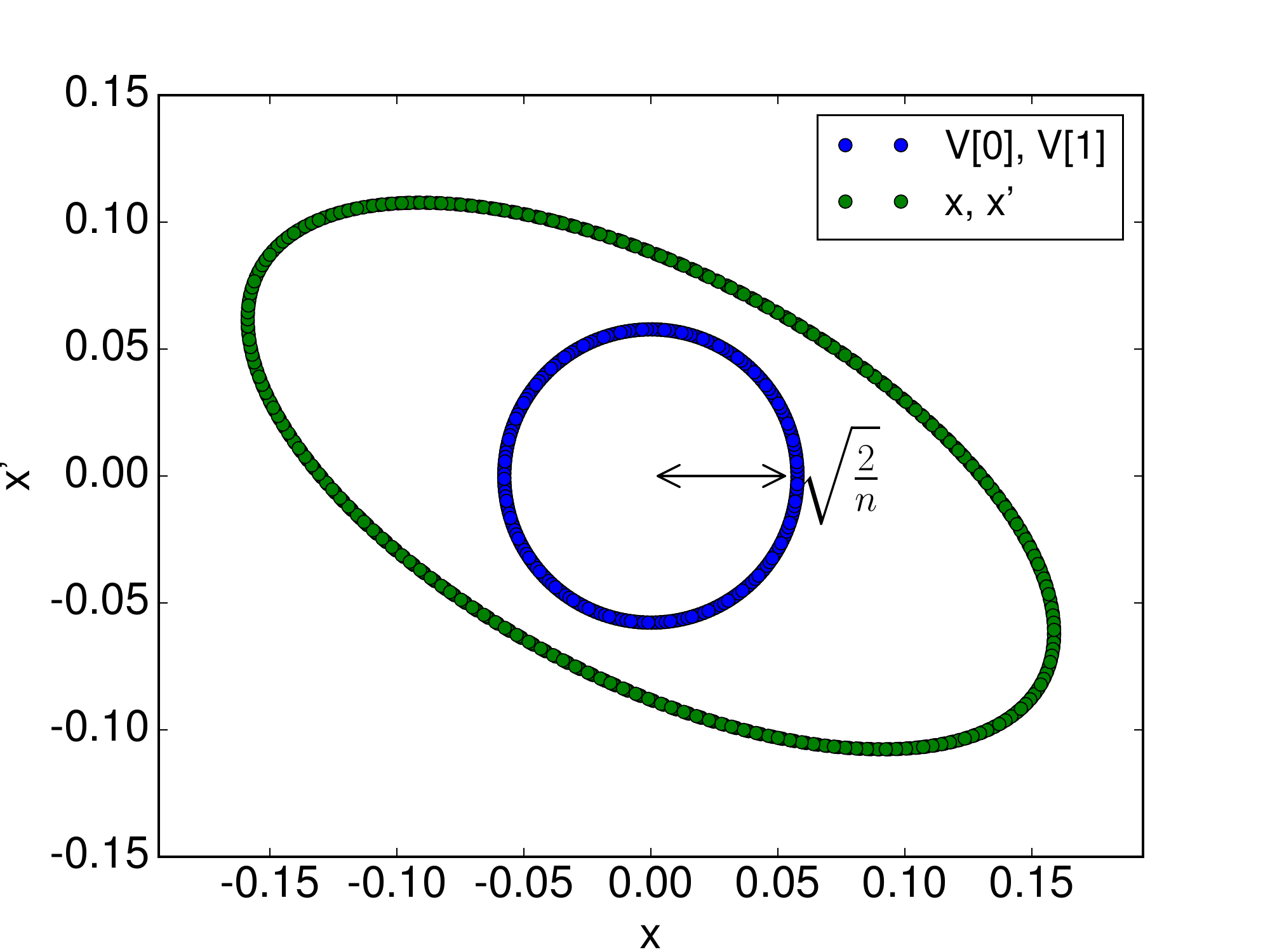}
  \caption{Illustration of the phase space ellipse with $\alpha=0.2$, $\beta=1.$
    and $\epsilon=2\times10^{-3}$ used in the Python code together with the first two right-singular modes of the $V$ matrix of the singular value
    decomposition of Eq.~(\ref{xx'SVD}). \label{V}}
\end{figure}

  $V_{2\times n}$ represents the turn-by-turn motion in a circle (like a  Normal form) in an arbitrary phase origin. Therefore, there must be a rotation $R(\theta)$ than can be inserted in the singular value decomposition as
\begin{equation}
  USV^T=USR(\theta)R^{-1}(\theta)V^T \ ,
\end{equation}
such that $R^{-1}(\theta)V^T$ corresponds to the Floquet Normal Form,
which has as main characteristic that the (1,2) element
in the transformation of Eq.~(\ref{Floquet}) is zero.
We have to solve the following equation,
  \begin{equation}
   \frac{1}{\sqrt{\det(S)}} USR(\theta) = \begin{pmatrix} \sqrt{\beta} & 0  \\ -\alpha/\sqrt{\beta} & 1/\sqrt{\beta} \end{pmatrix}\ ,
   \end{equation}
  where   $\theta$ is determined to make zero the element (1,2)
  of $USR(\theta)$. The normalization factor $\sqrt{\det(S)}$ is related
  to the single particle emittance as follows,
  \begin{equation}
  \epsilon=\frac{\det(S)}{n/2} \ ,
\end{equation}
  where $n$ is the number of turns. The following Python code
  implements the function $getbeta(x,px)$ that computes
  $\alpha$, $\beta$ and $\epsilon$ from turn-by-turn data together
  with an illustrative example. Figure~\ref{V} shows the $(x,x')$
  turn-by-turn data used in the code example together
  with the first two right-singular vectors of $V$.
  \begin{lstlisting}[language=Python]
#Computing alfa, beta and epsilon using SVD
import numpy as np
def getbeta(x,px):         # Function to return   betx, alfx, ex 
    U, s, V = np.linalg.svd([x,px])    # SVD 
    N = np.dot(U,np.diag(s))	         
    theta = np.arctan(-N[0,1]/N[0,0])   # Angle of R(theta)
    co = np.cos(theta); si = np.sin(theta)
    R = [ [co, si] , [-si, co] ]	       
    X = np.dot(N,R)                     # Floquet up to 1/det(USR)
    betx = np.abs(X[0,0]/X[1,1])        
    alfx = X[1,0]/X[1,1]
    ex=s[0]*s[1]/(len(x)/2.)         # emit = det(S)/(n/2)
    return betx, alfx, ex 

alpha = 0.2             #Example to use getbeta(x,px)
beta = 1.
ex = 2e-3
Q = 0.31
Nturns = 600
x = np.sqrt(beta*ex)*np.cos(2*np.pi*Q*np.arange(0,Nturns)) #easy tracking
px = -alpha*x/beta + np.sqrt(ex/beta)*np.sin(2*np.pi*Q*np.arange(0,Nturns))
betx, alfx, exc = getbeta(x,px)
\end{lstlisting}
A first version of this code was developed in~\cite{Gon}.

\subsection{Excitation techniques}
\begin{figure}\centering
  \includegraphics[trim = 20mm 0mm 0mm 1mm, clip,height=10cm, angle=-90]{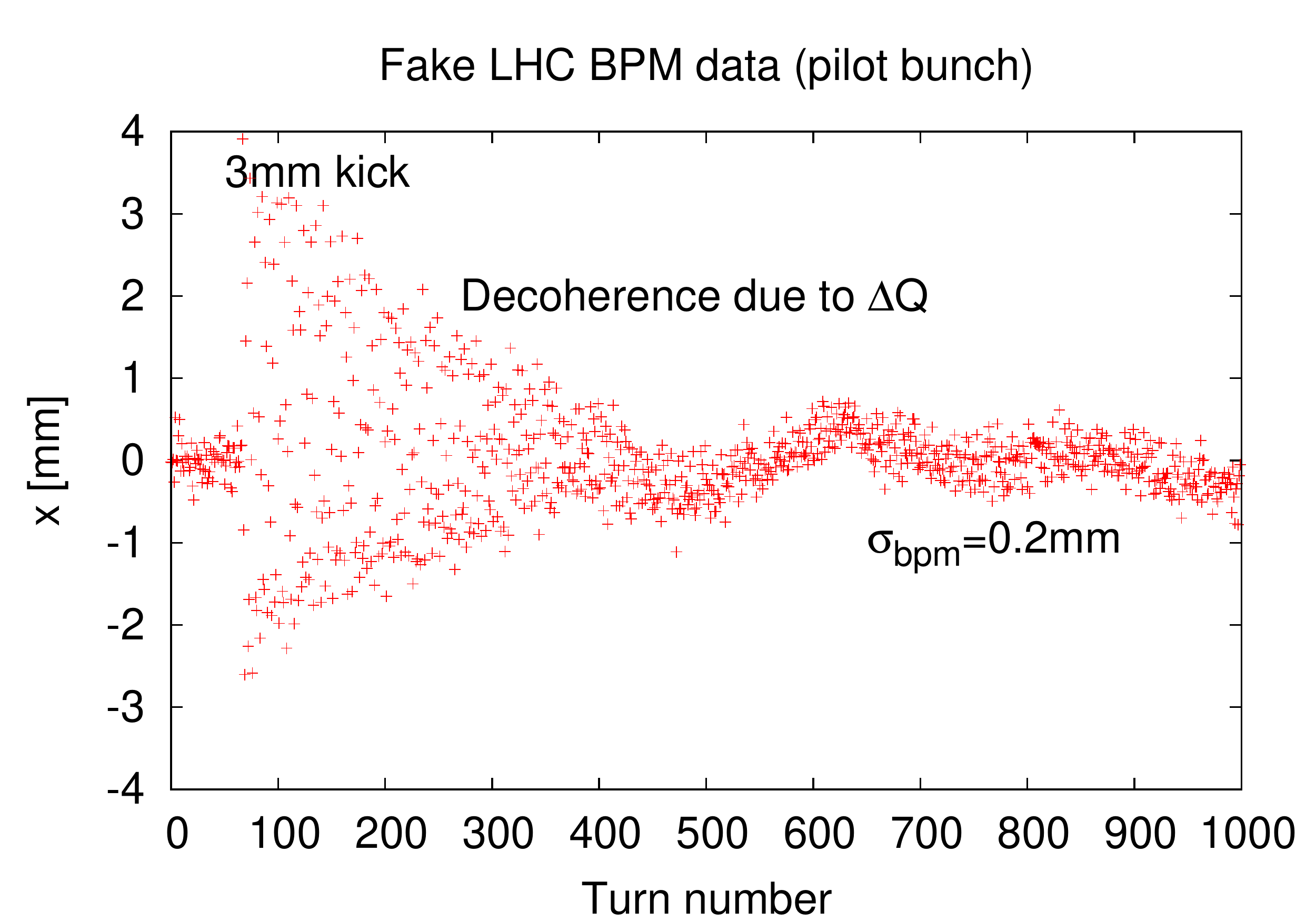}
  \caption{Illustration of turn-by-turn centroid data recorded by a BPM  undergoing
    decoherence after a transverse kick and with a BPM Gaussian noise of 0.2~mm.\label{singlekick}}
\end{figure}

In real accelerators 
Beam Position Monitors (BPMs) measure transverse
beam centroid position turn-by-turn, while the angle of the trajectory $x'$
is not easily accessible.
Betatron motion is usually excited  via applying a single
kick   or via a resonant excitation
The single kick technique has the limitation
that due to non-linearities not all the particles
in the bunch oscillate with the same tune the motion
eventually decoheres. The measured turn-by-turn data
following a single kick is illustrated in Figure~\ref{singlekick}.
The evolution of the decoherence process  is illustrated
in Figure~\ref{deco}, where the initial 1$\sigma$ envelope
after the kick is shown in green and the deformation of this envelope
with time emerges from the larger tune values for
particles with larger amplitude (amplitude detuning is typically
generated by sextupoles, octupoles or the beam-beam interaction).
The decoherence of the beam limits the number of turns
available for data analysis and therefore limits the accuracy
of beam-based optics measurements. 

\newcommand{\hdec}{5.7cm}
\begin{figure}\centering
  \includegraphics[trim = 25mm 0mm 25mm 0mm, clip, height=\hdec,angle=0]{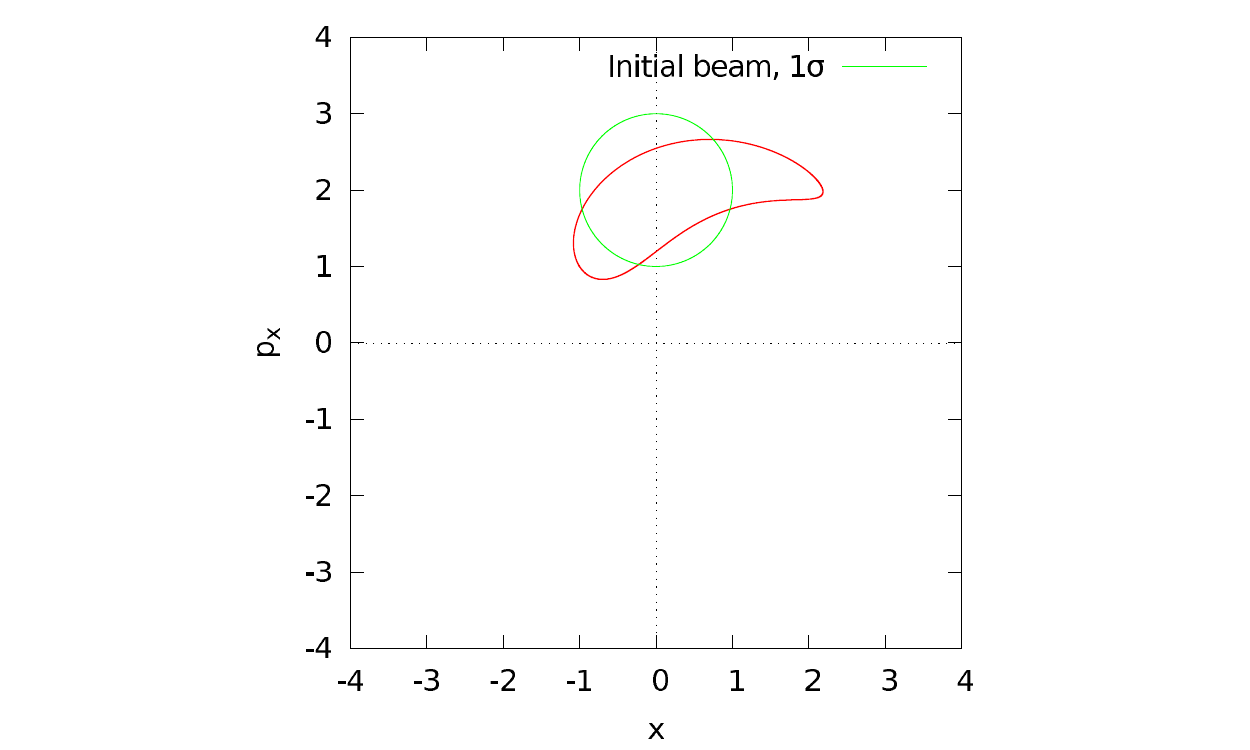}
  \includegraphics[trim = 25mm 0mm 25mm 0mm, clip, height=\hdec,angle=0]{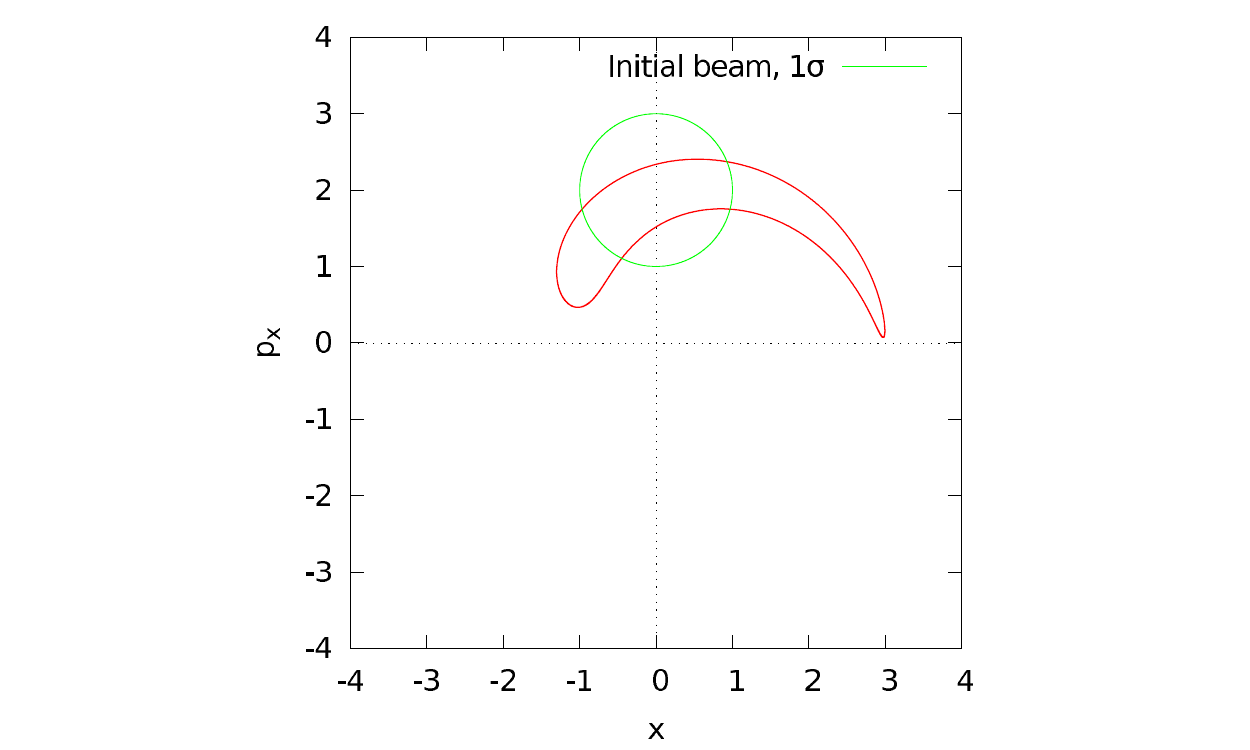}\\
  \includegraphics[trim = 25mm 0mm 25mm 0mm, clip, height=\hdec,angle=0]{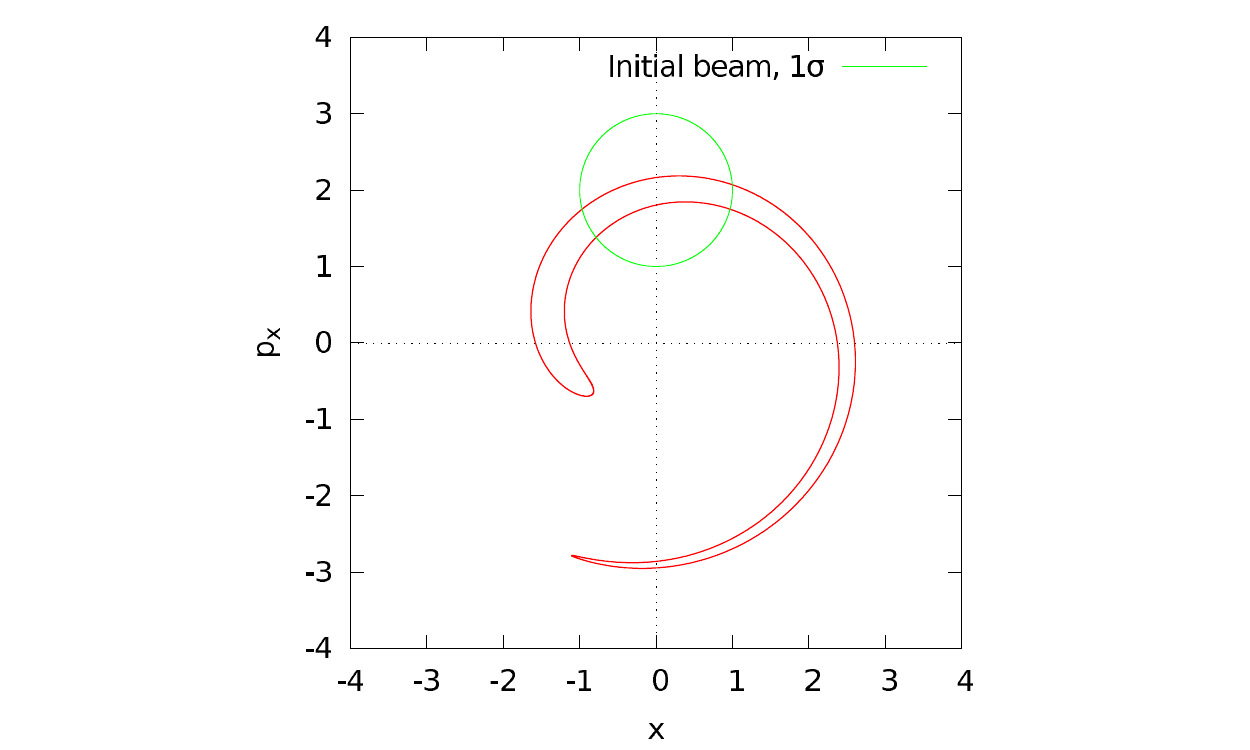}
  \includegraphics[trim = 25mm 0mm 25mm 0mm, clip, height=\hdec,angle=0]{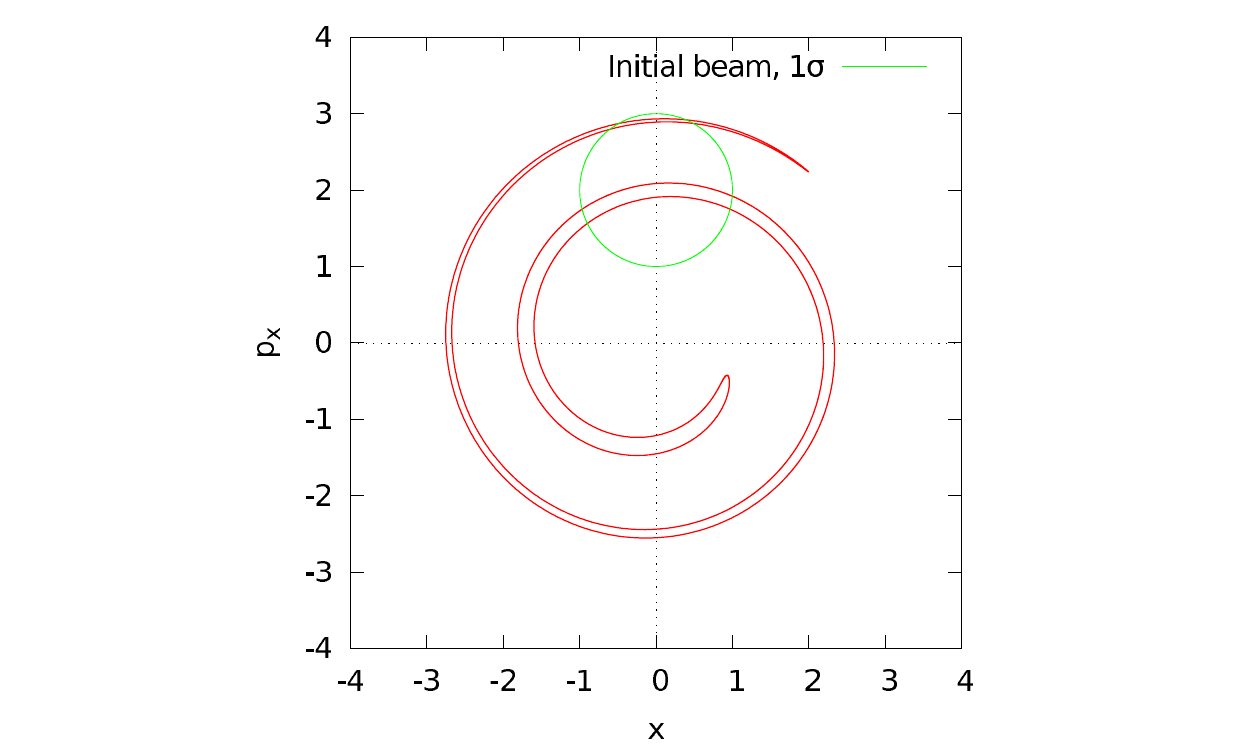}\\
  \includegraphics[trim = 25mm 0mm 25mm 0mm, clip, height=\hdec,angle=0]{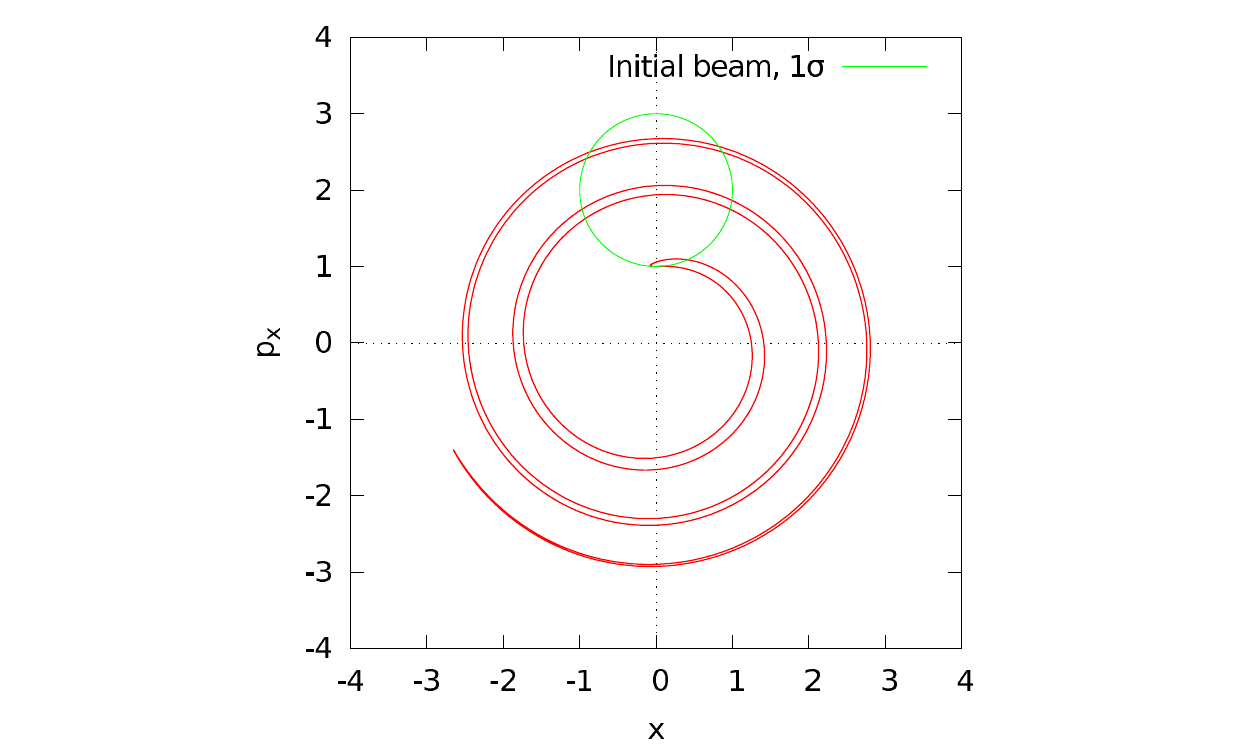}
  \includegraphics[trim = 25mm 0mm 25mm 0mm, clip, height=\hdec,angle=0]{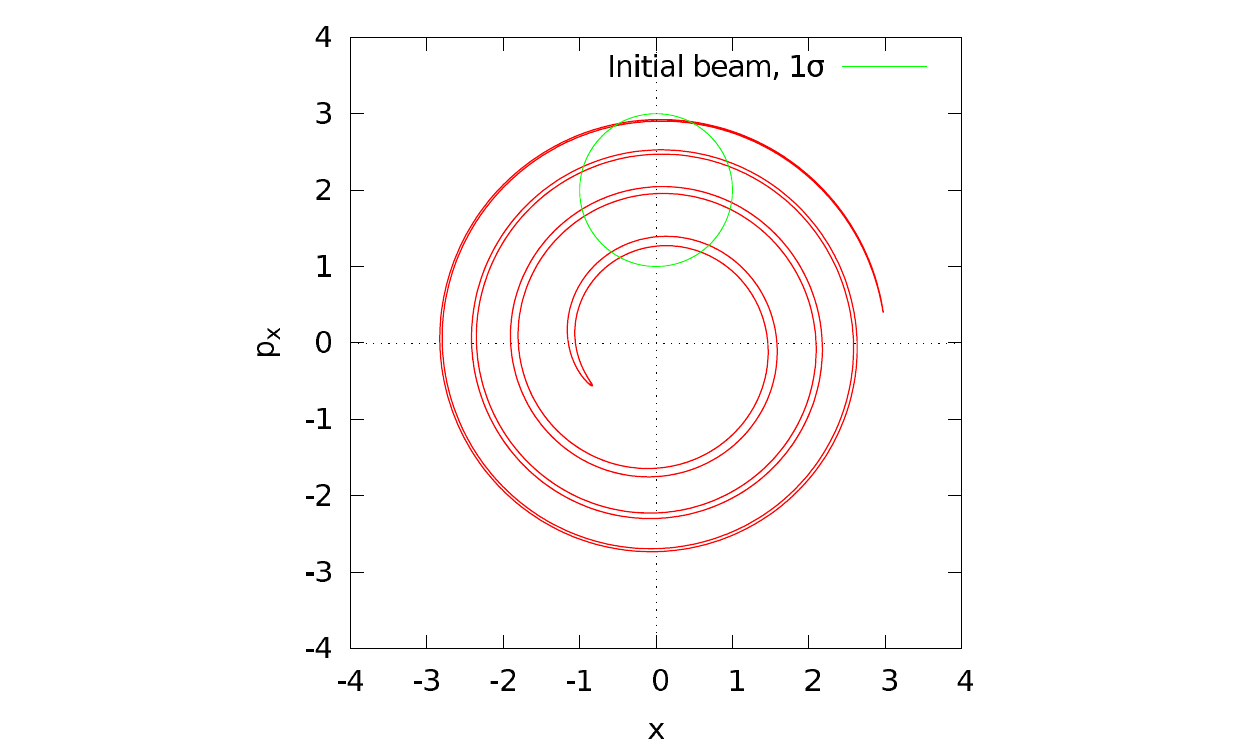}\\
  \includegraphics[trim = 25mm 0mm 25mm 0mm, clip, height=\hdec,angle=0]{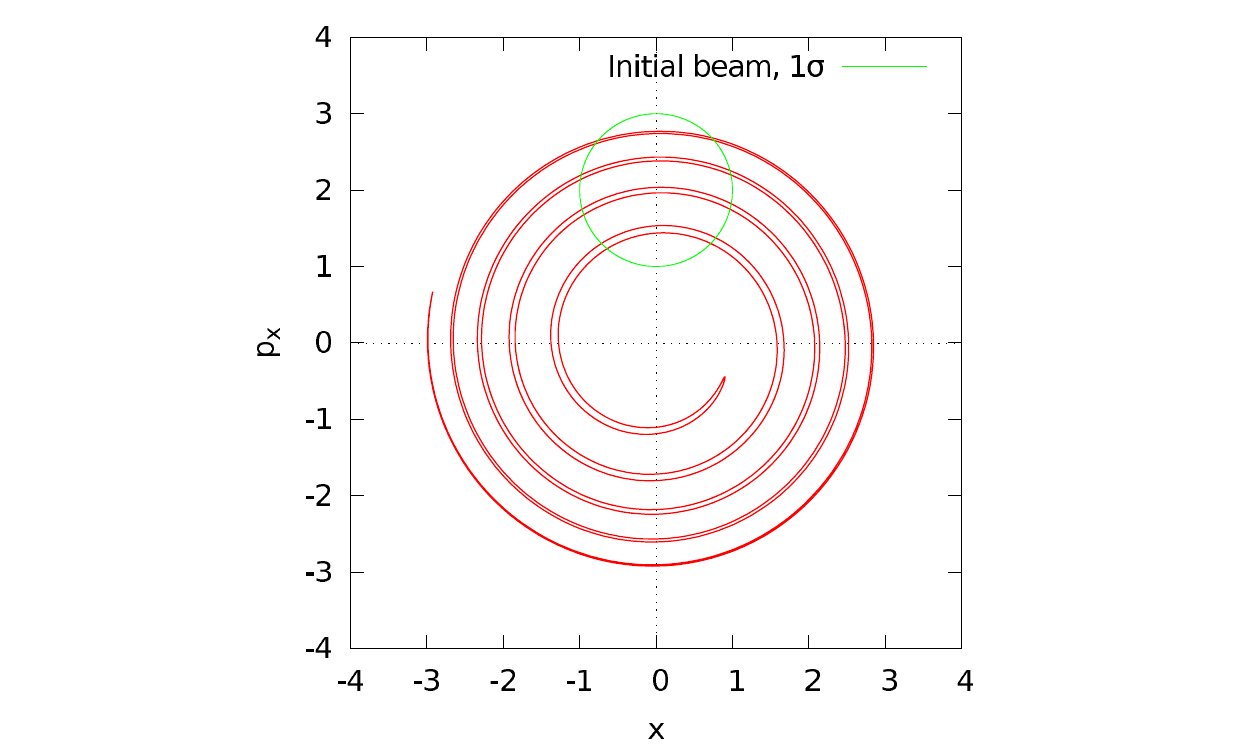}
  \includegraphics[trim = 25mm 0mm 25mm 0mm, clip, height=\hdec,angle=0]{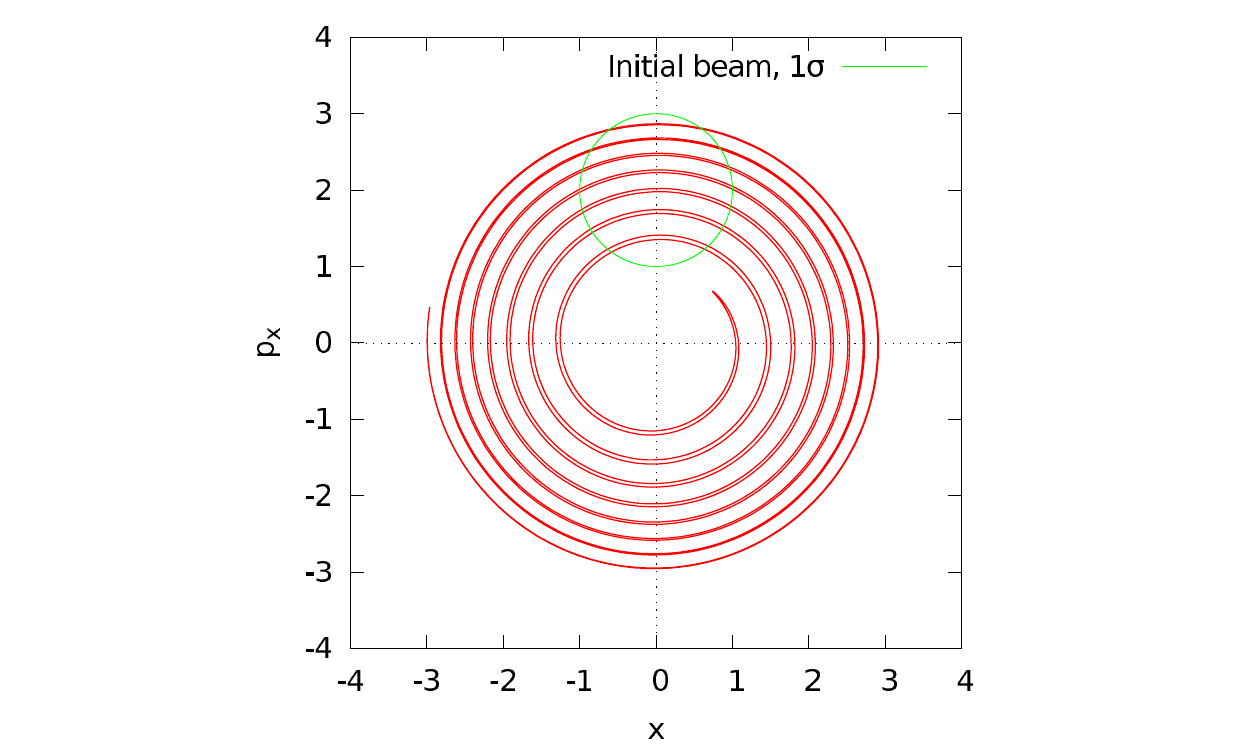}
  \caption{Illustration of the decoherence process.\label{deco}}
\end{figure}

\begin{figure}\centering
\includegraphics[height=12cm, width=3.5cm, angle=-90]{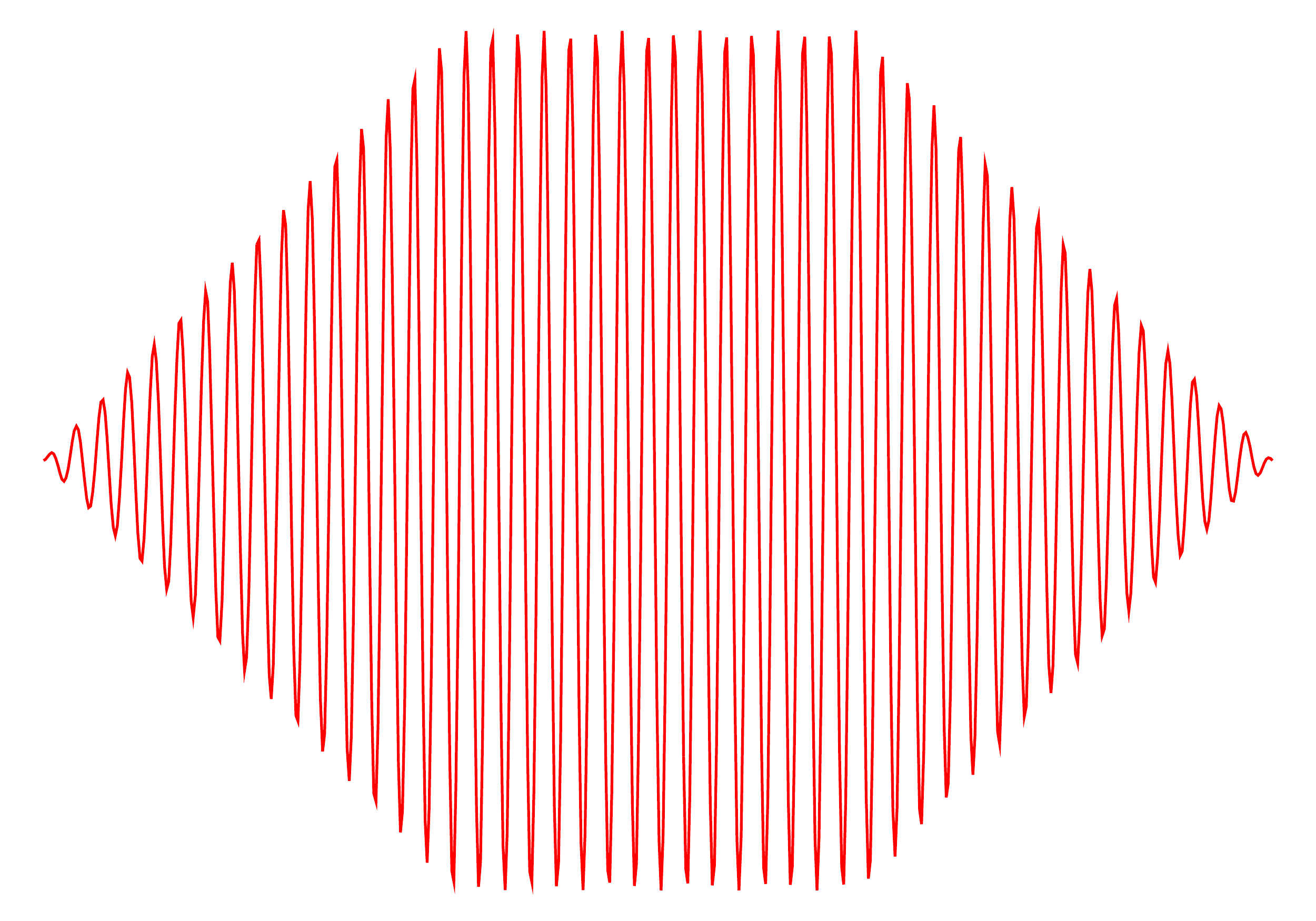}
\caption{Illustration of the AC dipole cycle including
  ramp-up, plateau and ramp-down.\label{acdip}}
\end{figure}

To avoid the limitation from decoherence it is possible
to drive betatron oscillations with an AC~dipole
with a frequency close to the tune. Furthermore this forced
oscillation can be ramped up and down adiabatically without causing
emittance growth. The AC dipole cycle is illustrated in Fig.~\ref{acdip}.

\begin{figure}\centering
    {\includegraphics[trim=30mm 0mm 35mm 0mm, width=\textwidth, angle=-0]{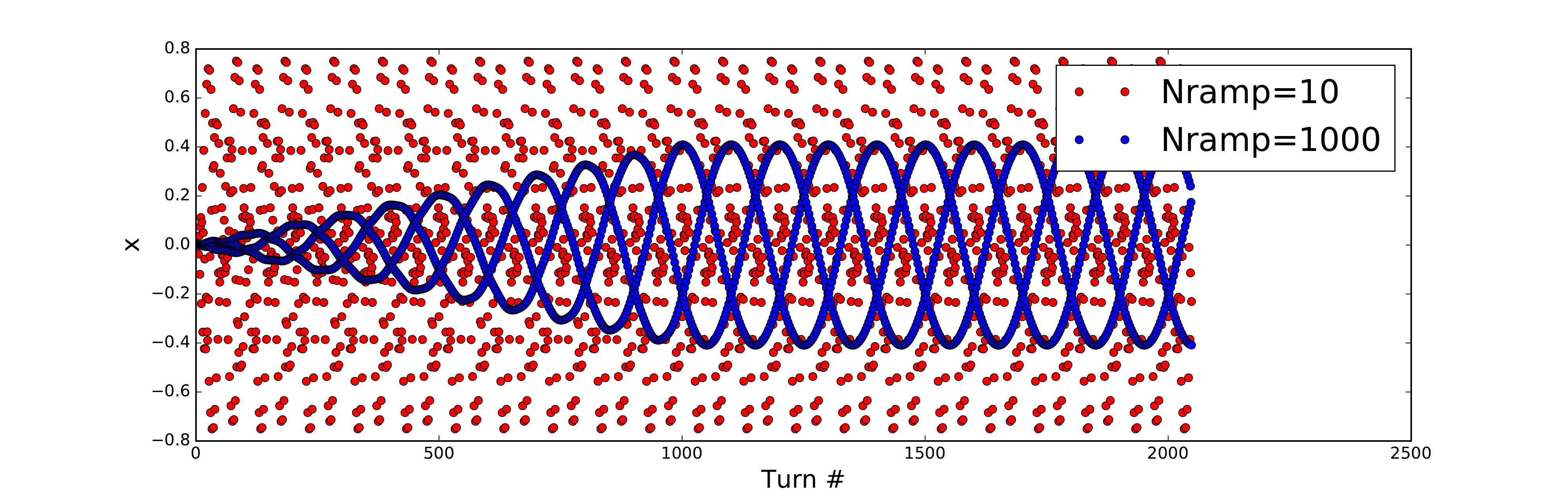}}
    \caption{Simulated turn-by-turn beam data during the AC dipole
      excitation for two different ramp lengths of 10 and 1000 turns, showing the relevance of an adiabatic excitation.\label{adiabaticity}}
\end{figure}

\begin{figure}\centering
    {\includegraphics[width=10.2cm, angle=-0]{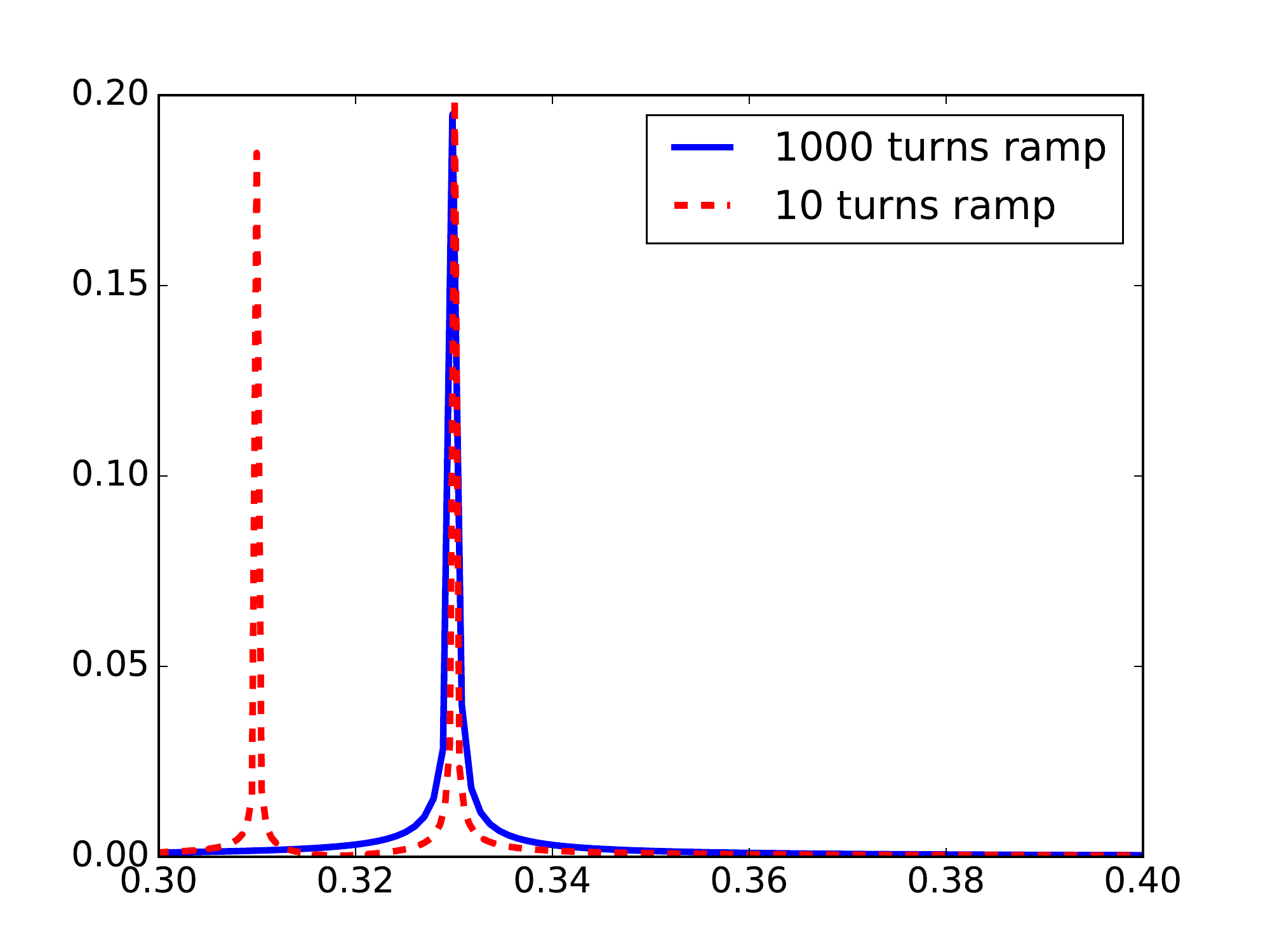}}
    \caption{Spectrum of the simulated turn-by-turn beam data during the AC dipole plateau following two different ramp lengths of 10 and 1000 turns, showing the appearance of the natural tune for the non-adiabatic excitation.
      AC dipole tune is 0.33 and natural tune is 0.31.\label{acspectra}}
\end{figure}

The adiabaticity of the ramping process of an AC dipole~\cite{adiabatic}
can be easily simulated with the following code, which is used to
produce the plot in Fig.~\ref{adiabaticity}
that compares the particle turn-by-turn motion for
two AC dipole ramp-up lengths, 10 and 1000 turns, showing
a lack of adiabaticity for the 10~turn ramp.
The lack of adiabaticity implies energy transfer to the natural
betatron motion with tune equal 0.31 as shown in the spectral
components of particle motion in Fig.~\ref{acspectra}
as computed in the following example code.

\begin{lstlisting}[language=Python]
#Simulating the AC dipole 
from numpy import *
import matplotlib.pyplot as plt

Q = 0.31         # Machine tune (fractional part)
Qac = Q + 0.02   # AC dipole tune
q = 2*pi*Q
R = array([[cos(q), -sin(q)],[sin(q), cos(q)]]) #1 turn map
x=[[0.,0.]]   # initial x, px
Nramp = 1000    # Number of turns to ramp up AC dipole strength
Nturn = 2048    # Number of turns to track

def ramp(j):  # define the AC dipole linear ramp
    return min(1, j*1.0/Nramp)  

for i in range(Nturn):   # tracking loop  R with AC dipole kick
   x.append( dot(R,x[-1]) + ramp(i)*array([0, 0.1*cos(Qac*i*2*pi)]))
F = fft.fft(array(x)[Nramp:].T[0])  # FFT data after AC ramp
\end{lstlisting}

\section{Measurement techniques and data analysis}\label{sec5}

\subsection{Cleaning experimental BPM data}
The BPM turn-by-turn data is  fundamental
to measure optics parameters around the accelerator.
Betatron oscillations represent highly correlated
signals among BPMs. This feature can be used to
reduce the BPM noise by discarding the signals with
low correlation levels. SVD is used for this purpose.
Imagine $R$ is the BPM matrix containing turn-by-turn data
for all BPMs and his SVD is given by,
\begin{equation}
R=U
\left( \begin{array}{ccc}
\sigma_1 & 0 & 0 \\
0 & \sigma_2 & 0 \\
0 & 0 & \sigma_3 \\
0 & 0 & 0 \end{array} \right)
V^T \ .
  \end{equation}
If $\sigma_3\ll\sigma_2\leq\sigma_1$, then we can neglect $\sigma_3$
by making $\sigma_3=0$
and reconstruct $R$ loosing a negligible amount of information.
Denoting the reconstructed matrix as  $R_{denoised}$, it is given
by the following equation,
  \begin{equation}
  R_{denoised}=U
\left( \begin{array}{ccc}
\sigma_1 & 0 & 0 \\
0 & \sigma_2 & 0 \\
0 & 0 & 0 \\
0 & 0 & 0 \end{array} \right)
V^T \ .
  \end{equation}

\begin{figure}\centering
\includegraphics[trim = 20mm 10mm 10mm 12mm, clip,height=7.1cm, angle=-0]{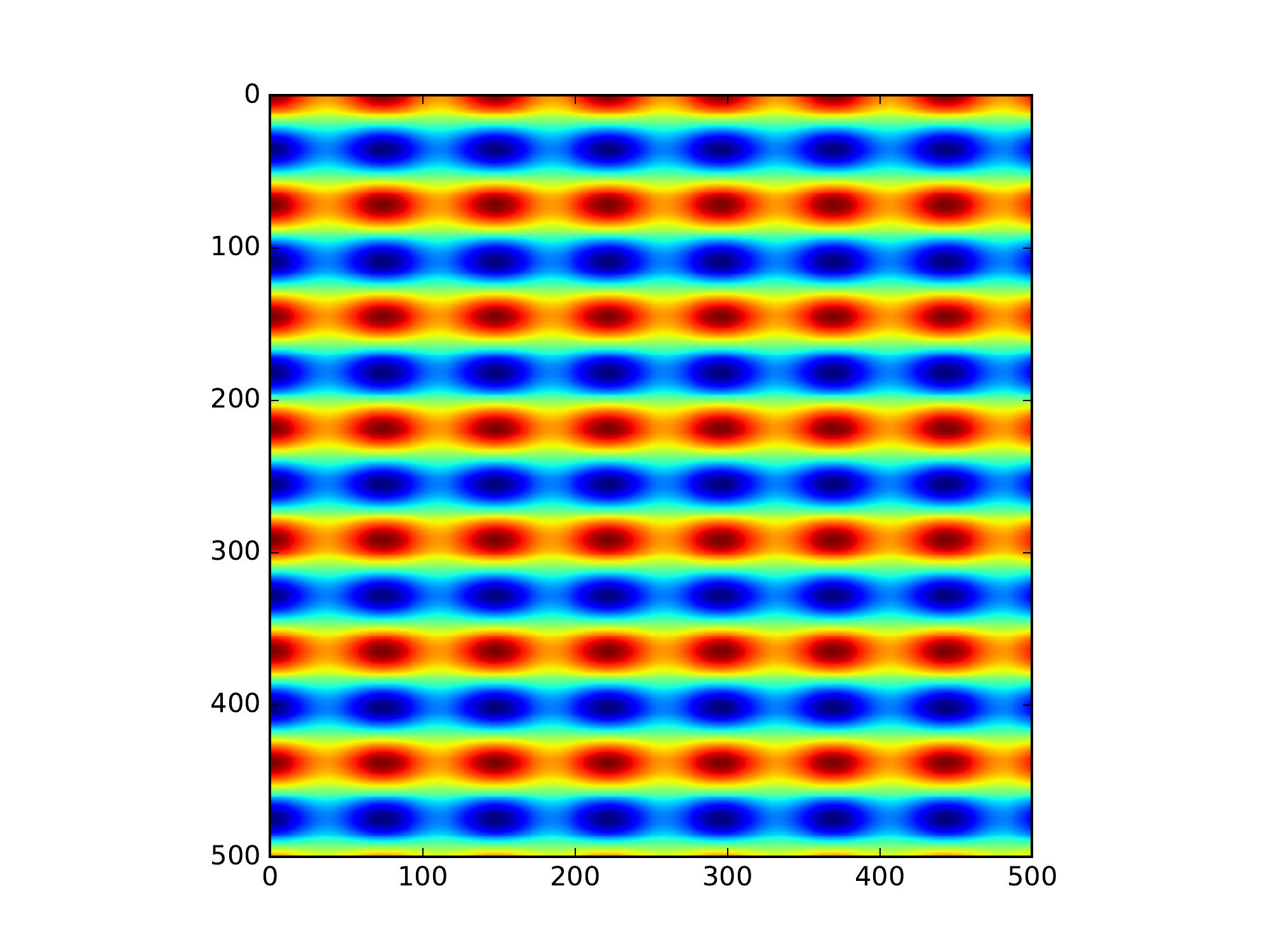}
\includegraphics[trim = 20mm 10mm 10mm 12mm, clip,height=7.1cm, angle=-0]{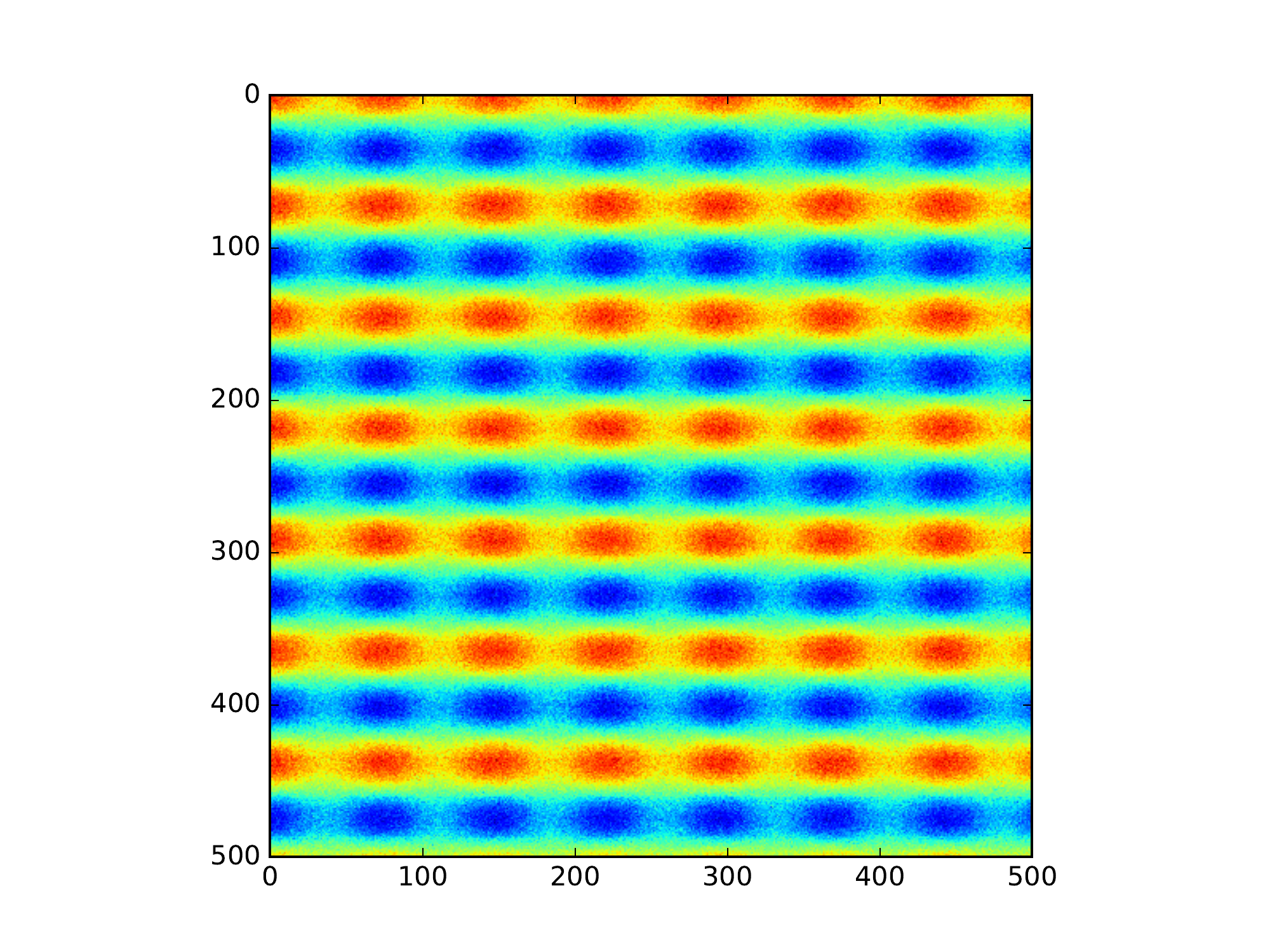}
\includegraphics[trim = 20mm 10mm 10mm 12mm, clip,height=7.1cm, angle=-0]{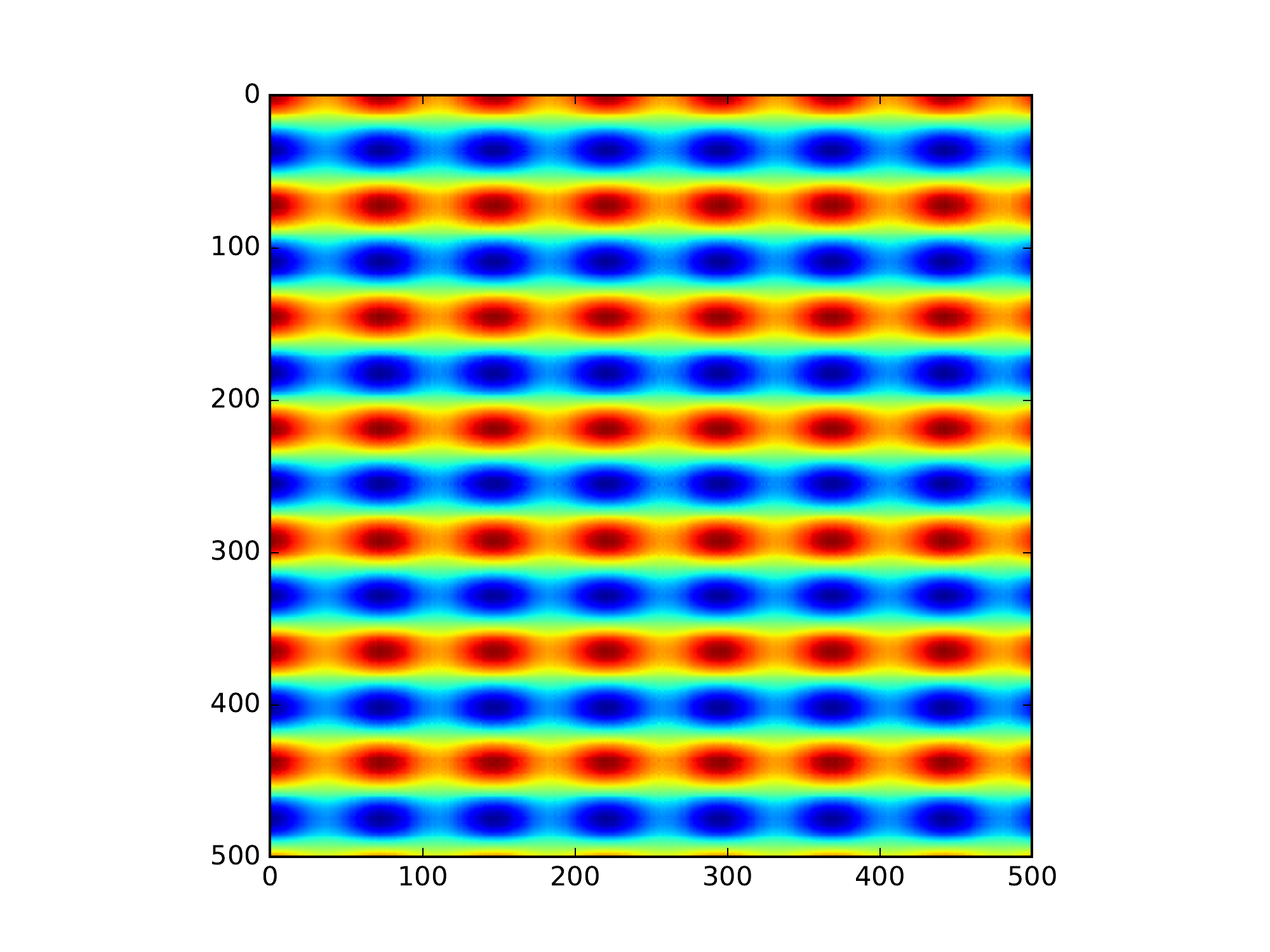}
  \caption{Ideal beam data versus turn number and versus longitudinal location (top), same data adding Gaussian noise (middle) and  after cleaning the noise with SVD (bottom).\label{svdclean}}
\end{figure}

This technique is illustrated with the following
Python code and in Fig.~\ref{svdclean}.
In the code  turn-by-turn data is simulated
with very low tunes to produce pictures that
can be easily visualized.
Random Gaussian noise is added to mimic
BPM noise with signal-to-noise ratio varying between
1:0.2 and 2:0.2. The SVD reconstruction is performed
by keeping only the two largest singular values.
The process is illustrated in Fig.~\ref{svdclean}
showing the 3 matrices in color code.
It is impressive that the reconstructed matrix
looks identical to the ideal one before adding the noise.
Actually this technique is equally used to denoise
digital pictures.

\begin{lstlisting}[language=Python]
# Denoising BPM signal
import matplotlib.pyplot as plt
from scipy import misc,ndimage
import numpy as np
from numpy.linalg import svd

#Generating ideal  Beam Position data
im = np.zeros((500, 500))
for i in range(500):
    for j in range(500):
       amplitudej=1+(np.cos(0.00678*j*2*np.pi)**2  
       im[i,j] = amplitudej * np.cos(i*0.0137*2*np.pi)

#Adding noise like measurement error
im = im + 0.2 * np.random.randn(*im.shape)

#Denoising  with Singular Value Decomposition
k=2
U,s,V=svd(im, full_matrices=False)
rim = np.dot(U[:,:k], np.dot(np.diag(s[:k]),V[:k,:]))
\end{lstlisting}

\begin{figure}\centering
    {\includegraphics[trim=10mm 20mm 10mm 20mm  ,width=15cm, angle=-0]{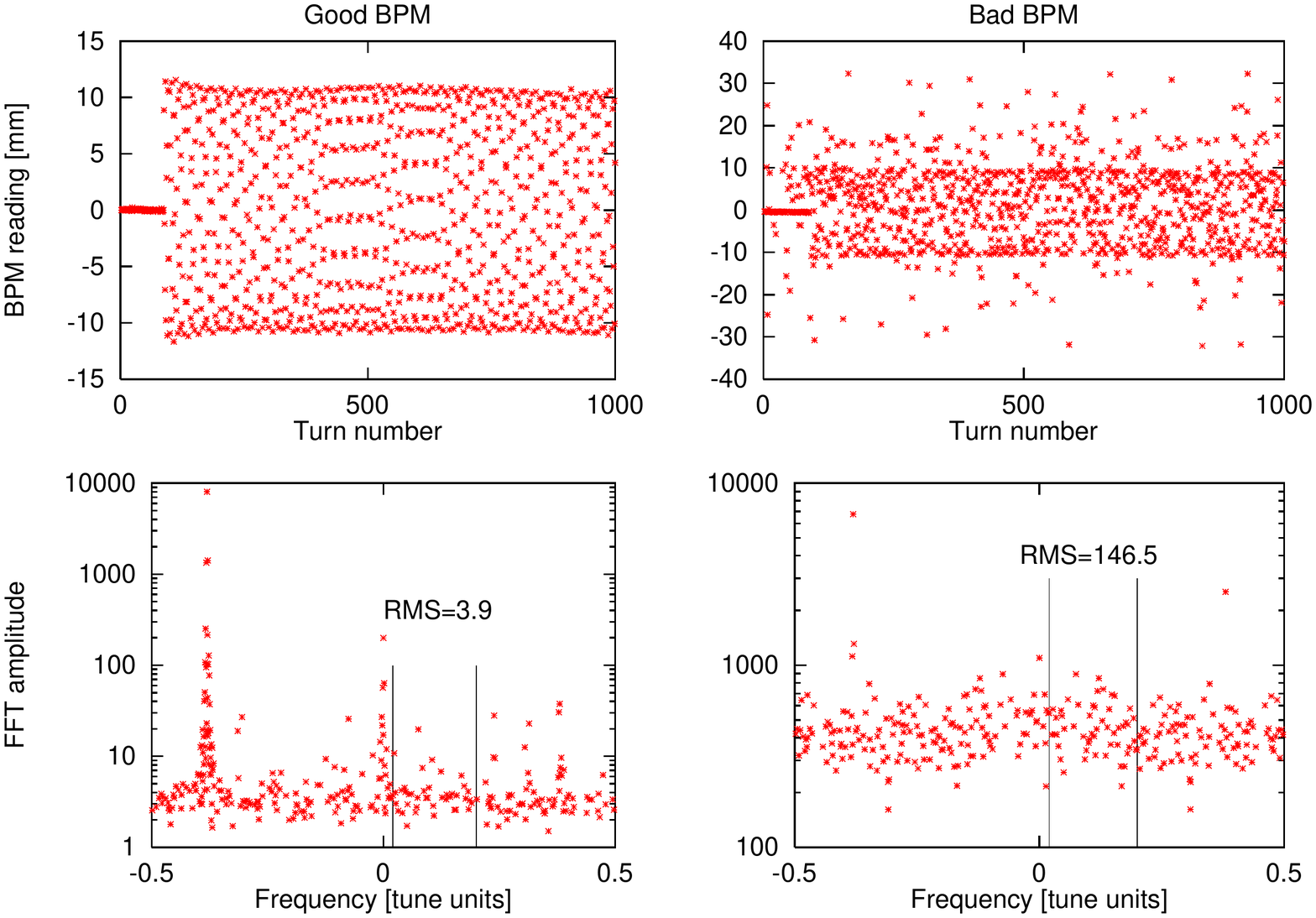}}
    \caption{Good BPM (left) and bad BPM (right) with corresponding spectra (bottom), from~\cite{ro}.\label{badbpms}}
\end{figure}

Large BPM systems always present some malfunctioning BPMs
that need to be removed before the analysis.
An example of good and bad BPMs is shown in Fig.~\ref{badbpms}
from the CERN SPS~\cite{ro}.
The plots in the bottom of the figure show how  the
Fourier spectrum can be used to identify bad BPMs
by looking in regions of the spectra where no beam signal
is expected. SVD has also been extensively used
to identify bad BPMs~\cite{MIA,rhicbpms}.

\begin{figure}\hspace{2.5cm}
  \includegraphics[trim = 35mm 98mm 35mm 108mm, clip,height=6cm, angle=-0]{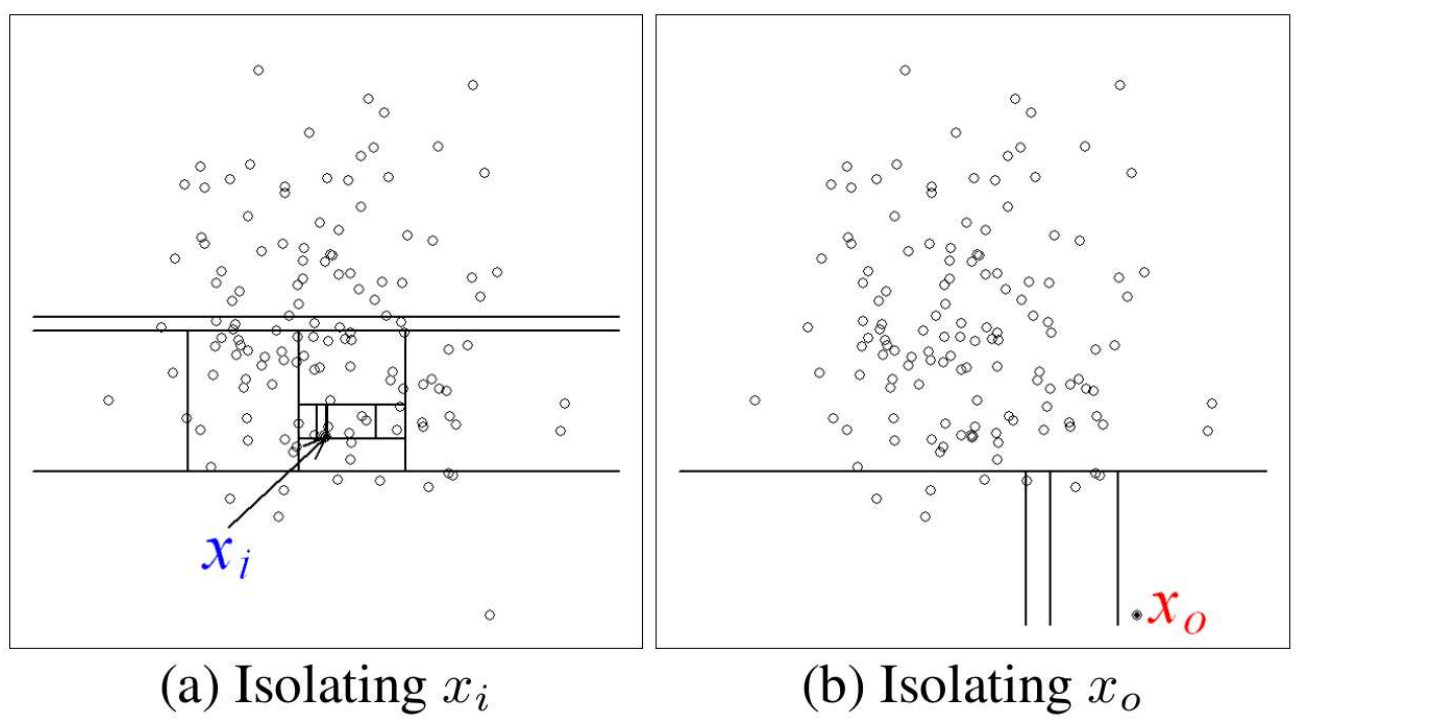}
  \caption{Illustration of the Isolation Forest algorithm
    applied to a normal data point (left) requiring many
    cuts to reach isolation and to an anomaly (right) with
    fewer cuts.\label{IF}}
\end{figure}

More recently Isolation Forest has been demonstrated
to be very effective at finding malfunctioning BPMs
as outliers within the distribution of selected features
of the BPM data~\cite{efolibic18,efolipac19}.
Figure~\ref{IF} illustrates the concept of the Isolation
Forest algorithm, where random cuts are applied
to the data
for one randomly selected feature at a time
until single data points are isolated.
The basic concept is that anomalies
require fewer number of cuts to reach isolation.
A decision function is established using this number
averaged over the number of trees.

\begin{figure}\centering
  \includegraphics[trim = 0mm 0mm 0mm 0mm, clip,height=7.1cm, angle=-0]{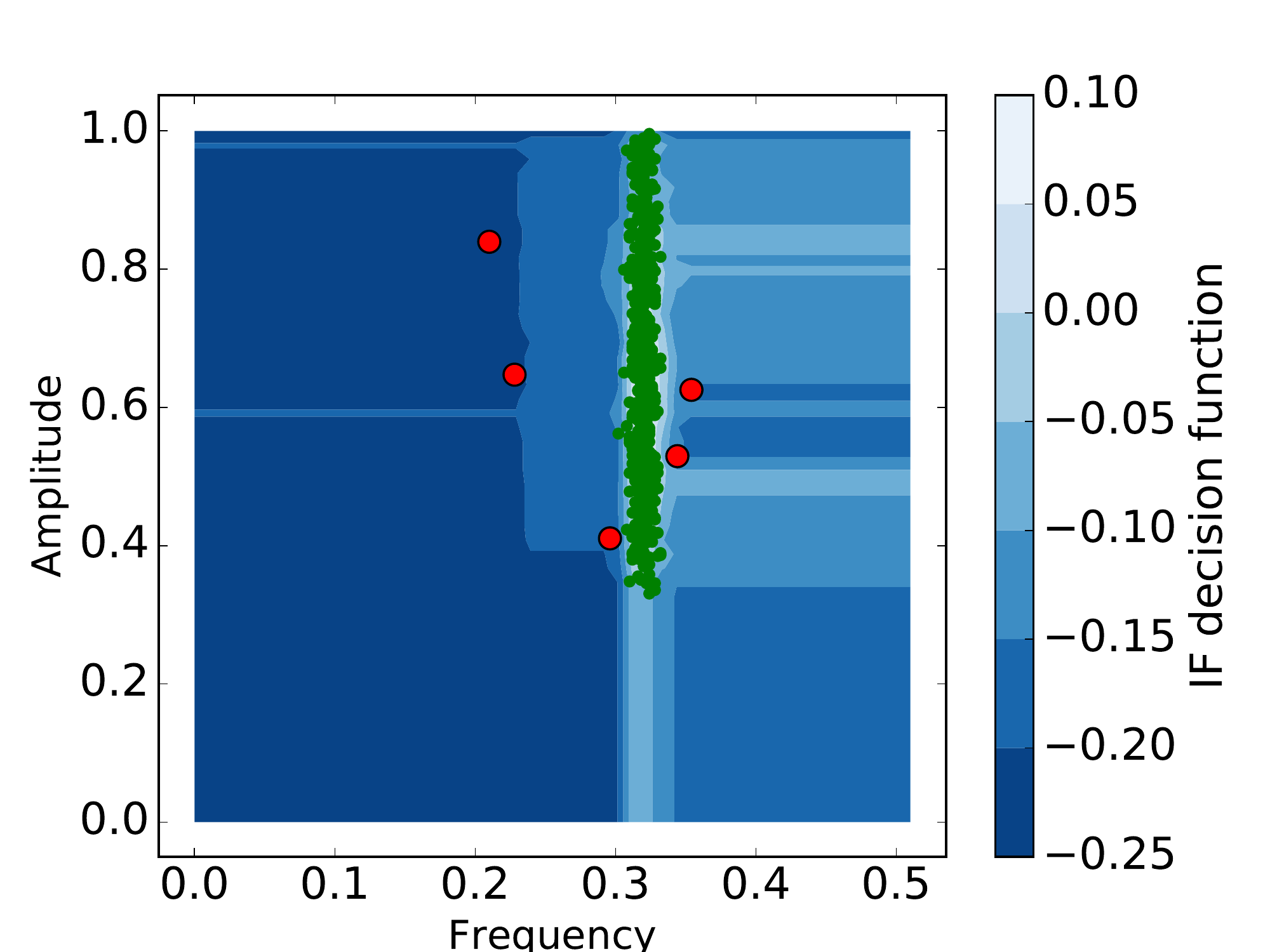}
  \caption{Isolation Forest applied to BPM data using frequency and amplitude of highest Fourier peak as features.\label{IFex}}
\end{figure}

The following Python code applies the Isolation Forest to
simulated turn-by-turn BPM data with Gaussian noise and five bad BPMs.
In this illustration the bad BPMs are chosen to have larger
Gaussian noise and a different tune.
The features chosen to compute the decision function
are amplitude and frequency of the main spectral line.
The features for all BPMs are shown in Fig.~\ref{IFex} together with the
decision function. The red BPMs are the BPMs identified as bad
by assuming a contamination factor of~1\%.

\begin{lstlisting}[language=Python]
# Applying Isolation Forest to detect bad BPMs
import numpy as np
from sklearn.ensemble import IsolationForest

N_TURNS = 500
N_BPMS = 500
# generate bpm data with some bad signal - different tune, additional noise
bad_bpms_idx = [1, 10, 20, 30, 40]
im = np.zeros((N_TURNS, N_BPMS))
for bpm in range(N_BPMS):
    err= 0.05 * np.random.randn()
    amp=(np.cos(0.00678 * bpm * 2 *np.pi) ** 2 + 1)  # sqrt(beta e)
    for turn in range(N_TURNS):  
        if bpm in bad_bpms_idx:   # A bad BPM with different tune and noise
          im[turn,bpm]=amp*np.cos(turn*(0.32+err)*2*np.pi)+0.3*np.random.randn()
        else:              # Good BPM 
          im[turn,bpm]=amp*np.cos(turn*(0.32+err/10)*2*np.pi) + 0.1*np.random.randn()

# extract frequency and amplitude - features - from bpm signal
amplitudes = [np.max(x) for x in np.abs(np.fft.rfft(im.T))/N_TURNS]
frequencies= np.array([np.argmax(x) for x in np.abs(np.fft.rfft(im.T))])*1.0/N_TURNS
features = np.vstack((frequencies, amplitudes)).T

# fit Isolation Forest model to the data and detect anomalies (contamination is the fraction of anomalies)
iforest = IsolationForest(n_estimators=10, contamination=0.01)
outlier_detection = iforest.fit(features).predict(features) # Bad BPMs ==-1
\end{lstlisting}

\subsection{Generic measurement cleaning}
Most of the measured quantities are assumed to be normally distributed, 
however in case of failure or an artefact in data processing; 
outlying values may be produced and should be removed from the data sample. 
Finite-sized samples of a normal distribution follow a t-student distribution, 
which is also parametrised by a number of degrees of freedom. 

An iterative cleaning procedure (developed in~\cite{lukas_thesis}) removes ``tails''
 which are more populated than in the same-sized normally distributed quantity. 
In each iteration, values are tested for a hypothesis of belonging to a sample of 
normal distribution given the mean value, standard deviation and sample size (t-distribution). 
The algorithm represented by the following code can also operate onto two linearly dependent sets. 
In such a case, the fitted dependency on a second dataset is subtracted in every iteration. 
\begin{lstlisting}[language=Python]
# Iterative cleaning
import numpy as np
from scipy.stats import t
import matplotlib.pyplot as plt

def filter_mask(data, x_data=None, limit=0.0, niter=20):
    mask = np.ones(len(data), dtype=bool)
    nsigmas = t.ppf([1 - 0.5 / len(data)], len(data))
    prevlen = np.sum(mask) + 1
    for _ in range(niter):  # iterate
        if not ((np.sum(mask) < prevlen) and (np.sum(mask) > 2)):
            break
        prevlen = np.sum(mask)
        if x_data is not None:  # linearly dependent data
            m, b = np.polyfit(x_data[mask], data[mask], 1)
            y, y_orig = data[mask] - b - m * x_data[mask], data - b - m * x_data
        else:  # independent data
            y, y_orig = data[mask], data[:]
        mask = np.abs(y_orig - np.mean(y)) < np.max([limit, nsigmas * np.std(y)])
    return mask

x_data = 100 * np.random.rand(1000)
y_data = 0.35 * x_data + np.random.randn(1000)  # create data
y_data[-100:] = y_data[99::-1]  # corrupt some of the data
x_data[:50] = 38 + np.random.randn(50)
mask = filter_mask(y_data, x_data=x_data)
plt.plot(x_data, y_data, 'ro')
plt.plot(x_data[mask], y_data[mask], 'bo')
\end{lstlisting}

\subsection{Fourier analysis}
The Fast Fourier Transform (FFT) of a turn-by-turn data sample with $N$ turns
has the following tune ($Q$), amplitude ($A$) and phase ($\phi$) resolutions, respectively
\begin{equation}
  \sigma_{Q} \leq \frac{1}{2N}\ ,\ \  \sigma_{A} \approx \sqrt{\frac{2}{N}}\sigma  \ , \ \  \sigma_{\phi}\approx\sqrt{\frac{2}{N}}\frac{\sigma}{A}\ ,
\end{equation}
where $\sigma$ is the BPM random error, assumed to follow a Gaussian
distribution.

Many interpolation techniques have been developed to improve the
frequency resolution of the FFT~\cite{laskar,sadi,bartolini}.
Zero-padding is a very simple approach that can significantly
improve the determination of fundamental frequencies but is
computationally expensive. A Python example using zero-padding follows.
\begin{lstlisting}[language=Python]
#FFT with zero padding
import numpy as np
N = 4096
i = 2 * np.pi * np.arange(N)
data = np.cos(0.134 * i) + np.cos(0.244 * i) + 0.01 * np.random.randn(N)
f_zeropad=np.abs(np.fft.fft(data, n=10*N)/(N))
\end{lstlisting}

\begin{figure}\centering
\includegraphics[trim = 0mm 0mm 0mm 10mm, clip,height=6.4cm, angle=-0]{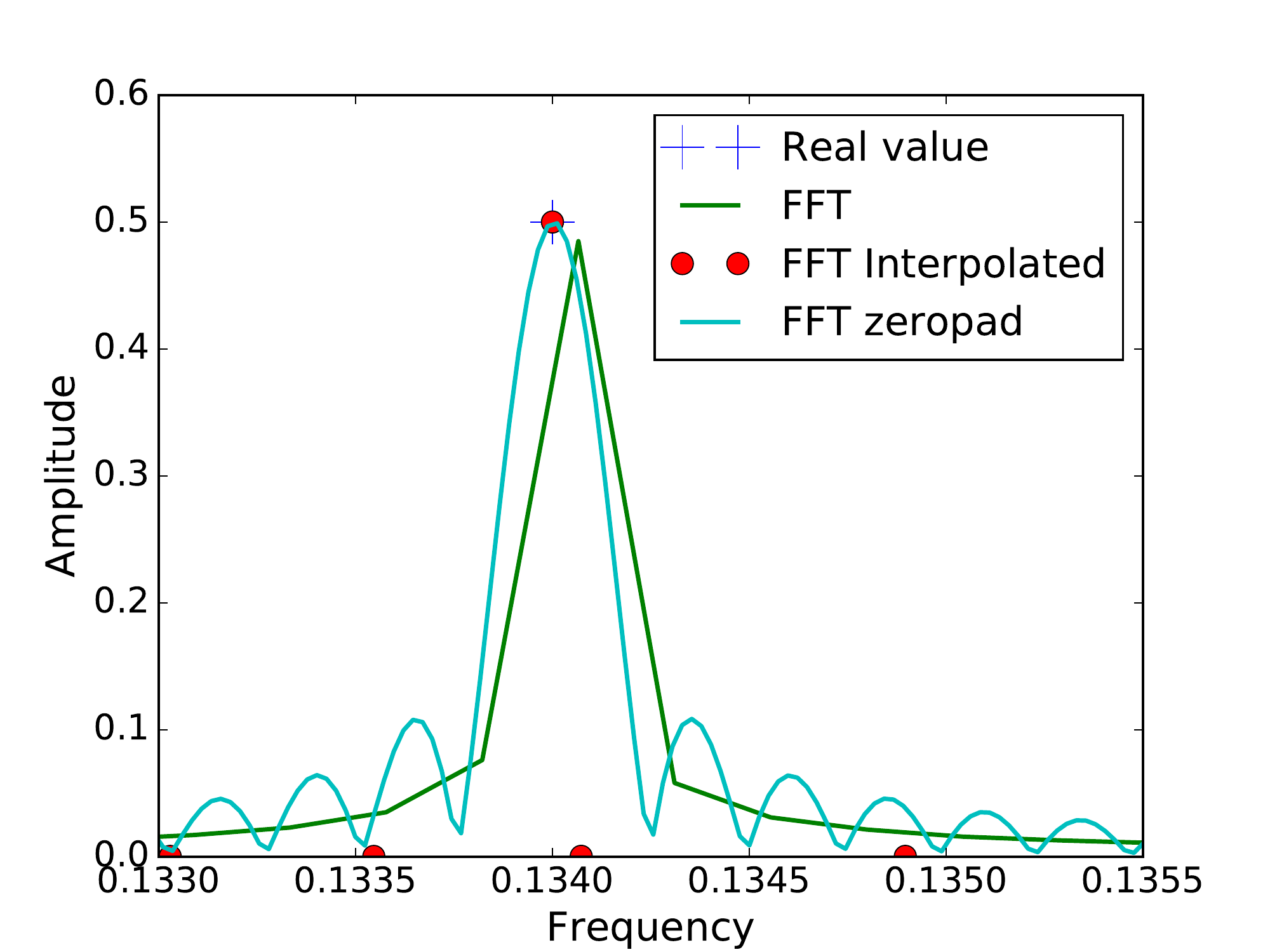}
\caption{Illustration of different algorithms to find the main spectral
  frequencies.\label{FTs}}
\end{figure}

The algorithm NAFF~\cite{laskar} finds the frequency $Q$ that maximizes $|\sum x(N)e^{i2\pi QN}|$, where $x(N)$ is the sample data, and continues
to find the next leading frequency after subtracting the found frequency component from $x(n)$ and iterating. Python and Fortran implementations of NAFF
can be found in~\cite{pynaff,sussix}.
The following code, first version of Harpy~\cite{harpy}, implements the NAFF algorithm but making
a 3~point interpolation (Jacobsen method~\cite{jacobsen}) rather than maximizing $|\sum x(N)e^{i2\pi QN}|$.
Figure~\ref{FTs} shows the spectrum of the signal in the Python
example computed with different approaches around the main frequency 0.134.
The plain FFT gives, as expected, the worst performance
in identifying the spectral line.
Interpolating with Jacobsen method ~\cite{jacobsen} or zero padding give similar results in this example.
The later version of Harpy~\cite{newharpy} implements zero padding and reduces the computational 
costs by a combination with SVD. Moreover, the combination with SVD allows to estimate errors in the frequency spectra.
\begin{lstlisting}[language=Python]
# First version of Harpy implementing NAFF with Jacobsen interpolation
import numpy as np
PI2I = 2 * np.pi * complex(0, 1)

def harpy(samples, num_harmonics):
    n = len(samples)
    int_range = np.arange(n)
    coefficients = []
    frequencies = []
    for _ in range(num_harmonics):
        frequency = _jacobsen(np.fft.fft(samples), n) # Find dominant freq.
        exponents = np.exp(-PI2I * frequency * np.arange(n))
        coef =  np.sum(exponents*samples)/n    # compute amplitude and phase
        coefficients.append( coef )
        frequencies.append(frequency)
        new_signal = coef * np.exp(PI2I * frequency * int_range)
        samples = samples - new_signal  # Remove dominant freq.
    coefficients, frequencies = zip(*sorted(zip(coefficients, frequencies),
        key=lambda tuple: np.abs(tuple[0]), reverse=True))
    return frequencies, coefficients

def _jacobsen(dft, n): # Interpolate to find dominant freq. 
    k = np.argmax(np.abs(dft))
    delta = np.tan(np.pi / n) / (np.pi / n)
    kp = (k + 1) % n
    km = (k - 1) % n
    delta = delta * np.real((dft[km]-dft[kp])/(2*dft[k] - dft[km] - dft[kp]))
    return (k + delta) / n

N=4096
i = 2 * np.pi * np.arange(N)
data = np.cos(0.134 * i) + np.cos(0.244 * i) + 0.01 * np.random.randn(4096)
freqs, coeffs = harpy(data, 300)
\end{lstlisting}

\subsubsection{Phase measurement}
 The phase advance between 2 BPMs $\phi_{ij}=\phi_j-\phi_i$
 is a fundamental optics observable, it is model and BPM calibration independent. Care with averaging several measurements is needed 
 due to periodicity, i.e. circular mean has to be used. 
For $n$ measurements of certain angle or phase, $\alpha_i$,
  the circular mean is defined as
  \begin{equation}
  \overline{\alpha} = {\rm atan2}\left( \frac{1}{n}\sum_i^n \sin\alpha_i\ ,\ \ \frac{1}{n}\sum_i^n \cos\alpha_i   \right)\ ,
  \end{equation}
  and in Python it is simply computed using an existing function
  as shown in the following example code by computing the circular
  mean between 0 and 2$\pi$, which is not $\pi$ but 0.
  \begin{lstlisting}[language=Python]
#Computing the circular mean of 0 and 2pi    
from scipy.stats import circmean
import numpy as np    
circmean([0., 2*np.pi])
\end{lstlisting}

  \subsection{\texorpdfstring{$\beta$}{Beta} from amplitude}

  The average of the $\beta$ function around the ring in presence of random errors is related to the rms $\beta$-beating via the following expression~\cite{average},
  \begin{equation}
\left\langle \frac{\Delta\beta}{\beta}\right\rangle =  {\rm rms}^2\left( \frac{\Delta\beta}{\beta}\right)\ \label{avebetaeq}
\end{equation}

\begin{figure}  \centering
\includegraphics[trim = 0mm 0mm 0mm 0mm, clip,height=6cm, angle=-0]{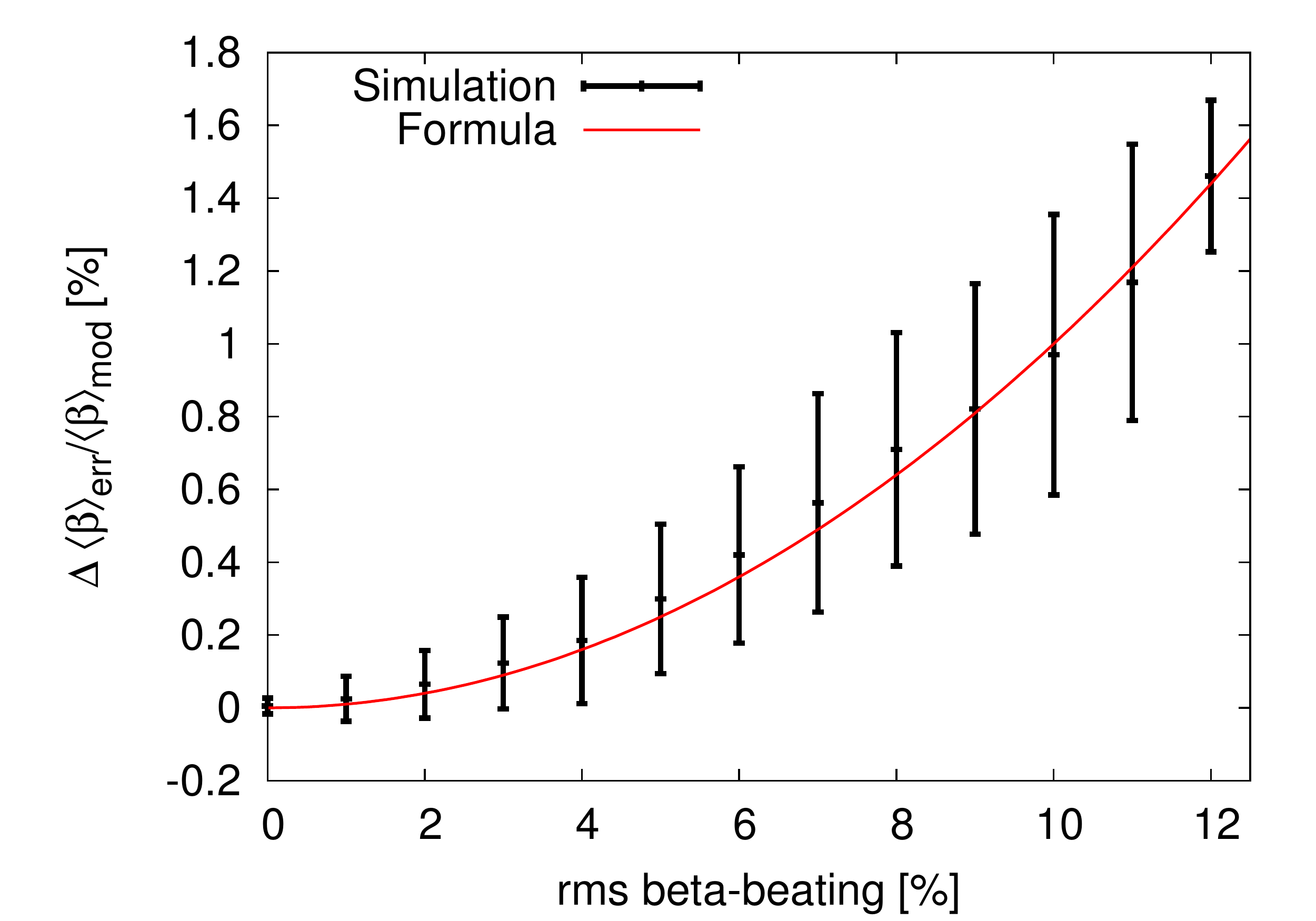}
\caption{Relative deviation of the average $\beta$ function in presence of random errors versus the corresponding $\beta$-beating together with the prediction from Eq.~(\ref{avebetaeq}). \label{fig:ringavebeta}}
\end{figure}

  Figure~\ref{fig:ringavebeta} shows the ring average $\beta$ function
versus its rms value 
for many realizations of the LHC with random errors.
In average random errors increase the beta-functions around
the ring implying that random errors are defocusing.
 
As shown in Eq.~(\ref{xNs}) the amplitude of betatron
oscillation at the location $s$  is $\sqrt{\beta(s)\epsilon}$.
  Having enough BPMs around the ring allows to compute the average and rms of $\beta\epsilon$ from the square of the FFT amplitude of the tune line.
   $\epsilon$ can be computed with
    \begin{equation}
      \epsilon \approx \frac{\left\langle \beta\epsilon \right\rangle}{\left\langle \beta_{\rm model} \right\rangle}\left(1-{\rm rms}^2\left(\frac{\Delta\beta}{\beta}\right)\right) \ ,
    \end{equation}
    where the numerator comes from measurement, the denominator is the
    average model $\beta$ function and the parenthesis corrects
    for the possible average  $\beta$-beating.
    Biggest limitation of this technique is BPM calibration errors.
    After computing $\epsilon$ it is possible to extract the
    $\beta$ function at every BPM using the amplitude of
    the tune line. The main limitation of this method is
    relying on a good gain calibration of BPMs.

 \subsection{\texorpdfstring{$\beta$}{Beta} from phase}
 It is possible to compute the $\beta$ function at one BPM by using
 the phase advances between that BPM and another 2 BPMs as follows\cite{castro},
 \begin{eqnarray}
       &\includegraphics[trim = 25mm 241mm 120mm 15mm, clip,height=2.9cm, angle=-0]{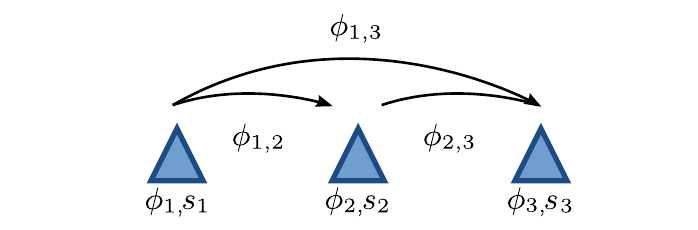}& \nonumber\\
  &\beta_1^{\rm meas}=\beta_1^{\rm mod}\frac{\cot \Delta\phi_{1,2}^{\rm meas}-\cot \Delta\phi_{1,3}^{\rm meas}}{\cot \Delta\phi_{1,2}^{\rm mod}-\cot \Delta\phi_{1,3}^{\rm mod}}& \nonumber \ .
\end{eqnarray}
 This is known as the 3 BPM method, which was later extended
 to N BPMs in~\cite{andy} relying on Montecarlo simulations
 and made fully analytical in~\cite{andreas} with a considerable
 improvement in speed.
    
  \subsection{Momentum reconstruction and resonance driving terms}
  BPMs only measure the centroid position. The angle
  of the trajectory can be computed from two BPMs separated by
  a drift, but there are usually very few BPMs placed in such configuration
  in an accelerator.
  Normalizing the turn-by-turn BPM signal by the amplitude of
  the tune line we define the normalized coordinate $\hat{x}$ , which
  for two nearby BPMs can be parametrized as follows,
\begin{eqnarray}
    \hat{x}_1(N)&=&\cos(2\pi Q_xN+\phi_1) \nonumber\ ,\\
    \hat{x}_2(N)&=&\cos(2\pi Q_xN + \phi_2)\nonumber \ .
\end{eqnarray}
  We can reconstruct the normalized $p_x$ at the first BPM  as
  \begin{eqnarray}
    \hat{p}_{x1}(N)&=& \sin(2\pi Q_xN+\phi_1) =  \frac{\hat{x}_2(N)}{\cos\delta} + \hat{x}_1(N)\tan\delta \ ,
  \end{eqnarray}
  with $\delta= \phi_2-\phi_1-\pi/2$. Note that when the phase advance
  between the 2 BPMs is $\pi/2$ then $ \hat{p}_{x1}(N)=\hat{x}_2(N)$,
  and when the phase advance is $\pi$ the equation diverges.
  $\hat{x}_1$ and $\hat{p}_{x1}$ can be used to plot the particle
  trajectory in the phase space up to a constant.
  Non-linearities deform this trajectories from ellipses
  to possibly very complex shapes.
  Using Normal Form the turn-by-turn motion  can be described in terms
  of the generating function terms $f_{jklm}$ as~\cite{frankS}
  \begin{eqnarray}
    \hat{x}_1-i\hat{p}_{x1}&=& e^{i2\pi Q_x N} -\nonumber\\
        &&   2i\sum j f_{jklm} \epsilon_x^{\frac{j+k-2}{2}}\epsilon_y^{\frac{l+m}{2}} e^{i2\pi N [(1-j+k)Q_x +(m-l)Q_y]+i\varphi}   \nonumber\ .
  \end{eqnarray}
  This equation allows characterizing the non-linear beam dynamics experimentally by  measuring the terms $f_{jklm}$ from the complex Fourier analysis of $ \hat{x}_1-i\hat{p}_{x1}$ as done in~\cite{bnl,ro,cern,andreaF}.

  \section{Farey sequences}\label{sec6}
  The Farey sequence $F_n$ of order n is the sequence of completely reduced fractions between 0 and 1 which, when in lowest terms, have denominators less than or equal to $N$, which corresponds to the resonances of order $N$ or lower (in one plane). The Farey sequence of order 5 is given by
\begin{equation}
F_5=\Big\{ \frac{0}{1} , \frac{1}{5} ,  \frac{1}{4} , \frac{1}{3}, \frac{2}{5},\frac{1}{2}, \frac{3}{5},  \frac{2}{3},\frac{3}{4},  \frac{4}{5},\frac{1}{1}    \Big\}
\end{equation}
Farey sequences have useful properties.
 The distance between neighbors in $F_n$ (aka two consecutive 
      resonances) $a/b$ and $c/d$ is equal to $1/(bd)$.
The next leading resonance in between two consecutive resonances
$a/b$ and $c/d$ is given by the mediant operation between these
two fractions,
  \[\frac{a+c}{b+d}\ .\]
The number of 1D resonances of order $N$ or lower tends asymptotically 
to $3N^2/\pi^2$. The Farey sequence is efficiently computed in Python as follows,
\begin{lstlisting}[language=Python]
# The Farey sequence of order n
def Farey(n):
    """Return the nth Farey sequence, ascending."""
    seq=[[0,1]]
    a, b, c, d = 0, 1,  1  , n 
    while c <= n :
        k = int((n + b)/d)
        a, b, c, d = c, d, k*c - a, k*d - b
        seq.append([a,b])
    return seq
\end{lstlisting}
  
\begin{figure}  \centering
\includegraphics[trim = 0mm 0mm 0mm 0mm, clip,height=8.2cm, angle=-0]{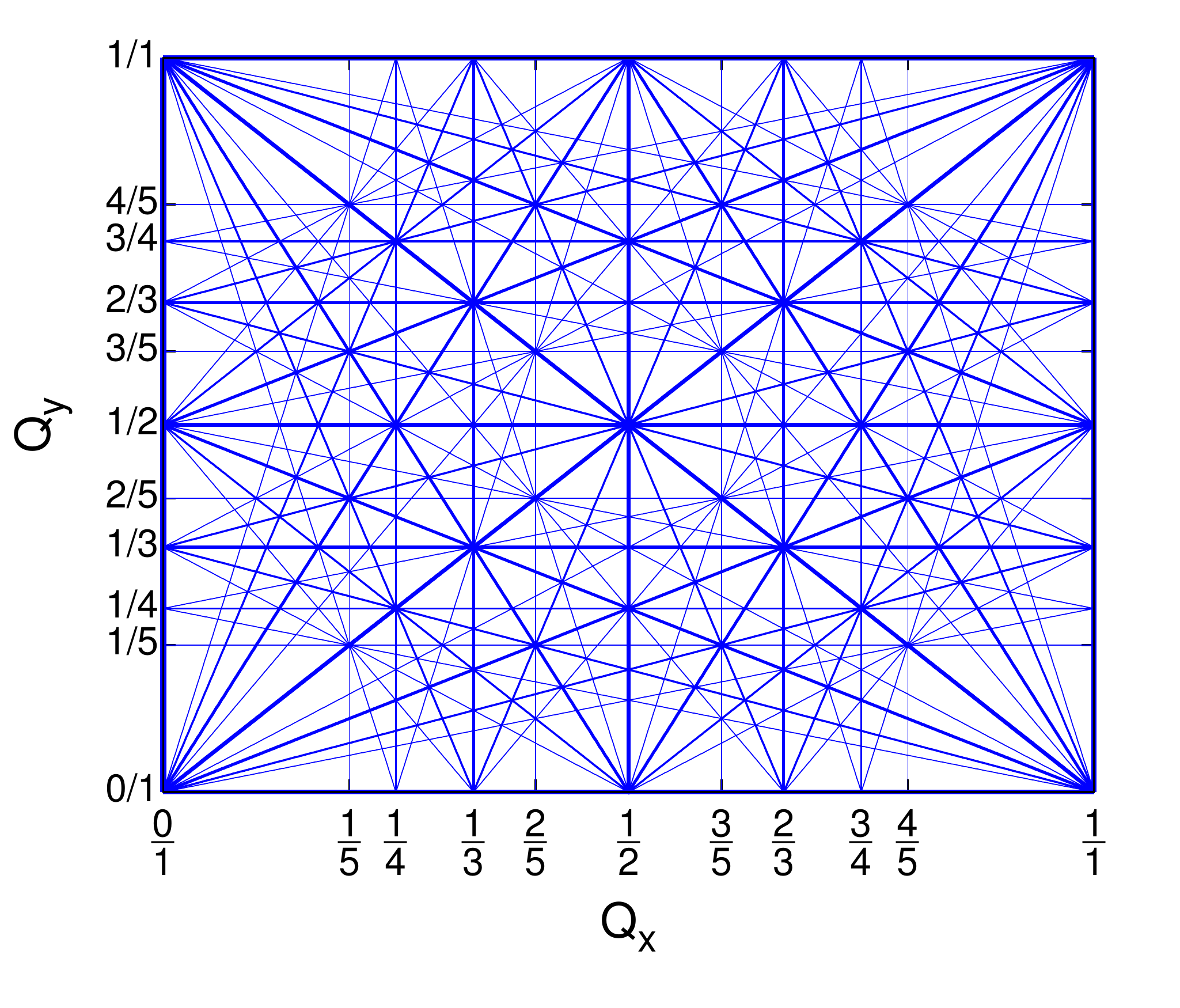}
\caption{Resonance diagram of order 5. \label{resdiag}}
\end{figure}
The 2D tune resonance diagram is defined by all solutions of the
following equation,
\[
a Q_x + b Q_y = p\ ,
\]
with $a$ and $b$   and $p$ integer numbers. These resonance lines
are to be avoided in normal operation as some resonance driving terms
diverge when approaching them. Figure~\ref{resdiag} shows
the resonance diagram of order 5. The resonance diagram is also connected
to the Farey sequence.
The lines going trough $Q_x=\frac{h}{k}$, $Q_y=0$ relate to the elements
  in $F_N$ between 0 and $\frac{1}{k}$~\cite{farey}.
The number of resonance lines in the 2D diagram is~\cite{asympt}
  \begin{equation}
\frac{2N^3}{3\zeta(3)} + O\left(\frac{N^3}{\log N}\right)\ ,
\end{equation}
  where $\zeta(3)\approx 1.20205$ is the Riemann zeta function evaluated at 3.
  The  relation between the 2D resonance lines and the Farey sequence is most easily explained in the following code example to
  plot the resonance diagram.
  \begin{lstlisting}[language=Python]
# Plotting the 2D resonance diagram with Farey sequences
import matplotlib.pyplot as plt
import numpy as np
fig = plt.figure()
ax = plt.axes()
plt.ylim((0,1))
plt.xlim((0,1))
x = np.linspace(0, 1, 1000)
FN = Farey(5)   # Farey function defined in the previous code example
for f in FN:
    h , k = f      # Node h/k on the axes
    for sf in  FN:
        p , q = sf    
        c=float(p*h)
        a=float(k*p)    # Resonance line  a Qx + b Qy = c linked to p/q  
        b=float(q-k*p)
        if a>0:
            plt.plot(x, c/a - x*b/a, color='blue')
            plt.plot(x, c/a + x*b/a, color='blue')
            plt.plot(c/a - x*b/a, x, color='blue')
            plt.plot(c/a + x*b/a, x, color='blue')
            plt.plot(c/a - x*b/a, 1-x, color='blue')
            plt.plot(c/a + x*b/a, 1-x, color='blue')
        if q==k and p==1:   # FN elements below 1/k
            break
plt.show()
\end{lstlisting}

\begin{figure} \centering   
   \includegraphics[height=8.2cm, trim= 35mm 5mm 31mm 5mm,clip,angle=-0]{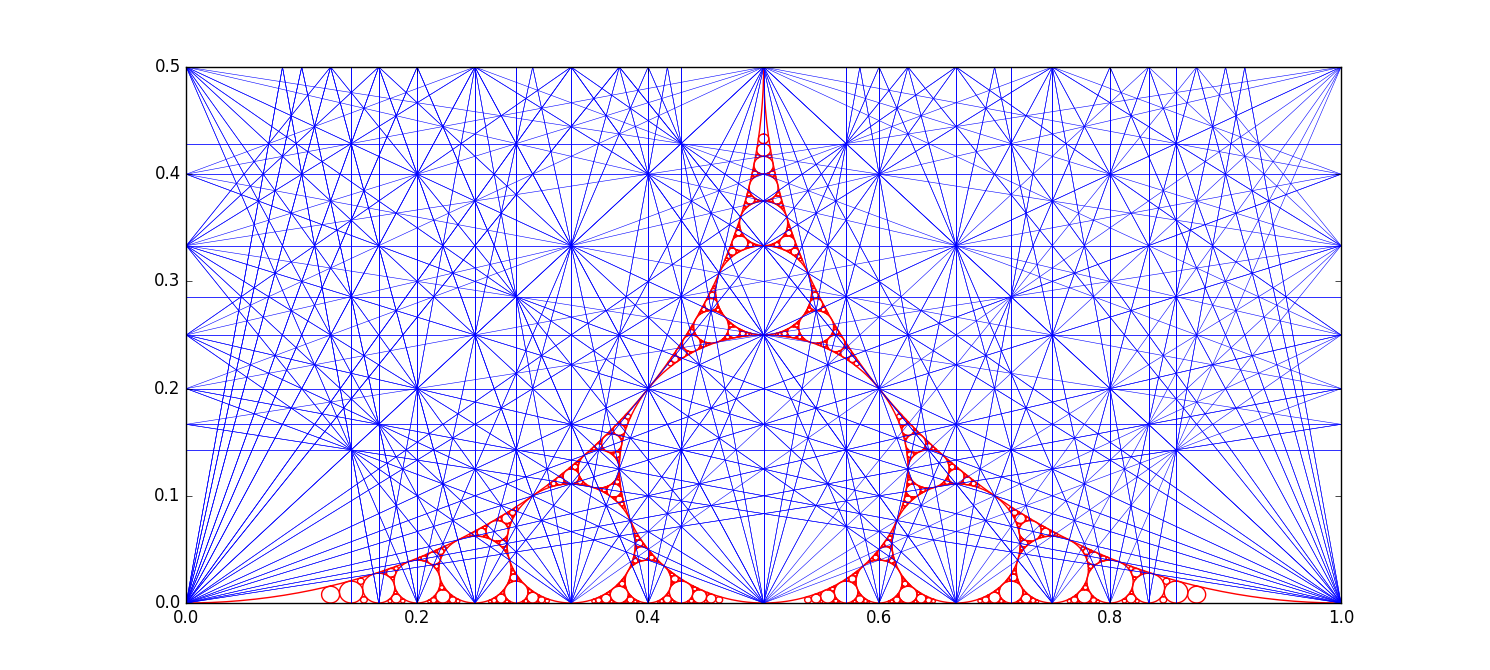}
   \caption{Lower half of the resonance diagram and the Apollonian gasket (0,0,1,1). \label{apollo}}
   \end{figure}

 The resonance diagram has also intriguing connections to
 the Apollonian gasket (0,0,1,1) as shown in Fig.~\ref{apollo}.

 \section{Corrections}\label{sec7}
 
 The goal of corrections is to bring machine
 optics parameters as close as possible to the design
 values to ensure machine safety and performance.
 Corrections are classified as local and global as
 discussed below.

 \subsection{Local correction}
 Local corrections are restricted
 within a predefined segment of the machine. They ensure
 that the perturbations from the errors within the segment
 are confined in the segment without significant leakage to
 the rest of the machine.

 The most effective local correction is identifying the source
 and fixing it. However this can only be applied exceptionally.

 In hadron colliders, it is fundamental to perform local corrections in the interaction regions. Two techniques have successfully demonstrated
these local corrections: action and phase jump~\cite{cardona,cardona2} and segment-by-segment~\cite{1st,2nd}.

\subsection{Global corrections}
Global corrections use distributed magnets around the ring
to minimize deviations of optics parameters from design values.
The simplest global corrections use a predefined set of magnets
to control a single optics parameters without affecting the others.
The set of magnets together with their strength variation is named
as {\it knob} and it is computed using the ideal optics design.
Precomputed knobs are primarily used to control orbit, tune and coupling
deviations.

The most general correction approach is based on a response
matrix between the available correctors and the optics parameters
to correct. For efficient use, the measured values are weighted by their errors as well as by quantity-based weights~\cite{journey}.
Phase beating, $\beta$-beating, dispersion deviations and tune errors 
can be put in a vector connected to the normal quadrupole gradient
changes $\vec{k}$ via the matrix $\mathbf{P_{\rm theo}}$.
In~\cite{claindepe} it is shown that using the normalized dispersion ($D_x/\sqrt{\beta_x}$) in the calculation of corrections improves the correction performance. 
Coupling resonance driving terms
and vertical dispersion connect to skew quadrupole changes $\vec{k}_{s}$
via the matrix $\mathbf{T_{\rm theo}}$. These two relations are given
in the following equations,
\begin{eqnarray} \nonumber
\left(\begin{array}{c}
         \Delta\vec{\phi}_x\\ 
         \Delta\vec{\phi}_y\\
         \frac{\vec{\Delta{\beta}_x}}{\beta_x}\\ 
         \frac{\vec{\Delta{\beta}_y}}{\beta_y}\\
         \Delta\vec{D}_x\\
         \Delta\vec{Q}
      \end{array}\right)_{\rm meas}
=\mathbf{P_{\rm theo}}\Delta\vec{k}
\ \ , \ \ \  \ \ \ \ \ \ \ \ \ \ \
\left(\begin{array}{c}
         \vec{f}_{1001}\\ 
         \vec{f}_{1010}\\
         \vec{D}_y\\
      \end{array}\right)_{\rm meas}
=\mathbf{T_{\rm theo}} \Delta\vec{k}_s \ .
\end{eqnarray}
$\mathbf{P_{\rm theo}}$ and   $\mathbf{T_{\rm theo}}$
can be computed by varying one gradient strength at a time and computing 
the new optics parameters or by collecting large statistics to train 
the linear regression model~\cite{elenaIPAC19corr}.
$\mathbf{P_{\rm theo}}$ and   $\mathbf{T_{\rm theo}}$ are pseudo-inverted
to compute corrections from the measured optics deviations.

\subsection{The best N corrector problem}
The best N corrector problem consists in finding the best $N$ correctors
among the full set of $M$ correctors, with $M>N$. This is very useful when 
the correction has some cost which increases with the number of correctors 
or when we aim to localize the error.
It is likely that the best 1 corrector is near the error source, although there
is no guarantee.
The time required  to find the exact solution to this problem  scales rapidly with $M$ and $N$ as
\[
t \propto \frac{M!}{(M-N)!}\ ,
\]
as all possibilities have to be explored and compared.
An approximation algorithm to solve this problem is known as Micado~\cite{mikado} by iteratively
finding a new best corrector in each step, starting
by finding the best 1 corrector, then finding the second best corrector keeping
the first one and so on.
This problem is also considered in signal theory as a way to decompose 
signals into a weighted sum of finitely many functions. The algorithm to solve
it is very similar to Micado and it is known as
matching pursuit or orthogonal matching pursuit (OMP)~\cite{OMP}.

\begin{figure}\centering
  \includegraphics[trim = 0mm 0mm 0mm 0mm, clip,height=7.8cm, angle=-0]{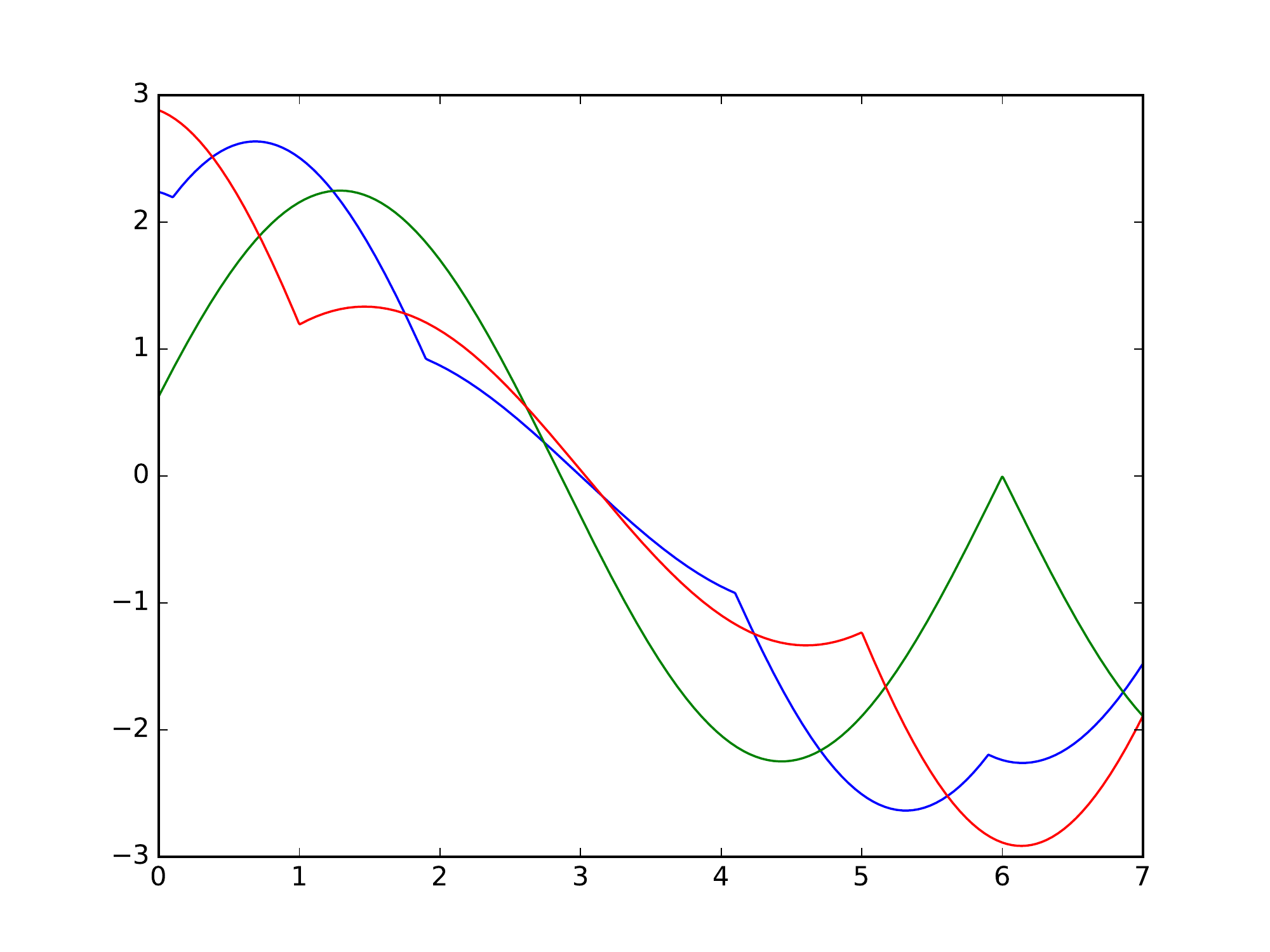}\put(-98.5,200){\color{red}*}\put(-229,200){\color{red}*}\put(-66,207){\color{green}*}\vspace*{-.2cm}
  \caption{Measured orbit versus longitudinal location (blue) together
    with resulting orbit using the best 1 corrector (green) and best 2 correctors (red). Location of best correctors is shown with the corresponding color. The best 1 corrector is not within the best two correctors.\label{bestNcorr}}
  \end{figure}

The following Python code implements the exact solution of the best $N$ corrector problem 
with 7 available correctors and for all values of $N$ below 7. 
The orbit generated by the $i^{\rm th}$ corrector is approximated by $\sin(|x-x_i|)$
and the target orbit to match is defined in the $measured\_orbit(x)$ function. 
\begin{lstlisting}[language=Python]
# Exact solutions of the best N corrector problem
import numpy as np
from scipy.optimize import least_squares
from itertools import product
import matplotlib.pyplot as plt

N_corrs=7
s=np.linspace(0, N_corrs, 1000)   # 1000 observation points

def corrs(x,i):   # Assume correctors at i=integer < N_corrs
    return np.sin(np.abs(x-i))

def model(x, c):  # Orbit at x from corrector strengths as c
    if len(x)==1:
        return sum(c*corrs(x,np.arange(N_corrs)))
    return [model([y],c) for y in x ]

def measured_orbit(x):  # Target Orbit
    return np.sin(np.abs(x-0.1)) + np.sin(np.abs(x-1.9)) - np.sin(np.abs(x-4.1)) - np.sin(np.abs(x-5.9))
    
def f(c):      #Figure of merit for given corrector choice encoded in mask 
    return  model(s, c*mask) - measured_orbit(s) 

best=1e16*np.ones(N_corrs+1) ; bestmask=np.zeros([N_corrs+1,N_corrs])
for mask in product([0,1],repeat=N_corrs):  # Try all corrector combinations
    res = least_squares(f, x0=np.ones(N_corrs))  #Orbit correction
    if res.cost < best[sum(mask)]:
        bestmask[sum(mask)]=mask*res.x ; best[sum(mask)]=res.cost
        
plt.plot(s, measured_orbit(s))
plt.plot(s, model(s, bestmask[1]))   #Plot best 1 corrector
plt.plot(s, model(s, bestmask[2]))   #Plot best 2 correctors
\end{lstlisting}
Figure~\ref{bestNcorr} shows the measured orbit together with the orbit generated by the best 1 and 2 correctors.
The problem has been chosen to show that in the exact solution the best 1 corrector is not necessarily within the
2 best correctors.

\begin{figure}[!tbh]\centering
\includegraphics[trim = 0mm 0mm 0mm 0mm, clip,height=7.8cm, angle=-0]{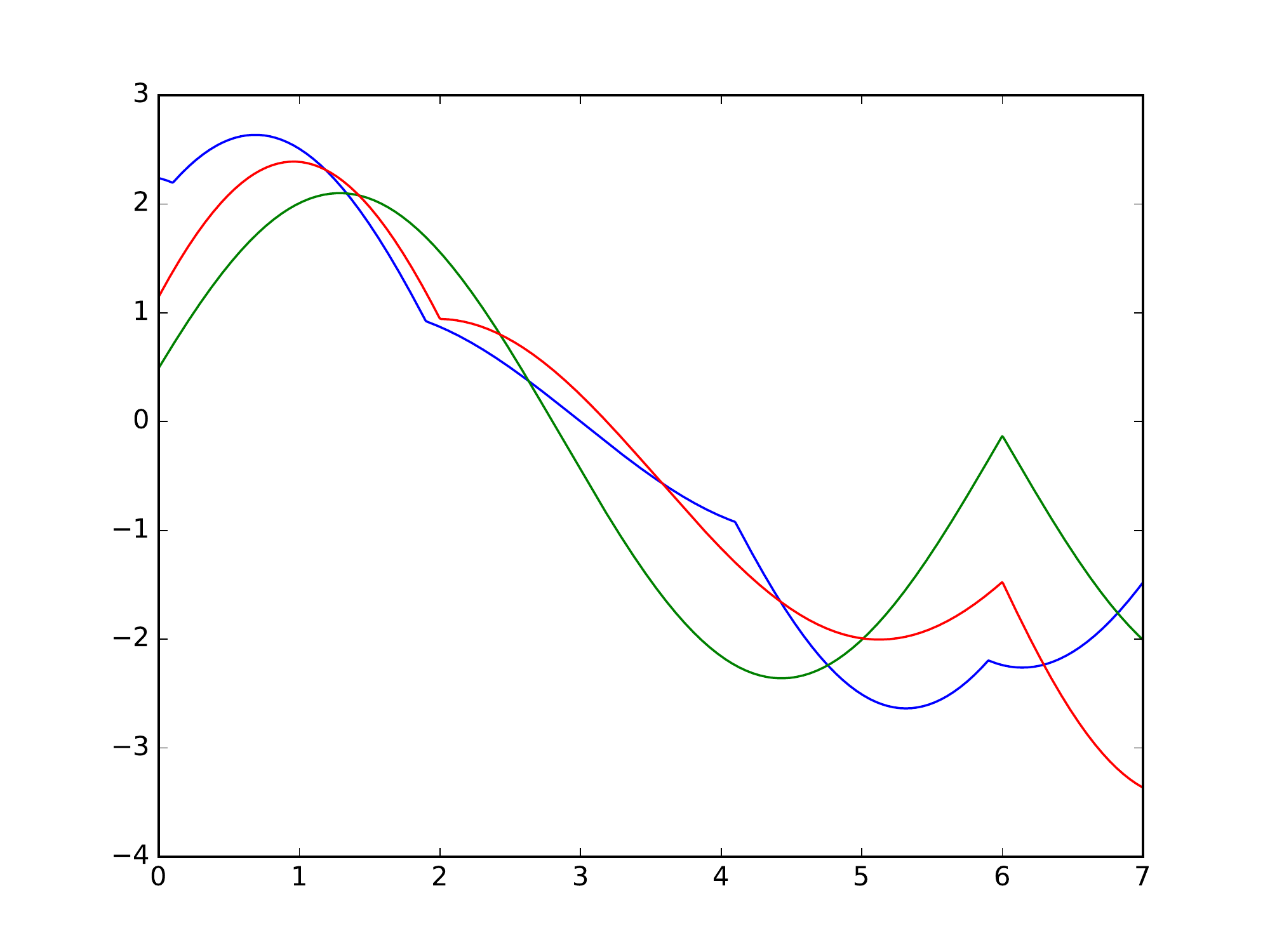}\put(-66,200){\color{red}*}\put(-196,200){\color{red}*}\put(-66,207){\color{green}*}
 \caption{Measured orbit versus longitudinal location (blue) together
    with resulting orbit after using the best 1 corrector (green) and  best 2 correctors (red) using Orthogonal Matching Pursuit. Location of best correctors is shown with the corresponding color. The best 1 corrector is within the best two correctors as this is the main approximation of the algorithm.\label{bestNcorrOMP}}
  \end{figure}

The following Python code solves the same problem as above by implementing 
the OMP algorithm using existing Python libraries.
The OMP results are shown in Fig.~\ref{bestNcorrOMP}, to be compared to the previous exact solution in Fig.~\ref{bestNcorr}.
Now the best 1 corrector is included in the best 2 correctors, as this solution is only an approximation. 
\begin{lstlisting}[language=Python]
# Best N corrector problem with Orthogonal Matching Pursuit
from sklearn.linear_model import OrthogonalMatchingPursuit
import numpy as np
import matplotlib.pyplot as plt

N_corrs=7
N_BPMs=1000
s=np.linspace(0, N_corrs, N_BPMs)  # 1000 BPMs

def corrs(x,i):       
    return np.sin(np.abs(x-i))   

def measured_orbit(x):
    return np.sin(np.abs(x-0.1)) + np.sin(np.abs(x-1.9)) - np.sin(np.abs(x-4.1)) - np.sin(np.abs(x-5.9))

################   New part for OMP ###############  

X=[]
for  i in range(N_BPMs): # Prepare response matrix for OPM
    X.append(corrs(s[i],np.arange(N_corrs)))
y= measured_orbit(s) 
reg = OrthogonalMatchingPursuit(n_nonzero_coefs=1).fit(X, y)  #Run OMP for best 1 corr
print reg.coef_  # coefficient of best 1 corr 
plt.plot(s, reg.predict(X))
\end{lstlisting}

\end{document}